\documentclass[fleqn]{aa}
\usepackage{ifpdf}
\ifpdf
  \usepackage[pdftex]{graphicx}
\else
  \usepackage{graphicx}
\fi
\usepackage[varg]{txfonts}
\usepackage[sort]{natbib}

\bibpunct{(}{)}{;}{a}{}{,}

\newcommand{\kms}{km\,s$^{-1}$}
\newcommand{\Msun}{M$_{\sun}$}

\newcommand{\Lsun}{L$_{\sun}$}

\newcommand{\Mdot}{M$_{\sun}$\,yr$^{-1}$}

\newcommand{\hb}{H$\beta$}
\newcommand{\ha}{H$\alpha$}

\newcommand{\teff}{T_\mathrm{eff}}

\newcommand{\nii}{[N\,{\sc ii}]}
\newcommand{\oii}{[O\,{\sc ii}]}
\newcommand{\oiii}{[O\,{\sc iii}]}

\defcitealias{perinotto.04}{Paper~I}
\defcitealias{schoenetal.05a}{Paper~II}
\defcitealias{schoenetal.05b}{Paper~III}
\defcitealias{schoenetal.07}{Paper~IV}
\defcitealias{kings.94}{KB94}
\def\changed{}
\newcommand{\changedIII}{}

\begin{document}
\title{The evolution of planetary nebulae}
\subtitle{VII. Modelling planetary nebulae of distant stellar systems}

\titlerunning{The evolution of planetary nebulae. VII.}

\author{D. Sch\"onberner \and R. Jacob \and C. Sandin \and M. Steffen}

\institute{Astrophysikalisches Institut Potsdam, An der Sternwarte 16, 14482 Potsdam,
           Germany\\
\email{deschoenberner@aip.de, rjacob@aip.de, csandin@aip.de, msteffen@aip.de}}

\offprints{M. Steffen}

\date{Received 6 October 2009 / Accepted 30 August 2010}

\abstract{}
   {By means of hydrodynamical models we do the first investigations of how the 
    properties of planetary nebulae are
    affected by their metal content and what can be learned from spatially
    unresolved spectrograms of planetary nebulae in distant stellar systems.
    }
   {We computed a new series of 1D radiation-hydrodynamics planetary nebulae model 
    sequences with central stars of 0.595~\Msun\ surrounded by initial envelope structures 
    that differ only by their metal content.  
    At selected phases along the evolutionary path, the hydrodynamic terms 
    were switched off, allowing the models to relax for fixed radial structure
    and radiation field into their equilibrium state with respect to energy 
    and ionisation.  The analyses of the line spectra emitted from both the 
    dynamical and static models enabled us to systematically study
    the influence of hydrodynamics as a function of metallicity and evolution.  
    We also recomputed selected sequences already used in previous publications, but now
    with different metal abundances.  These sequences were used to study the expansion
    properties of planetary nebulae close to the bright cut-off of the planetary nebula
    luminosity function.
    }
   {Our simulations show that the metal content strongly influences the expansion of planetary nebulae: 
    the lower the metal content, the weaker the pressure of the stellar wind bubble,
    but the faster the expansion of the outer shell because of the higher electron temperature. 
\changed{This is in variance with the predictions of the interacting-stellar-winds 
         model (or its variants) according to which only the central-star wind is thought
         to be responsible for driving the expansion of a planetary nebula.}
   Metal-poor objects 
\changed{around slowly evolving central stars} 
    become very dilute and are prone to depart from thermal equilibrium because 
\changed{then adiabatic expansion contributes to gas cooling.}   
\changed{We find indications that photoheating and line cooling are not fully balanced
         in the evolved planetary nebulae of the Galactic halo.}
    Expansion rates based on widths of volume-integrated line profiles 
    computed from our radiation-hydrodynamics models compare very well  
    with observations of distant stellar system. 
    Objects close to the bright cut-off of the planetary nebula luminosity function consist 
    of rather massive central stars ($>\! 0.6$~\Msun) with optically thick (or nearly thick) 
    nebular shells.  The half-width-half-maximum velocity during this bright phase 
    is virtually independent of metallicity, as observed, but somewhat depends on the final 
    AGB-wind \changed{parameters}.  
   }
   {The observed expansion properties of planetary nebulae in distant stellar systems with
    different metallicities are explained very well by our 1D radiation-hydrodynamics models.
    This result demonstrates convincingly that the formation and acceleration of a planetary
    nebula occurs mainly because of ionisation and heating of the circumstellar matter by the 
    stellar radiation field, and that the pressure exerted by the shocked stellar wind is 
    less important. 
\changed{Determinations of nebular abundances by means of photoionisation modelling may 
         become problematic for those cases where expansion cooling must be considered.}
   }


\maketitle

\section{Introduction}\label{intro}
  The use of planetary nebulae (PNe) has become an important tool for
  investigating the properties of unresolved stellar populations in galaxies.
  Applications are kinematic studies for probing the gravitational
  potential, luminosity functions for distance estimates,
  and, most importantly, abundance determinations by means of plasma diagnostic
  or photoionisation modelling.

  It has not been thoroughly investigated to date
  whether results based on PNe are generally trustworthy.
  For instance, by applying standard plasma diagnostics, it is
  implicitly assumed that a PN is a homogeneous object in which all
  physical processes are in equilibrium.  In reality, however, a PN is a
  highly structured expanding gas shell, subject to the dynamical pressure
  of a fast central-star wind and the thermal pressure of the photo-heated gas.
  Only if radiation processes dominate heating and cooling of the gas
  hydrodynamics can be neglected.   The questions whether and by how much
  PNe properties, and especially thermal equilibrium, depend on their metal content,
  has not been posed yet.

  The recent progress in modelling the evolution of PNe by means of
  radiation-hydrodynamics opens the possibility of a better appreciation
  of time-dependent effects.   \citet{marten.93a,marten.95}
  demonstrated that non-equilibrium may become
  important for rapidly expanding, low-density nebular shells, and also for
  cases with quickly evolving central stars. Well-known examples
  are the hot PN haloes that
  are extremely out of thermal and ionisation equilibrium \citep{marten.93b, sandetal.08}.

  Later \citet{perinotto.98} used hydrodynamical models to check the
  influence of hydrodynamics on the abundance determination by plasma diagnostics.
  The authors found no significant non-equilibrium effects, but note that large
  abundance errors may occur in more advanced evolutionary stages
  because of obviously inappropriate ionisation
  correction factors.  However, since the number of model sequences used by the
  authors was only small and restricted to a metallicity typical for Galactic disk
  objects, the case concerning the reliability of abundance
  determinations by plasma diagnostics is far from being settled.  
  In a forthcoming paper we will discuss the reliability of ionisation correction factors 
  presently used in more detail by means of the hydrodynamical simulations
  presented in this paper (Sch\"onberner et al., in prep.).

  While density and velocity structures of planetary nebulae are clearly the
  result of the dynamics, the case is different for the electron temperature.
  Deviations from the equilibrium value may occur by dynamical effects,
  i.e.\ by local gas compression/expansion, and/or by radiative effects, i.e.\ by
  heating/cooling and ionisation/recombination.   The radiative effects occur either
  by rapid stellar evolution and/or by changes of the optical depth \citep{marten.95}.
  Since temperature deviations from its equilibrium value cannot unequivocally be
  attributed to only one of these physical processes, we simply speak of
  ``non-equilibrium'' effects.

  To date, existing radiation-hydrodynamics simulations used the
  typical chemical composition of Galactic disk PNe.
  In the context of the grown interest in extragalactic studies, also of PNe
  populations, it is of great importance to investigate how the evolutionary 
  properties of PNe will change if they originate from a metal-poor population.

  The metal content is expected to influence the PN evolution  twofold: 
  Firstly, the cooling
  efficiency of the gas decreases with metallicity, leading to a higher electron
  temperature.  Secondly, the strength of the central-star wind decreases with 
  metallicity because of the reduced line opacity which drives the wind. 
  PNe in stellar populations with 
  low metal content are thus expected to expand faster than their Galactic disk 
  counterparts since the expansion speed is roughly proportional
  to the sound speed, i.e.\ to the square root of the electron temperature
  \citep[][hereafter Paper II]{schoenetal.05a}.
  Together with a reduced wind power, metal poor PNe will be substantially
  more dilute than their Galactic counterparts and may thus quickly run into 
  non-equilibrium since the line cooling efficiency of the gas depends on the
  density squared and may become less important for the nebular energy budget than 
  expansion cooling. 
  The radiation field of the central star, and to some extend also its luminosity,
  are in principal metal dependent, too.  Both effects are usually not considered and
  are neglected here for simplicity.´

  Already from the sample of planetary nebulae of the Galactic halo one can get
  observational hints that low-metallicity PNe are prone to deviate from thermal
  equilibrium. This sample comprises objects in different evolutionary stages with
  metallicities below solar at various degrees.  The study of halo PNe appears very 
  promising since they can be used as proxies for extragalactic nebulae which are 
  generally too faint and too small for spatially resolved investigations 
  with high spectroscopic resolution.

  \citet{howard.97} analysed the then known 9 PNe
  of the Galactic halo in a very homogeneous way by collecting all the available
  line data from the literature and by fitting the spectra with photoionisation
  models 
\changed{such that ``\ldots\ a reasonable match between model and observations is
         achieved".}
  Closer inspection of the results listed in
  \citet[][Table~2 therein]{howard.97} reveals clearly that the objects 
  with the highest electron temperatures  
  show a remarkable inconsistency between the observed and predicted ratios of the
  temperature sensitive  [\ion{O}{iii}] lines,
\changed{i.e. for these cases the procedure of a general line-strength fitting 
         from the ultraviolet to the optical spectral range did 
         not lead to a consistent determination of a (mean) electron temperature!}

\begin{figure}[t]
\vskip -4mm
\includegraphics[width= \linewidth]{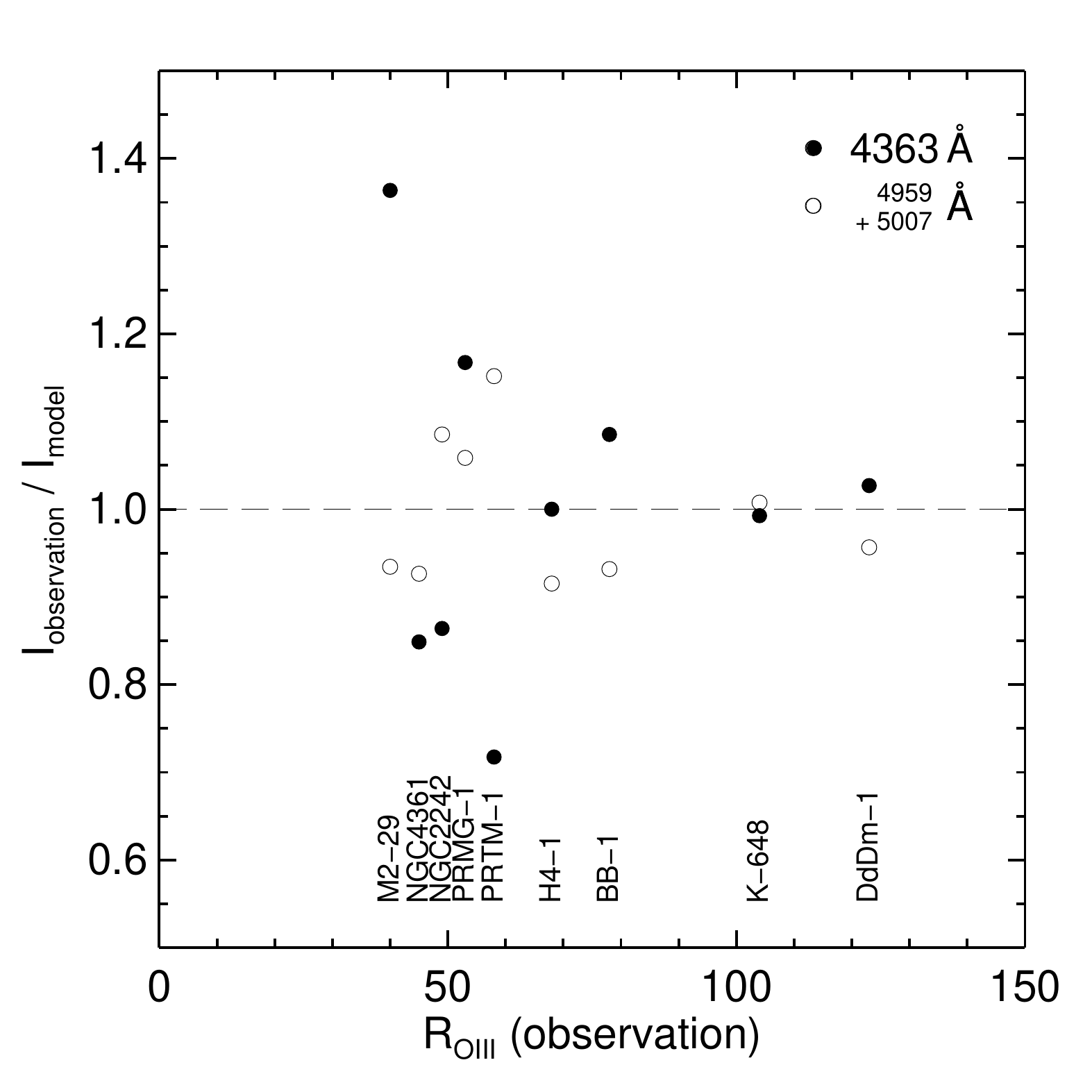}
\vskip-2mm
\caption{\changed{Ratio of observed and modelled strengths of relevant \oiii\ lines
         (indicated in the legend) vs. the temperature sensitive quantity 
         ${R_{\rm O\,III}=(I_{5007}+I_{4959})/I_{4363}}$ for the 9 Galactic Halo PNe
         investigated by \citet{howard.97}. }
}
\label{roiii}
\end{figure}

\changed{The case is illustrated in
        Fig.\,\ref{roiii} where we plotted the ratio of observed and modelled \oiii\
        lines against observed values of the temperature indicator 
        ${R_{\rm O\,III}=(I_{5007}+I_{4959})/I_{4363}}$.
        We see that for lower electron temperatures (larger $R_{\rm O\,III}$) the 
        photoionisation models are able to reproduce the observed \oiii\ line ratios 
        rather well within the 10\,\% error level.} 
  However, for ${R_{\rm O\,III}\la 60}$ (or ${T_{\rm e} \ga 16\,000}$~K)
  discrepancies between observations and predictions become very large. 

\changed{These differences are especially seen for the temperature sensitive 
         $\lambda$~4363 \AA\ line which is off by up to 30--40\,\% for 
         ${R_{\rm O\,III}\la 60}$.}\footnote
{\changed{We have no explanation why the modelled $\lambda$ 4363 \AA\ line is 
          sometimes stronger and sometimes weaker than the observed 
          line. It is probably the outcome of the fitting procedure.}}  
  We interprete \changed{this behaviour} at low $R_{\rm O\,III}$ as a clear indication for 
  the failure of photoionisation modelling because of \changed{a possible} breakdown 
  of thermal equilibrium at low metallicities.%

  Concerning the expansion properties of PNe and their expected dependence on 
  metallicity, the present situation is also unclear.  \citet{richer.06} studied
  the expansion behaviour of PNe in Local Group galaxies and could not find any
  significant dependence on the metal content:  the brightest PNe always have,
  on the average, a HWHM velocity (in the \oiii\ $\lambda$5007 \AA\ line) of
  ${\simeq\!18}$ \kms\ \citep[see also][]{richer.07}.  A similarly low value 
  (16.5 \kms) has recently
  been found for the brightest PNe in the Virgo cluster by \citet{arnaboldi.08}.
  This finding is puzzling since theory predicts, as outlined above, a clear
  correlation of the PN's speed of expansion with metallicity.

  Because of the importance of planetary nebulae 
  for extragalactic studies, we decided to investigate  
  how planetary nebulae evolution depends on
  metallicity.  In particular, we are interested in the expansion velocities of the
  brightest PNe and how these correlate with the metallicity of the parent stellar
  population, and in the question under which conditions standard photoionisation models 
  may fail. 

  We start with
  describing the setup of our 1D radiation-hydrodynamic models in the next section,
  and continue in Sect.~\ref{results} with a detailed presentation of our
  results.  In Sect. \ref{distant} we use our models to interprete the
  expansion velocities observed in distant stellar systems in terms of metallicity 
  and evolutionary stage.  
  The paper is concluded with Sects. \ref{disc} and \ref{concl}.
  Part of the results presented in Sect. \ref{results} can be found 
  in \citet{schoenetal.05c}.

\section{The models}\label{models}
   Our method of modelling the evolution of planetary nebulae by means of 1D
   radiation-hydrodynamics has been documented in previous publications and
   shall not be repeated here \citep[see][and references therein; henceforth called
   Paper I]{perinotto.04}. 
   A very detailed comparison of our models with observed 
   nebular structures can be found in \citet{SS.06}.
   We add only that all hydrodynamical sequences used in this work are based on a 
   new parallelised version of the code which also includes thermal conduction  
   as described in detail by \citet{SSW.08}.  
   Already existing sequences have been recomputed with this
   new version in order to achieve a homogeneous set of evolutionary sequences.
   We emphasise that, once the initial envelope structure and the central star,
   whose radiation field and wind represent the time-dependent inner boundary
   conditions, have been selected, the whole evolution is consistently determined
   by the hydrodynamics.  We note that all the central-star models used in
   this work burn hydrogen, which means that the case of nebulae around 
   hydrogen-deficient, i.e. Wolf-Rayet, central stars is thus \emph{not} 
   considered here. 
  
   Concerning the elements to be considered in the nebular envelopes, we included,
   next to hydrogen and helium, only the most important coolants, i.e.\ carbon, 
   nitrogen, oxygen, neon, sulphur, chlorine, and argon.
   For each individual element up to 12 ionisation stages are taken into account.
   We emphasise here that ionisation, recombination, heating and cooling are treated
   fully time dependently, and the (radiative) cooling function for each volume
   element is composed of the contributions of all individual ions
   \citep[see also][]{marten.97}.  Our reference abundance distribution 
   for the Galactic disk PNe, $Z_{\rm GD}$, is listed in Table \ref{tab.element}. 
   These abundances are the same as those used in our previous hydrodynamical 
   computations and are, apart from carbon and nitrogen, very close to the 
   most recent solar values \citep[see, e.g.,][]{asplund.09}.

\begin{table}[t]           
\caption{Distribution of the chemical abundances typical for Galactic disk PNe.
         The abundances, $\epsilon_i$, are given in the usual manner as (logarithmic)
	 number fractions relative to hydrogen, i.e.
	 $\epsilon_i = \log\, (n_i/n_{\rm H}) + 12$.}
\label{tab.element}
\centering
\tabcolsep=5.8pt
\begin{tabular}{ccccccccc}
\hline\hline\noalign{\smallskip}
  H    &    He  &   C    &   N   &    O   &   Ne  &   S   &   Cl   &   Ar  \\[2pt]
\hline\noalign{\smallskip}
 12.00 &  11.04 &  8.89  &  8.39 &  8.65  &  8.01 &  7.04 &  5.32  &  6.46  \\[2pt]
\hline
\end{tabular}
\end{table}

  We considered in general six cases with scaled Galactic disk abundances distributions,
  $Z_\mathrm{GD}$, of C, N, O, Ne, S, Cl,
  and Ar, covering the range from ${Z=3\,Z_\mathrm{GD}}$ 
  to ${Z=Z_\mathrm{GD}/100}$ in steps of 0.5 dex. 
  For simplicity, we did \emph{not} 
  consider metallicity-dependent variations of abundance ratios in this pilot study.
\changed{The range of metal contents considered here covers the observed degrees of
         metallicities in galaxies: While giant galaxies have metal contents within a
         factor of 3 of the solar one, dwarf galaxies have less metals, usually about 
         1/10 of the solar case.  The metal-poorer sequences are thought to demonstrate 
         non-equilibrium effects more clearly.  They are useful for interpreting
         objects with very low metallicity, like, for instance, the Galactic halo object 
         \object{PN\,G135.9+55.9} which has a mean metal content below 1/10 solar 
         \citep[][]{PT.05, stasetal.10, sandetal.10}.
        }

  The wind is treated in the same manner as originally devised by \citet{MS.91}, but
  its dependence on 
  metallicity, more correctly on the elements C, N, and O, is approximately
  accounted for by simple correction factors, i.e.\ ${\dot{M}\propto Z\,^{0.69}}$
  \citep{VKL.01} and $v_\infty \propto Z\,^{0.13}$ \citep{LRD.92}.  The outcome is  
  a nearly linear dependence of the wind luminosity with metallicity,
  $L_\mathrm{wind}\equiv 0.5\,\dot{M}\,v^2_\infty \propto Z\,^{0.95}$.  

  The surface of
  the star is always assumed to radiate like a black body, i.e. a dependence of the
  radiation field on the chemical composition of the stellar atmosphere is, 
  for simplicity, not considered here.  The influence of the metal content on details
  of the stellar evolution is expected to be small and is thus also neglected.  
  Throughout this pilot study we employed for all metallicity cases our standard set 
  of central-star models already used in our previous work.
  For the wind power calculations, the central star is assumed to have the same
  metal content as the nebular gas. 

  In every case we used the same initial envelope structures, independent of their 
  metal content, thereby ignoring the possible influence of the metallicity on 
  final AGB mass-loss rates and wind velocities.  Our present observational and
  theoretical knowledge about the 
  AGB wind properties as a function of metallicity is still too meager as to allow
  a more detailed description of initial configurations in terms of their metal
  content, although progress, 
\changed{theoretical and observational, is encouraging \citep{woodetal.92, loonetal.05,
         wachteral.08, sand.08, maretal.04, mattetal.08, sandetal.09}. 
         There are clear indications that the wind speed is somewhat reduced at lower
         metallicity, but the dependence of the mass-loss rate on metallicity, and
         on stellar parameters, remains still unclear.}  
\changed{It must also be noted that the particular evolutionary stage during which
         the remnant leaves the AGB, which is also the only relevant here, is not
         covered by any of these studies.}

\begin{figure}[!t]
\includegraphics[bb= 0.6cm 3.0cm 7.8cm 6.5cm, width=\linewidth, clip=true]
                {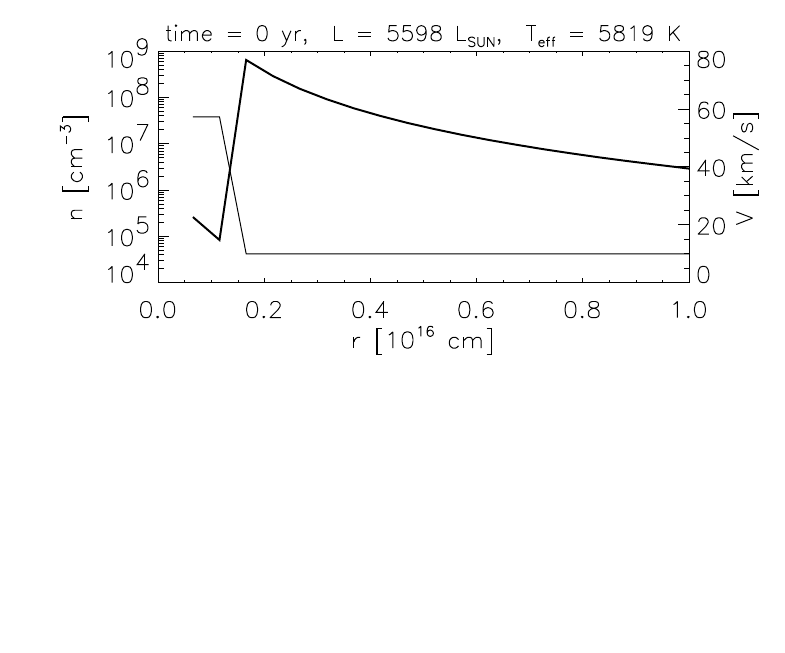}
\caption{Radial run of ion density (thick) and gas velocity (thin)
         for an initial nebular configuration with ${\alpha=3}$ density profile 
         at age zero.  The outer boundary is
         at $r=2.8\times 10^{18}$~cm (outside the graph).
         The stellar parameters are given at the panel's top and refer to the
         0.595~\Msun\ post-AGB track shown in Fig.\,\ref{schoen.star}.
         Density and velocity at $r=1.0\times10^{16}$~cm
         correspond to an AGB mass-loss rate of $1.3\times10^{-4}$~\Mdot.
         At $r\simeq0.1\times10^{16}$~cm, i.e.\ at the inner boundary of the
         envelope, density and velocity are set to the actual
         stellar wind properties for age zero, $\dot{M}=2.5\times10^{-7}$~\Mdot\
         and $v=60$~\kms\ (for $Z=Z_{\rm GD}$, see text).
        }
\label{schoen.init}
\end{figure}

   For a first assessment of the influence of the metal content, 
   we selected a circumstellar envelope structure with initially
   ${\alpha=3}$ and ${v=10}$~\kms\ from our set of initial models with power-law radial
   density profiles, ${\rho \propto r^{-\alpha}}$, used in
   \citetalias{schoenetal.05a}.

   There it was shown that the expansion speeds seen in Galactic disk PNe with 
   round/elliptical shapes imply radial density gradients with power-law exponents
   between 2.5 and $>$3, where the larger exponent is typical for the more evolved 
   objects. The choice of a density gradient with ${\alpha=3}$ for all metallicities 
   reflects just a compromise in order to make the model grid as simple as possible 
   without losing its significance for practical applications.

\begin{figure}[t]              
\vskip -5mm
\includegraphics[width=\linewidth]{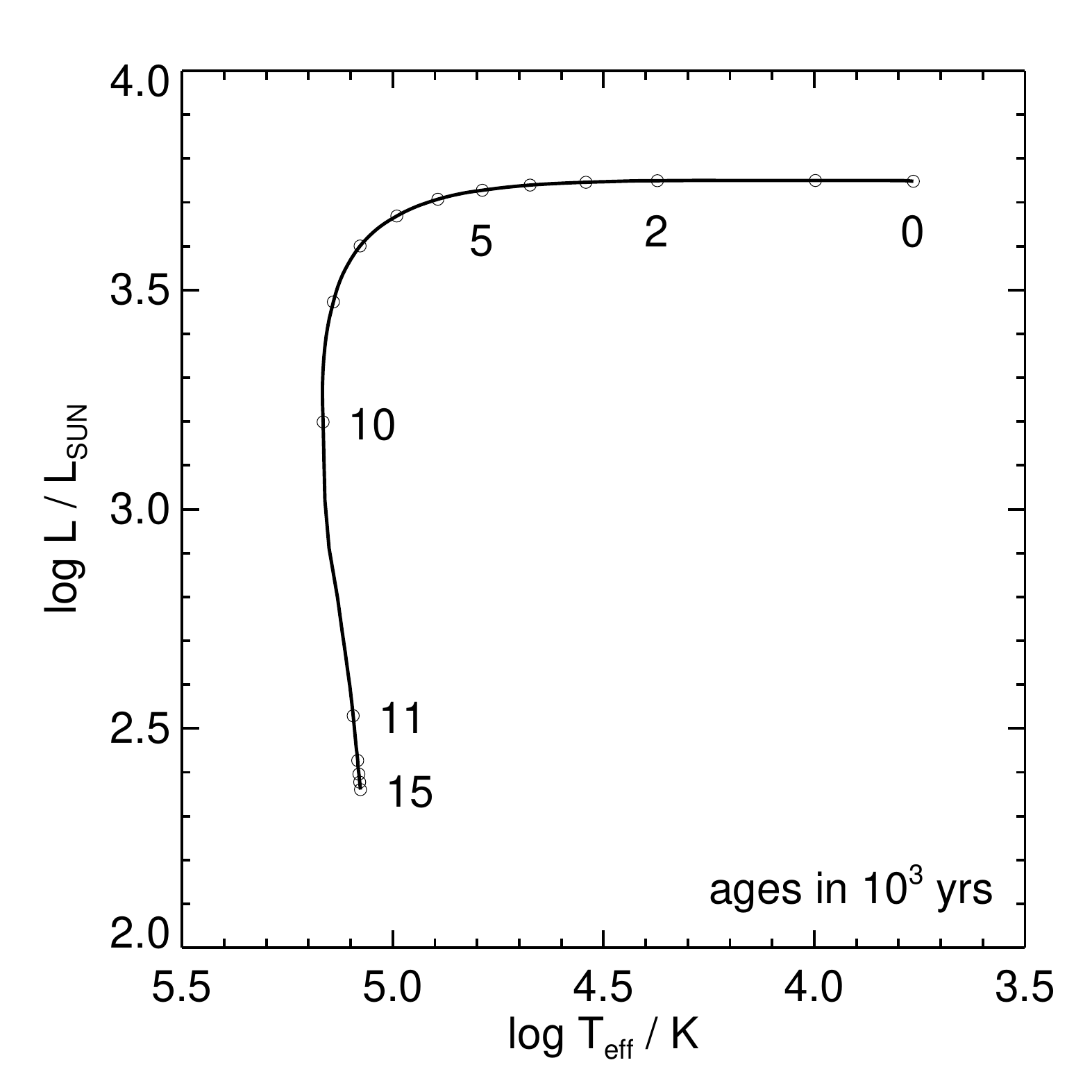}
\caption{Evolutionary path of the 0.595~\Msun\ post-AGB model used for the  
         ${\alpha =3}$ hydrodynamical sequences of this work.
         Post-AGB time marks (circles) are separated by $10^3$ years. 
        }
\label{schoen.star}
\end{figure}

\begin{figure}[t]              
\includegraphics*[bb= 0.8cm 0.4cm 15.4cm 15.0cm, height=7.0cm, width=\linewidth]
                 {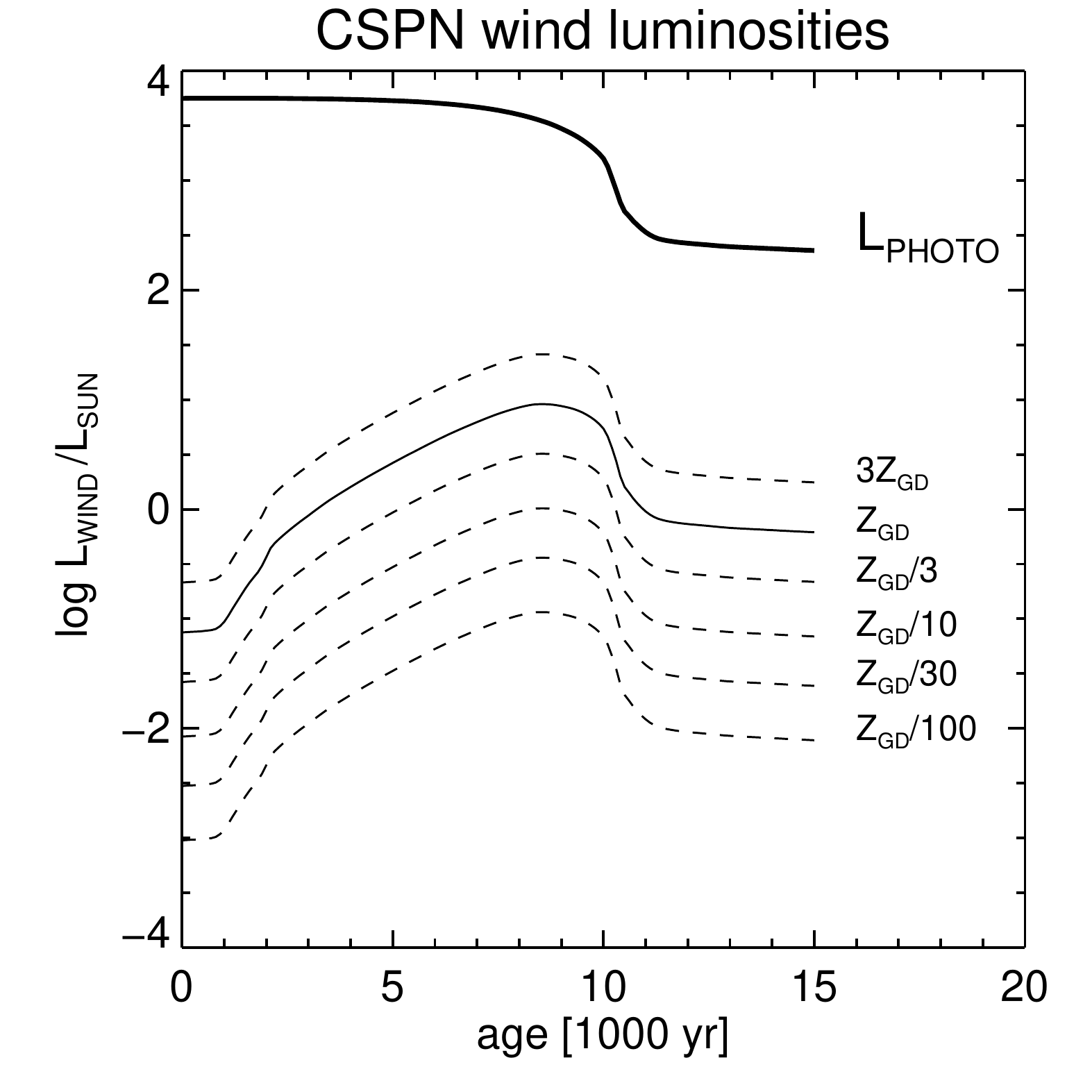}
\caption{Wind luminosities, $L_\mathrm{wind}= 0.5\,\dot{M}\,v^2_\infty$,
         for different metallicities as a function of the (post-AGB) age, compared with the
         photon luminosity of the 0.595~\Msun\ post-AGB model shown in
         Fig.\,\ref{schoen.star}.
        }
\label{schoen.windmodel}
\end{figure}

   Figure~\ref{schoen.init} illustrates the structure of the initial models in some detail.
   This envelope model was coupled to a hydrogen-burning post-AGB model of 0.595~\Msun\
   used in \citetalias{schoenetal.05a} whose evolution in the
   Hertzsprung-Russell diagram is depicted in Fig.\,\ref{schoen.star}.
  The actual variations of the central star wind, based on the adopted metallicity
  range and the approximations introduced above, is illustrated in
  Fig.\,\ref{schoen.windmodel}.  There are two facts which emerge from this figure:
\begin{itemize}
\item[(i)]  The wind power is very weak at the early phases of evolution but increases
            by about two orders-of-magnitude ($L_\mathrm{wind}\propto v^2_\infty$)
            before it decreases again with the fading central star. 
\item[(ii)] Even for the case of the most powerful wind ($3\,Z_\mathrm{GD}$) its 
            mechanical power is only a very small fraction of the photon luminosity.
\end{itemize}

  The evolution of the whole system, star and circumstellar envelope, is followed
  across the Hertzsprung-Russell diagram towards the white-dwarf cooling path,
  employing our 1D radiation-hydrodynamics code.
  An important aspect of our hydrodynamical simulations is that we can
  switch off the hydrodynamics \changed{by setting the gas velocity to zero}
  at any time in the simulation.  The model is then able to settle into
  its equilibrium state for fixed density profile and radiation field 
\changed{which is reached if the electron temperature distribution remains constant 
         throughout the entire computational domain.} 
\changed{This kind of model corresponds then to a standard photoionisation 
         model, but with the additional feature of having a velocity field and 
         also a halo.}
  We are thus able to estimate in a self-consistent manner possible systematic
  errors introduced by imposing equilibrium conditions for systems which are in
  reality \changed{influenced} by dynamics.

\begin{table}[!t]
\caption{\label{sequences}
         Overview of the model sequences used in this work. The 
	 table lists the central star masses (Col.~1), the 
	 AGB-wind properties (Cols.~2--4), the different 
	 metallicities covered for each model (Col.~5), 
	 and references to detailed discussions of the inital models (Col.~6).} 
\tabcolsep=4.8pt
\begin{tabular}{cccccl}
\hline\noalign{\smallskip}
 $M$     & Type       & $\dot{M}_{\rm agb}$                   & $v_{\rm agb}$  & Metallicity$\,^{\rm a}$        & Reference  \\[1.5pt]
[\Msun]  &            & [\Mdot]                               & [\kms]         & [$Z_{\rm GD}$]                 &            \\[1.5pt]
\hline\noalign{\smallskip}
 0.595   & $\alpha=3$ & $1.3\times10^{-4}$\rlap{$\,^{\rm b}$} &   10           & $3 \ldots 1/100$               & \citetalias{schoenetal.05a}\\
 0.595   & \multicolumn{2}{c}{Hydrodynamics sim.}             &   $\simeq$12~~ & $1$                            & \citetalias{schoenetal.05b} \\[5pt]
 0.605   & \multicolumn{2}{c}{Hydrodynamics sim.}             &   $\simeq$12~~ & $3, 1$                         & \citetalias{schoenetal.07} \\
 0.605   & $\alpha=2$ & $1.0\times10^{-4}$                    &   10           & $1$                            & \citetalias{perinotto.04}  \\[5pt]
 0.625   & $\alpha=2$ & $1.0\times10^{-4}$                    &   15           & $3 \ldots 1/10$                & \citetalias{perinotto.04}  \\
 0.625   & $\alpha=2$ & $0.5\times10^{-4}$                    & ~~7\rlap{.5}   & $1/3$                          &                            \\[5pt]
 0.696   & $\alpha=2$ & $1.0\times10^{-4}$                    &   15           & $6, 3 \ldots 1/10$             & \citetalias{perinotto.04}  \\
 0.696   & $\alpha=2$ & $1.0\times10^{-4}$                    &   20           & $1$                            &                            \\
 0.696   & $\alpha=2$ & $1.0\times10^{-4}$                    &   10           & $1, 1/3$                       &                            \\
 0.696   & $\alpha=2$ & $0.5\times10^{-4}$                    &   15           & $1$                            &                            \\
 0.696   & $\alpha=2$ & $2.0\times10^{-4}$                    &   15           & $1, 1/3$                       &                            \\[1.5pt]
\hline
\end{tabular}
\\[3pt]
$^{\rm a}$ Metallicity ranges are are filled in steps of 0.5 dex. 
\\
$^{\rm b}$ Value at a distance ${r=1.0\times10^{16}}$ cm from the central star. 
\end{table}

  Central stars of about 0.6 \Msun\ are not luminous enough as to harbour the
  brightest (and thus easiest observable) planetary nebulae in a distant stellar
  system.  We supplemented thus the already existing hydrodynamical sequences with
  central stars of 0.625 and 0.696 \Msun, both with $Z=Z_{\rm GD}$,  
  which were discussed in \citetalias{perinotto.04} and 
  \citet[][Paper IV hereafter]{schoenetal.07}, by additional sequences with 
  other metal contents: \changed{${Z=6\,Z_{\rm GD}}$ (0.696 \Msun\ only)},
  \ $3Z_{\rm GD},\ Z_{\rm GD}/3$, and $Z_{\rm GD}/10$.
  Except for the metallicity, the initial models are identical with those used in
  \citetalias{perinotto.04}: ${\alpha=2}$ with ${\dot{M}_{\rm agb}=10^{-4}}$ \Mdot, 
  ${v_{\rm agb}= 15}$ \kms.  In these cases the choice of a 
  constant mass outflow (i.e. ${\alpha=2}$) appears
  quite reasonable since these rather massive central stars have a very short 
  post-AGB phase during which the PN remains very compact and is expected not 
  to `see' larger variations of the circumstellar density gradient while it is
  expanding.

\begin{figure*}[t]
\sidecaption
\includegraphics*[bb= 0cm 0cm 17.0cm 11.8cm, width= 13.0cm]
      {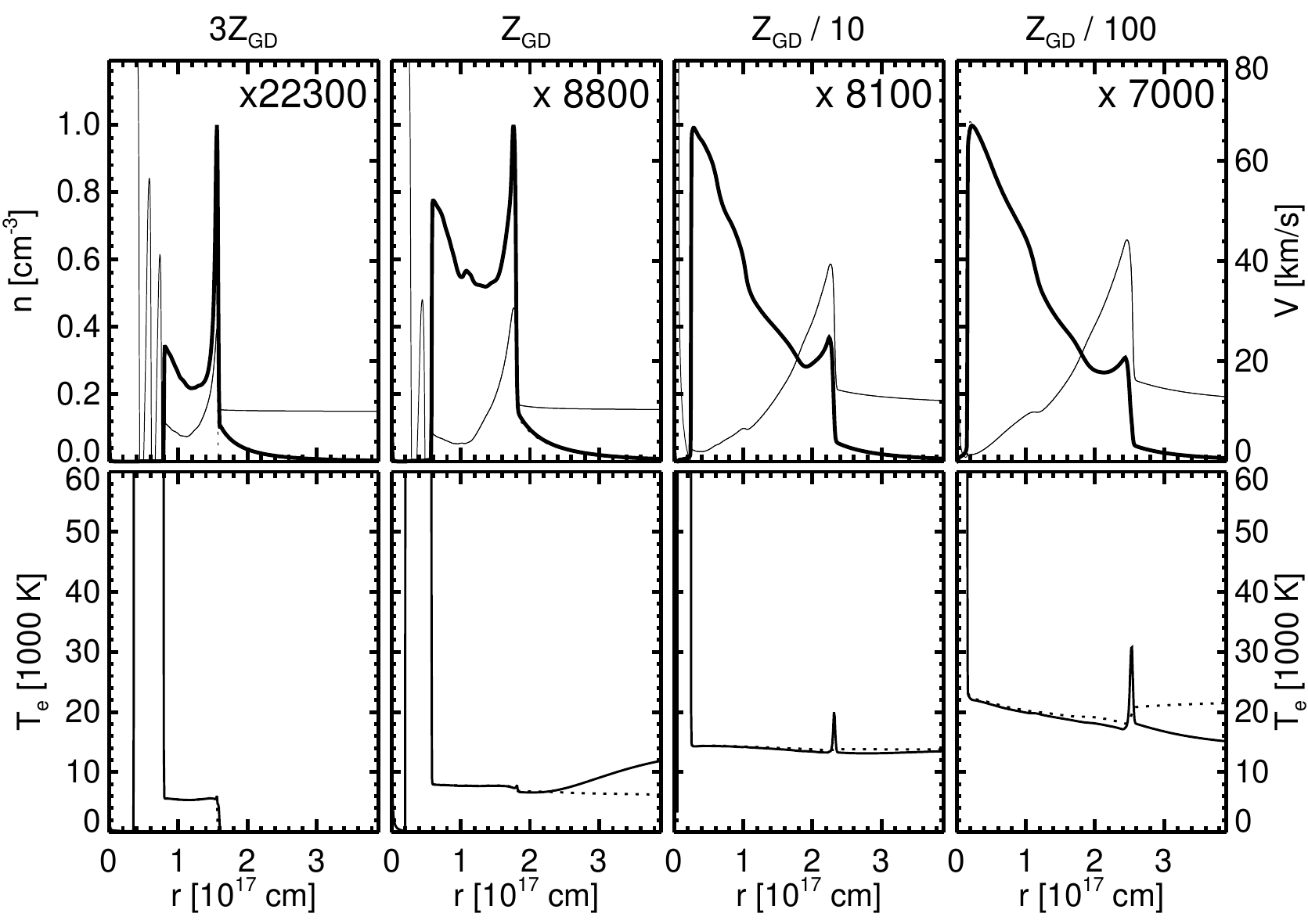}
\caption{Density, velocity and temperature profiles of models from the ${\alpha=3}$
         sequences, selected after $\simeq$3500 years of post-AGB evolution, and for the
	 metallicities indicated above each panel.
         The central star is at the origin, and its parameters are
	 ${\teff\simeq 41\,000}$\,K and ${L=5534}$\,\Lsun.
\changed{The velocity oscillations seen in some early models are due to a temporary
         numerical instability in the hot bubble caused by our heat conduction 
         routine, but have no impact on the model dynamics.} 
         \emph{Upper row}: the radial profiles of heavy particle densities
         (thick solid), electron densities (dotted), and gas velocities
         (thin solid).  The particle densities
	 are normalised and must be multiplied by the factors
	 given in the individual panels to get the true densities.  \emph{Lower row}:
	 the radial run of the electron temperatures, where the dotted
	 lines represent the equilibrium temperatures (see text for the details, and
         also Sects. \ref{T.electron} and \ref{equi.model}).
      }
\label{evol.1}
\end{figure*}

\changed{In order to estimate possible influences of different, albeit constant, mass-loss 
         rates and outflow velocities on the nebular kinematics we computed several 
         additional sequences for the 0.696 \Msun\ post-AGB model with AGB mass-loss rates 
         and wind velocities changed. 
         For the 0.625 \Msun\ post-AGB model, we
         computed one additional $Z_{\rm GD}/3$ sequence with 
         $\dot{M}_{\rm agb}=0.5\times10^{-4}$ \Mdot\ and ${v_{\rm agb}= 7.5}$ \kms.
         An overview of all the sequences used in this paper is presented in 
         Table~\ref{sequences}.}

  In the following sections we discuss in more detail implications which follow from
  our simulations and which are important in interpreting PNe in distant stellar
  systems with varying metallicities.  We will thereby distinguish between the
  ${\alpha=2}$ and ${\alpha=3}$ model sequences.

\section{Results}\label{results}
\subsection{The $\alpha=3$ sequences}\label{alpha}
\begin{figure*}
\sidecaption
\includegraphics*[bb= 0cm 0cm 17.0cm 12cm, width= 13.0cm]
      {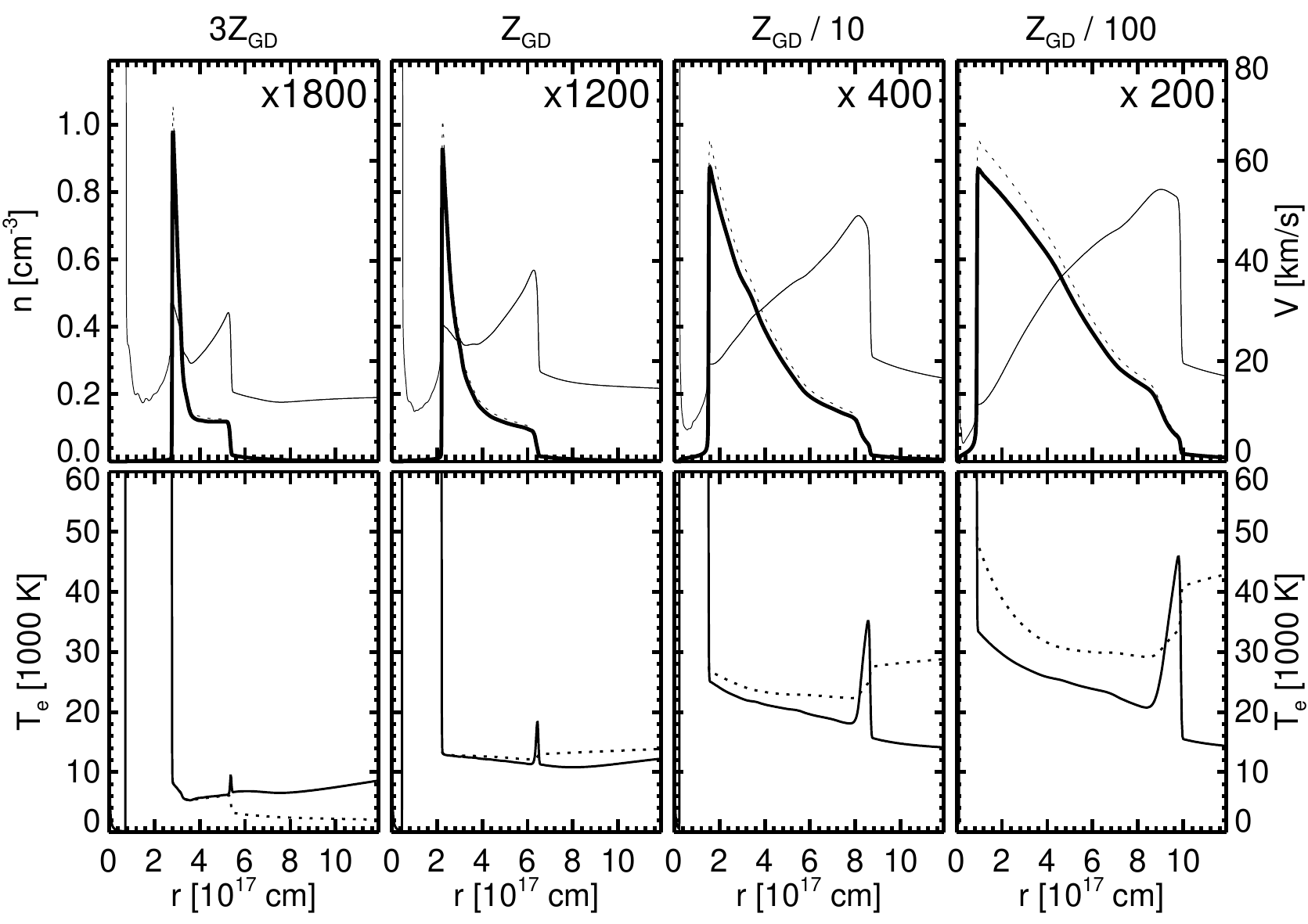}
\caption{The same as in Fig.~\ref{evol.1} but after $\simeq$7150 years.
         the $3Z_{\rm GD}$ model is now also optically thin.
         The stellar parameters are $\teff\simeq 100\,900$~K and $L=4580$~\Lsun.
        }
\label{evol.2}
\end{figure*}
\begin{figure*}
\sidecaption
\includegraphics*[bb= 0cm 0cm 17.0cm 12cm, width= 13.0cm]
      {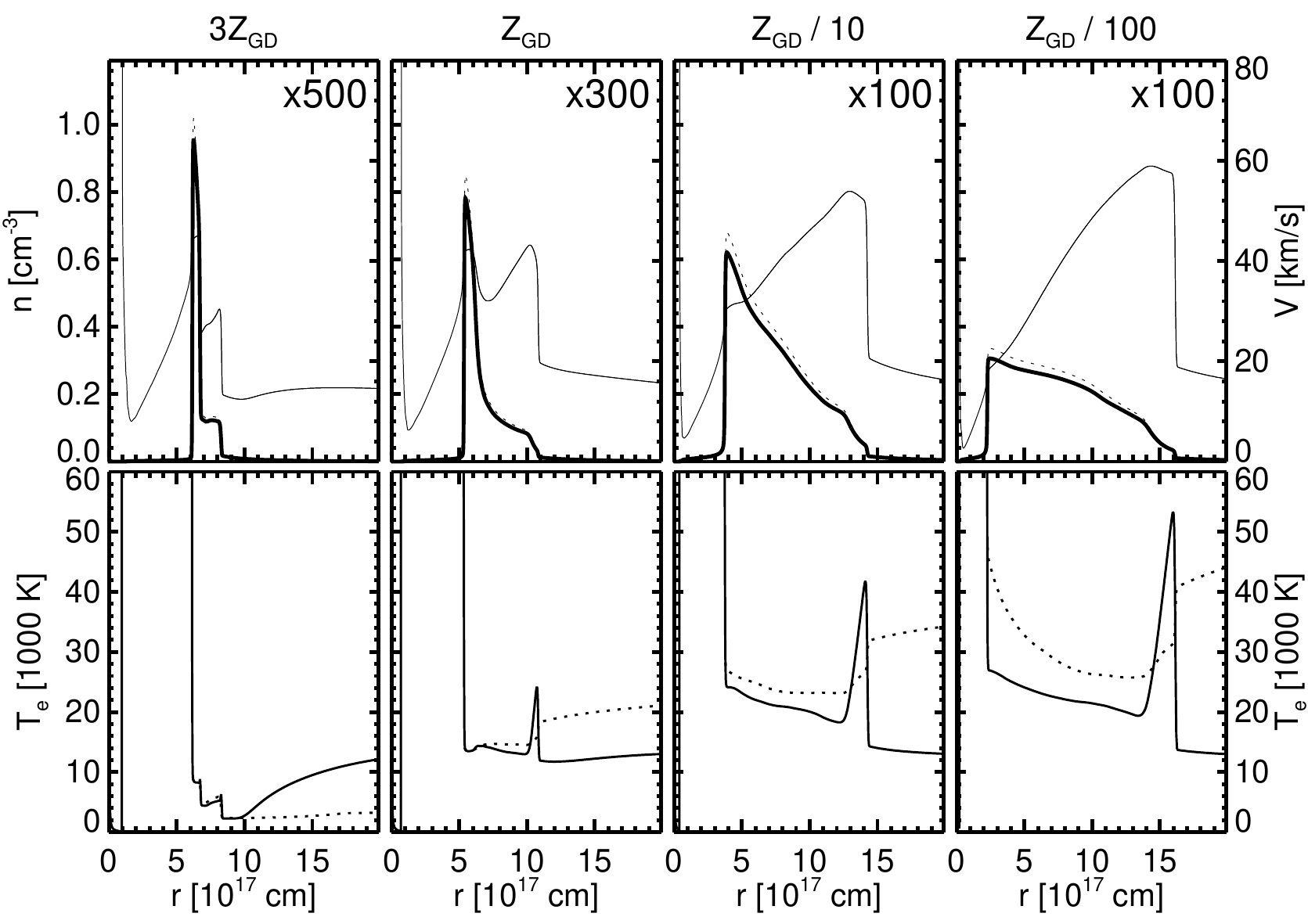}
\caption{The same as in Fig.~\ref{evol.1} but after $\simeq$9840 years.
         The stellar parameters are now $\teff\simeq 146\,870$~K and $L=1845$~\Lsun,
\changed{corresponding to the turn-around point of the stellar track seen in Fig. 
         \ref{schoen.star}. }
        }
\label{evol.3}
\end{figure*}
\begin{figure*}
\sidecaption
\includegraphics*[bb= 0cm 0cm 17.0cm 12cm, width= 13.0cm]
      {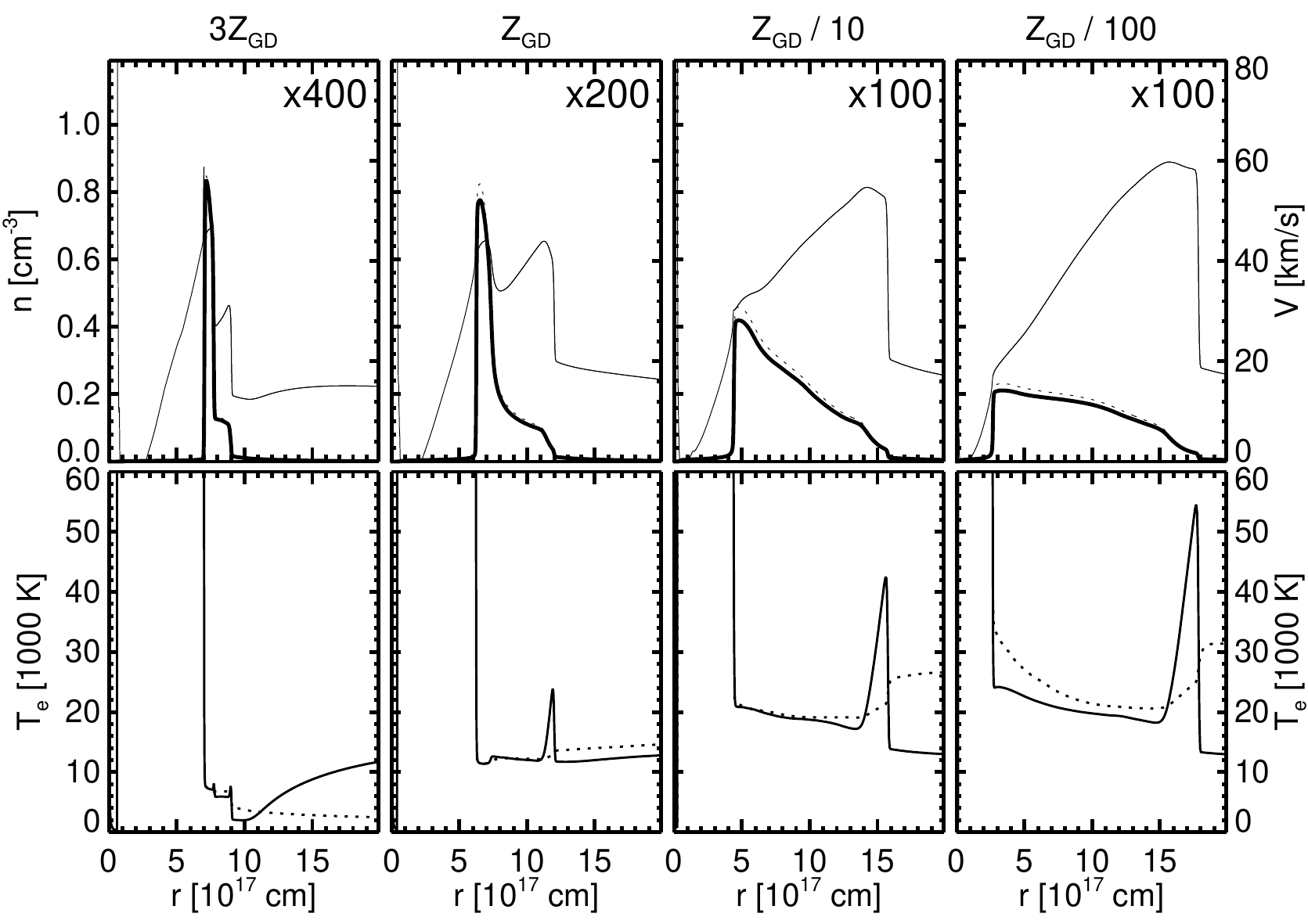}
\caption{
\changed{The same as in Fig.~\ref{evol.1} but after $\simeq$10\,530 years. The stellar
         parameters are ${\teff\simeq 131\,000}$~K and ${L\simeq500}$~\Lsun, right
         after the end of the recombination phase.}
        }
\label{evol.3a}
\end{figure*}

\begin{figure*}
\sidecaption
\includegraphics*[bb= 0cm 0cm 17.0cm 12cm, width= 13.0cm]
      {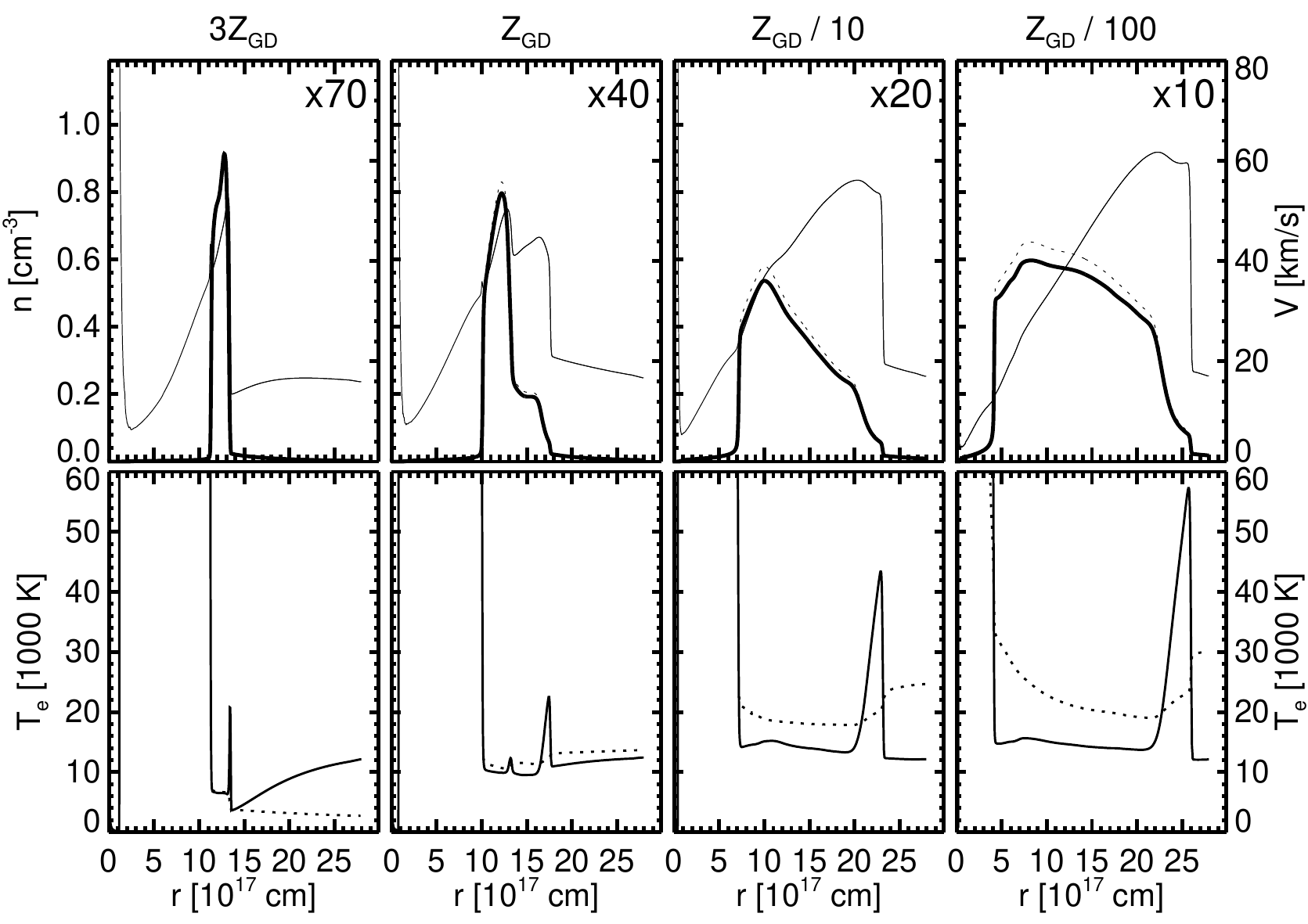}
\caption{The same as in Fig.~\ref{evol.1} but after $\simeq$13\,850 years.
         The stellar parameters are now $\teff\simeq 119\,800$~K and $L=240$~\Lsun,
\changed{typical for the reionisation stage when the nebula expands around a central
         star with virtually constant luminosity and temperature.}
        }
\label{evol.4}
\end{figure*}
  
  Figures.~\ref{evol.1}--\ref{evol.4} illustrate how the general evolution of the
  model nebulae is influenced by their metal content.  All sequences
  have the same central star of 0.595~\Msun\ and start with the same initial model
  depicted in Fig. \ref{models}.  The figures show \hbox{$5\times4$} snapshots along
  the stellar track,  viz.\ for five positions 
  and for four metallicities each, ${Z=3Z_\mathrm{GD}},\ Z_\mathrm{GD},\
  Z_\mathrm{GD}/10,\ Z_\mathrm{GD}/100$.  The positions selected are at the early
  ionisation phase at about \hbox{$\teff\simeq 40\,000$~K} (Fig.~\ref{evol.1}),
  at the high-ionisation stage at about 100\,000~K (Fig.~\ref{evol.2}),
  at the maximum stellar effective temperature (Fig.~\ref{evol.3}), 
\changed{during nebular recombination at a stellar luminosity of about 
         500 \Lsun\ (Fig. \ref{evol.3a}),}
  and at an even lower luminosity of only 240~\Lsun\ after reionisation has started
  (Fig.~\ref{evol.4}).  Note the different scaling of each individual panel
  within each figure, and also the different radial ranges of the figures.

  An extensive discussion of the principles of PN evolution based on
  radiation-hydrodynamics simulations has been given in
  \citetalias{perinotto.04},
  which the reader is referred to for details.  We concentrate here only on those
  aspects relevant in connection with the metal content of the models.

\subsubsection{Structures and kinematics of the models}
\label{struc.kin}
  The upper panels of Figs.~\ref{evol.1}--\ref{evol.4} demonstrate how 
  structure and kinematics evolve with time.  We see a strong trend with metallicity:
  the lower the metal content, the larger and thicker become the nebular shells
  and the smaller the wind-blown cavities.  Responsible
  for this behaviour are two factors, both depending on the metal content.
  The first is the expansion speed of the hot, ionised gas which scales with the
  sound velocity $\propto\!\sqrt{T_{\rm e}}$ \citepalias[cf.][]{schoenetal.05a} and
  becomes higher at lower metallicities because of the reduced line cooling
  efficiency.  
  The second is the wind power which decreases with metallicity as
  outlined in Sect.~\ref{models} (Fig.~\ref{schoen.windmodel}) and loses its
  ability to compress and accelerate the inner nebular parts.  
  This is seen in the two metal-poor sequences where the nebular density falls off
  radially more gradually with a nearly linear slope (Figs. \ref{evol.1}--\ref{evol.3a},
  top panels).  A sharp density contrast between the rim and the shell, 
  typical for the metal-richer models, 
\changed{is therefore an indication of a strong wind. } 

  Once the central star has passed the position of its maximum effective temperature
  in the Hertzsprung-Russell diagram (Fig.~\ref{evol.3}) and starts to fade quickly, 
  also the wind power becomes progressively weaker in all cases 
  (see Fig.~\ref{schoen.windmodel}).
\changed{Figure \ref{evol.3a} displays a moment during the recombination phase as the
         central star fades, and}
  Fig.~\ref{evol.4} a stage after reionisation has begun to dominate the inner nebular 
  regions, 
\changed{forcing there the matter to develop a positive density gradient, very 
         similar to the typical situation behind a D-type shock during the first 
         ionisation.}

\paragraph{Structures}
\label{struc}

\begin{figure}[t]
\includegraphics*[bb= 0.1cm 0.5cm 15.8cm 15.1cm, width=\linewidth]
                {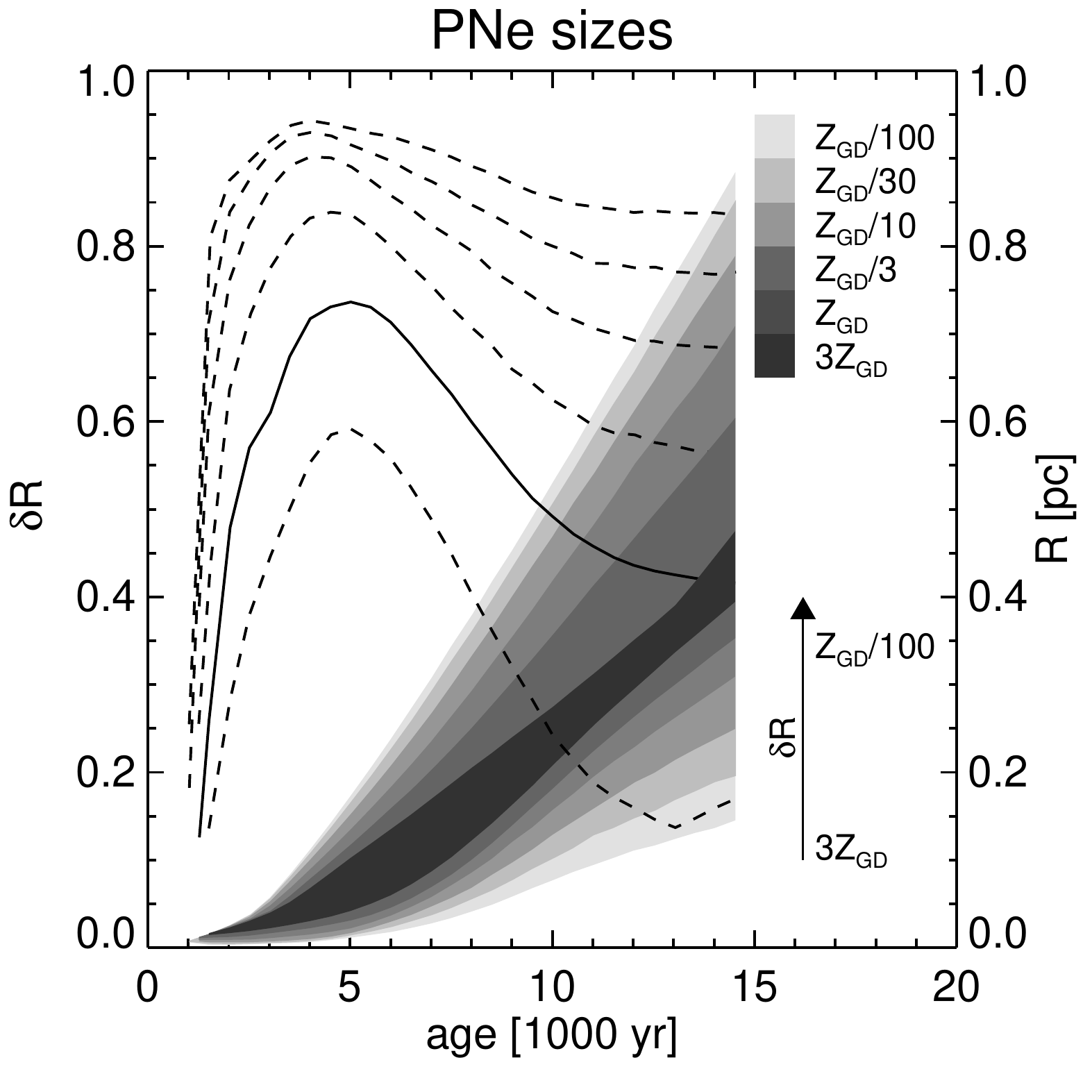}
\caption{Development of nebular sizes with time for the $\alpha=3$ sequences with various
         metallicities.  \emph{Right ordinate}: inner ($R_{\rm cd}$)
         and outer radii ($R_{\rm out}$), with the different metallicities indicated by 
         various (overlapping) gray shades; see legend. \emph{Left ordinate}: relative 
         thicknesses, $\delta R=(R_{\rm out} - R_{\rm cd})/R_{\rm out}$, for the same 
         sequences. Metallicity decreases monotonically from the bottom curve towards the 
         top curve.
        }
\label{sizes}
\end{figure}

  Figure~\ref{sizes} summarises the complete size evolution with time for all six
  metallicities investigated and illustrates clearly the importance of the
  wind interaction, 
\changed{given by the pressure of shocked wind (bubble) gas, as compared to the thermal 
         pressure of nebular matter, and how these pressures}
  change during the course of evolution.  
  From this figure it becomes evident that wind interaction 
  is \emph{not} the main driver for the nebular expansion, although its importance
  becomes larger during the end phases of evolution and generally with metallicity.
  Rather, the expansion is initiated by thermal pressure differences, caused by 
  photo-heating of the former neutral circumstellar gas, and continues independently 
  of the stellar wind.  Wind interaction is only responsible for compressing the 
  inner parts of the expanding gas and preventing the shell from collapsing.  
\changed{This last statement is also true during the final evolution along the white-dwarf
         sequence: Even at the lowest metallicity considered here the hot bubble is still 
         expanding, although the stellar wind power is reduced by about one order-of-magnitude
         (cf. Fig. \ref{schoen.windmodel}).}

  For the metal-richer models with low expansion rates and more powerful 
  central-star winds, wind interaction becomes more important 
  and leads to very (geometrically) thin nebular shells.
\changed{However, even for the highest metallicity considered here ($3Z_{\rm GD}$)
         the double-shell structure remains visible during the whole high-luminosity part of
         evolution. The rim succeeds in overtaking the shell not before the end of evolution 
         (cf. Figs. \ref{evol.3} to \ref{evol.4}). }
  Neglecting photo-heating would lead to \changed{single-shell (i.e. rim only) 
  structures in \emph{all} cases.}

  From the observer's point of view, only the relative sizes $\delta R$  (also shown in 
  Fig. \ref{sizes}) are of interest because only these can be measured distance-independently.  
  In general, the rapid increase of $\delta R$ at the beginning of the evolution reflects 
  the dynamics of the increasing thermal pressure due to ionisation, and the decrease of 
  $\delta R$ later on is due to the increased nebular size and the stronger stellar wind.
  At the lowest metallicity, ${Z=Z_\mathrm{GD}/100}$, the relative sizes are 
  close to 0.9 for most of the time, in contrast to the metal-rich models,
  ${Z=3Z_\mathrm{GD}}$, which are much more compressed and whose relative thicknesses
  do not exceed $\approx$0.6.
  The Galactic disk composition favours medium thick objects with $\delta R$ between
  $\simeq$0.5\ldots0.7, consistent with the observations of round/elliptical PNe
  \citepalias[cf. Fig. 6 in][]{schoenetal.07}. 

\begin{figure}[t]
\includegraphics*[width=\linewidth]{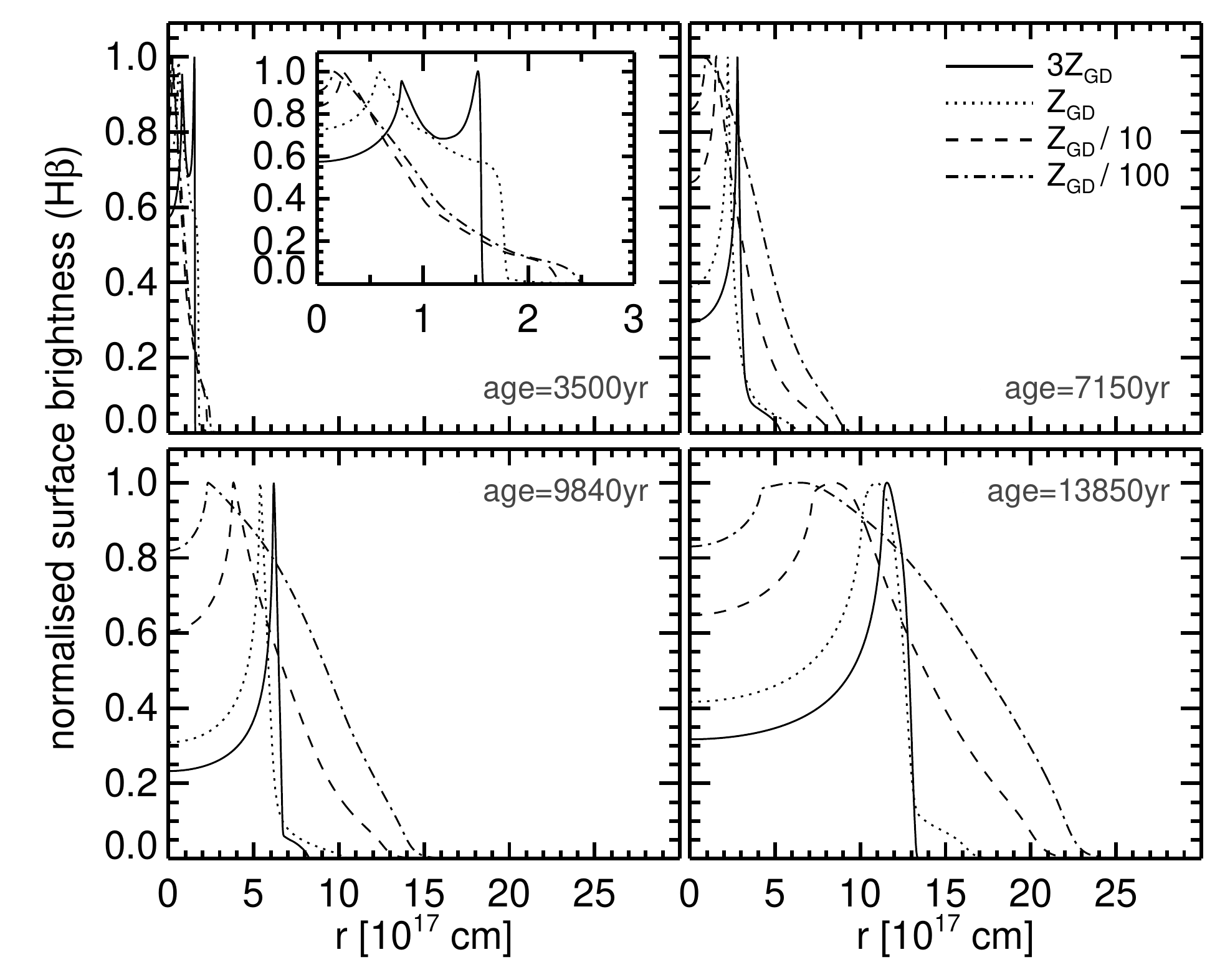}
\vskip-1mm
\caption{\label{sb}
         Normalised radial surface brightness (intensity) profiles in H$\beta$ of the 
         models shown in Figs. \ref{evol.1}--\ref{evol.3} and \ref{evol.4}. 
         For clarity, the inset 
         in the \emph{top left} panel shows the surface brightnesses at an enlarged scale.
        }
\end{figure}

  The very different density profiles which our models develop during their course of  
  evolution are, of course, also reflected in their surface brightness distributions.
  The case is illustrated in Fig. \ref{sb} where the H$\beta$\ surface brightness 
  profiles of the models shown in Figs. \ref{evol.1}--\ref{evol.3} and \ref{evol.4} 
  are displayed.

  One sees a clear trend with metallicity: The metal-rich models ($Z\ga Z_{\rm GD}$)
  develop a deep central cavity and a pronounced rim-shell structure where the shell 
  brightness is only about 10\,\% or less of the maximum rim brightness.  
  The metal-poor models display a more gradual brightness decline with distance from
  the star, with a nearly linear (negative) slope for the lowest metallicity investigated 
  (${Z= Z_{\rm GD}/100}$). The rim-shell dichotomy known from most PNe is no
  longer existent.  Additionally, the central cavity becomes much smaller and is nearly 
  filled up by the projection effect.

\paragraph{Kinematics}
\label{kin}

  The velocity field is very similar in all cases. Once ionisation has started,
  the gas velocity increases nearly linearly with radius. The post-shock speed
  (i.e. the gas velocity immediately behind the outer edge of the nebular shell) 
  increases with time in line with
  the shock acceleration (Figs.~\ref{evol.1}--\ref{evol.4}). 
  Note that during the early \changed{stage of nebular evolution the innermost 
  part of the ionised shell expands} \emph{slower} than the former AGB wind 
\changed{because it is decelerated by the high thermal pressure.}   
  This fact holds also 
  during the whole evolution in the low-metallicity cases (top panels in Figs. 
  \ref{evol.1}--\ref{evol.4}).
  The post-shock speed is usually also the maximum expansion velocity within the
  whole nebula, with the
  exception of the more metal-rich models with their strong
  stellar winds which create dense rims
  and accelerates them to velocities greater than the post-shock one
  (Figs.~\ref{evol.3}--\ref{evol.4}).
  An extreme case occurs for ${Z=3Z_\mathrm{GD}}$ where the rim becomes very dense,
  thin, and fast, and swallows finally the outer shell
  (see Figs.~\ref{evol.3}--\ref{evol.4}).

  The extremely metal-poor models reach quite high post-shock (gas) velocities,
  up to 60~\kms, which is twice the value achieved by the models of the
  $3Z_\mathrm{GD}$ sequence.
  However, the most conspicuous features seen in Figs. \ref{evol.1}--\ref{evol.4} 
  (bottom panels) are the strong shocks that these models develop with time
   where the shell interacts with the largely undisturbed AGB wind.
  The post-shock temperatures reach extremely high values,
  up to 50\,000--60\,000 K, and the whole post-shock thickness can be substantial
  for the most metal-poor cases. Despite of this no observed signatures are produced
  by these shocks because of the very low matter densities involved.

\begin{figure*}[t]
\sidecaption
\includegraphics*[bb= 0.7cm 0.5cm 15.4cm 15.1cm, width=0.35\linewidth]
                {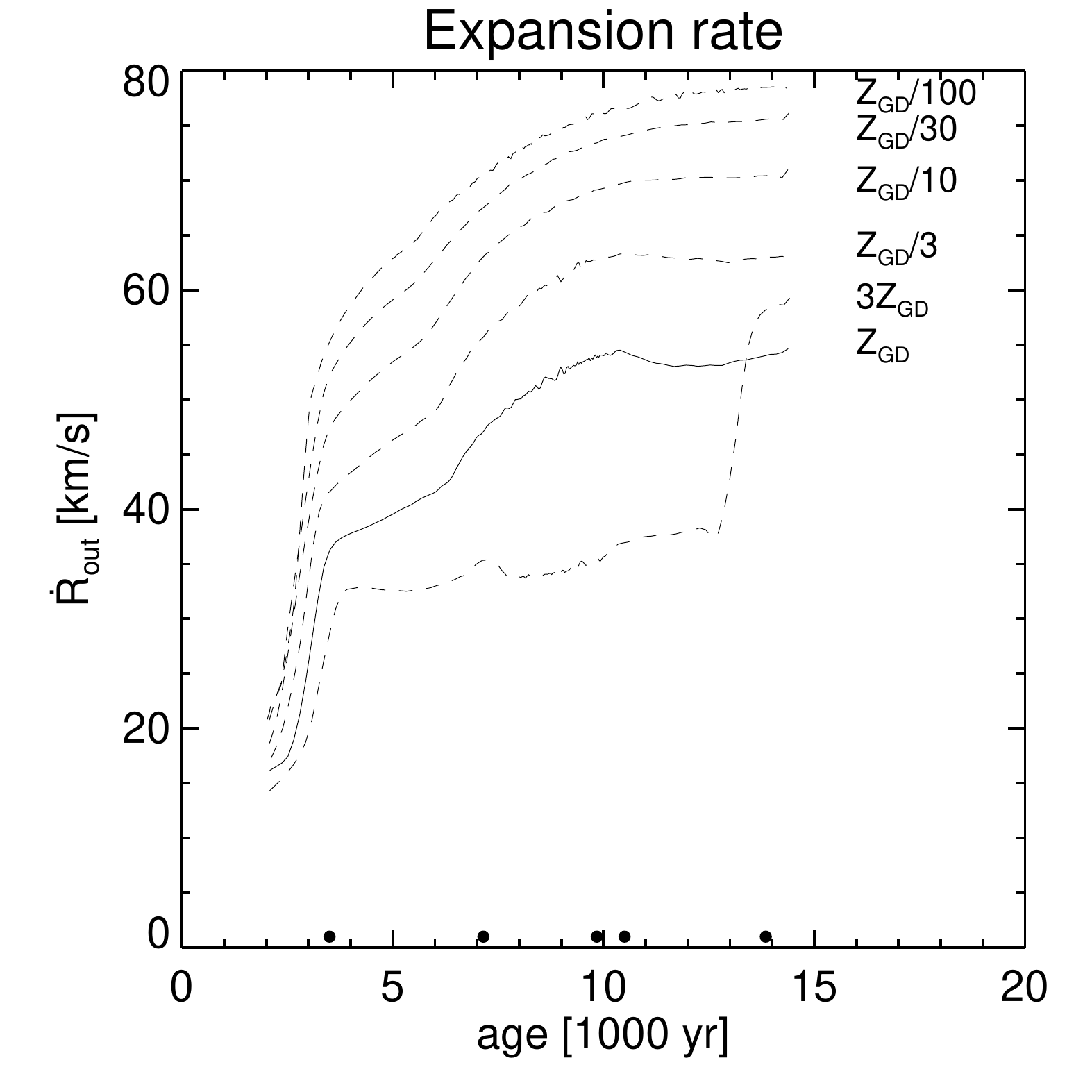}
\includegraphics*[bb= 0.7cm 0.5cm 15.4cm 15.1cm, width=0.35\linewidth]
                {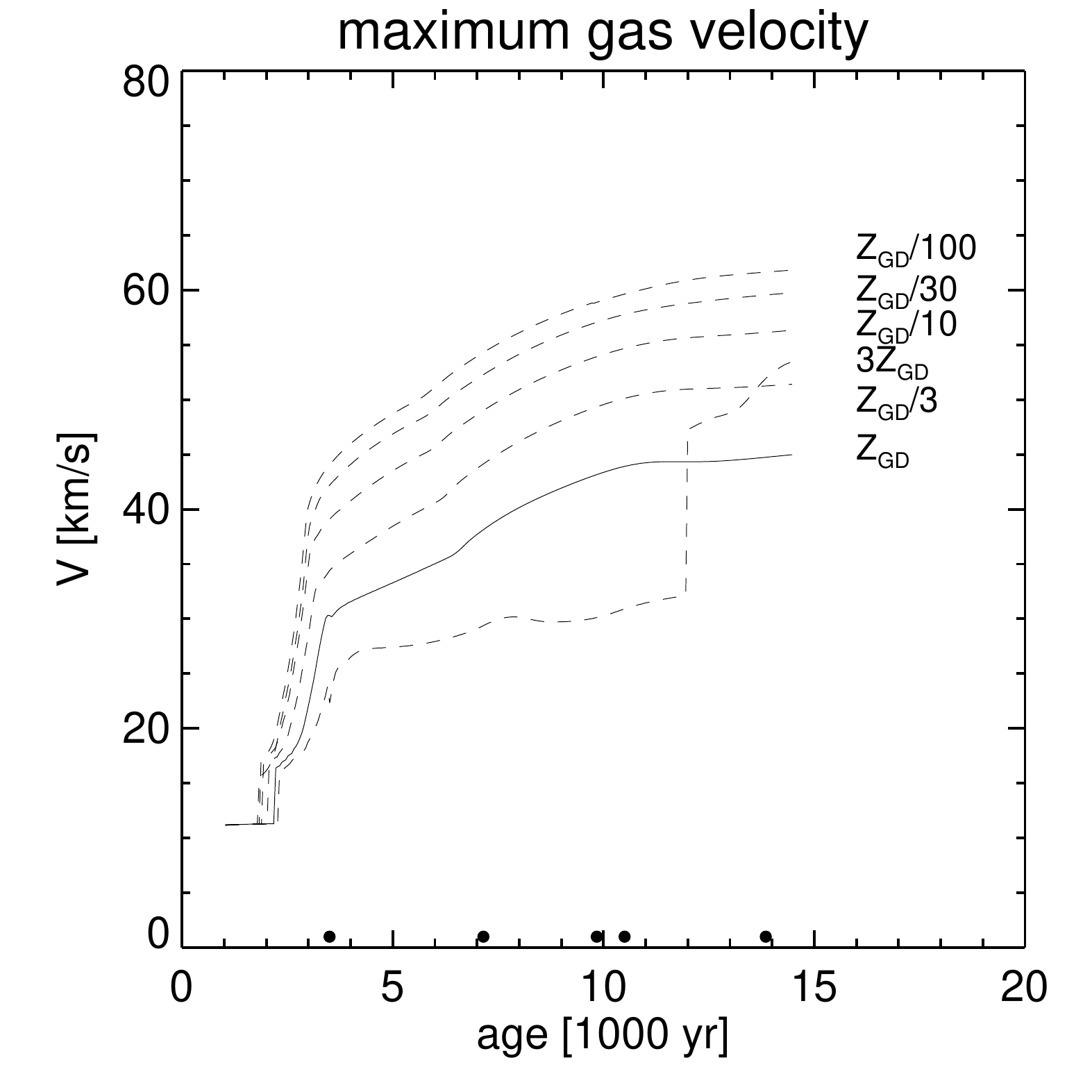}
\caption{Expansion properties of the model nebulae of the six ${\alpha=3}$ sequences
         with different metallicities vs. the post-AGB age. The five black dots along
         the abscissas refer to the snapshots depicted in Figs.
         \ref{evol.1}--\ref{evol.4}.  The initial velocities of the models are 10 \kms\ 
         (cf. Fig.\,\ref{schoen.init}).
         \emph{Left}:  propagation rates of the outer shocks,
	 $\dot{R}_{\rm out}$, which define the nebular sizes.  The
	 velocity jump of the $3Z_{\rm GD}$ model at ${t\simeq 13\,000}$ yr marks
	 the moment when the faster expanding rim starts overtaking the outer shock, 
         and ${\dot{R}_{\rm out} \equiv \dot{R}_{\rm rim}}$ afterwards.
	 \emph{Right}: 
	 the gas velocities immediately behind the outer shock.  Again, the velocity
	 jump of the $3Z_{\rm GD}$ model, now already at ${t\simeq 12\,000}$~yr, is 
         due to the disappearing shell.
        }
\label{vmax}
\end{figure*}
\begin{figure*}[t]
\sidecaption
\includegraphics*[bb= 0.7cm 0.5cm 15.4cm 15.1cm, width=0.35\linewidth]
                {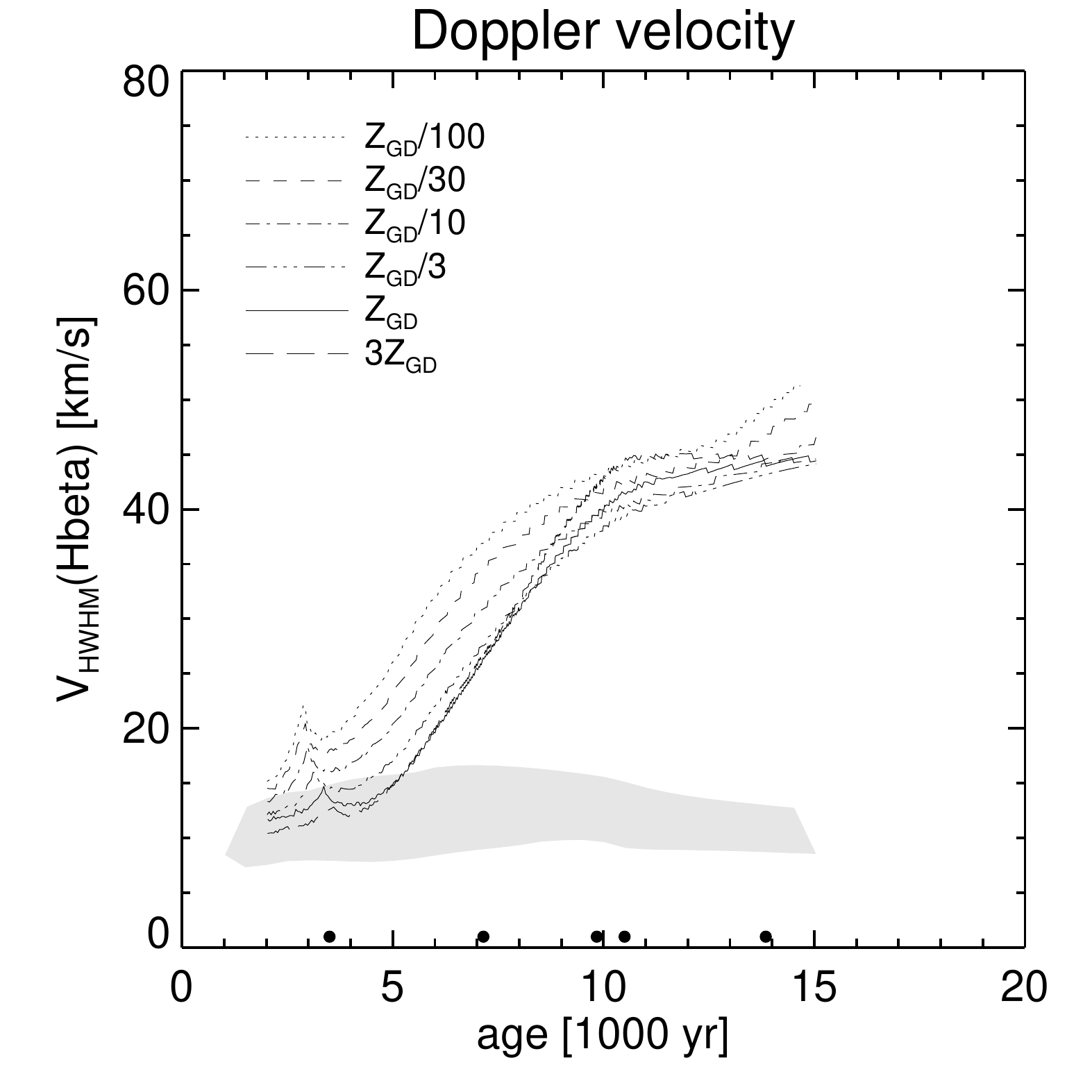}
\includegraphics*[bb= 0.7cm 0.5cm 15.4cm 15.1cm, width=0.35\linewidth]
                {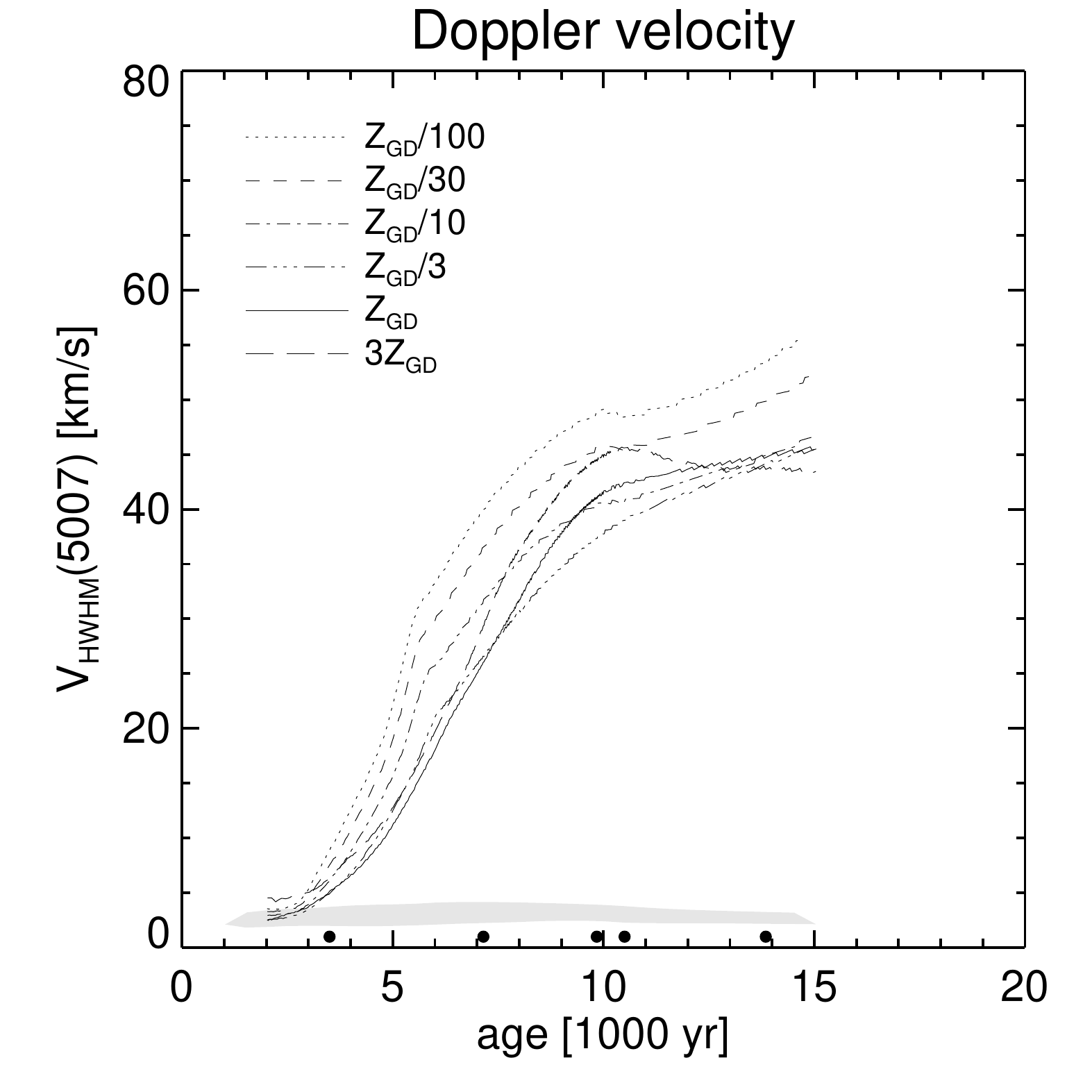}
\caption{Expansion \changed{properties as} derived from the half width at half maximum 
        (HWHM) of the spatially in\-teg\-rated line profiles for the same six  ${\alpha=3}$ 
         sequences shown in Fig.\,\ref{vmax} vs. the post-AGB age, for H$\beta$ 
         (\emph{left}) and \oiii\ 5007 \AA\ (\emph{right}). Again, the five black dots 
         along the abscissas refer to the snapshots depicted in Figs.
         \ref{evol.1}--\ref{evol.4}.  The shadowed areas indicate the thermal
         Doppler broadening, computed by means of the mean electron temperatures (see
         next section).  The lower boundary refers to the models with the highest
         metallicity, $3Z_{\rm GD}$, the upper boundary to those with the lowest
         metallicity, $Z_{\rm GD}/100$.
        }
\label{HWHM}
\end{figure*}

  The overall nebular expansion properties and their dependence on
  metallicity is summarised in Fig.~\ref{vmax} (left) where the propagation speeds,
  $\dot{R}_{\rm out}$, of the shock, which defines the outer edge of the model PNe,
  are plotted.  Starting at rather low values 
  the shock propagation speed increases in all cases very rapidly until the
  nebular shell becomes optically thin.  Afterwards the shock 
  \changed{speed (relative to the upstream flow) is determined by the 
   density profile, ${\rho\propto r^{-3}}$, and the electron temperature 
  (or sound speed)} which becomes higher as the central star gets hotter and 
  the ionising photon flux more energetic
  \citepalias[cf.][Sect.\ 3.1 and Fig. 7 therein]{schoenetal.05a}.  
  For instance, the sudden  
  velocity increase between 6000 and 7000 years seen in all sequences is due to the
  second ionisation of helium, causing a fast growth of the electron temperatures
  (see also Fig.~\ref{etemp.1}).  

  As expected, the shock velocities 
  increase with decreasing metallicity, reflecting the values of the nebular
  electron temperature (cf.\ Figs.~\ref{evol.1}--\ref{evol.4},
  and also Fig.~\ref{etemp.1}).  The effect becomes smaller for the lowest
  metallicities just because also the electron temperature increase levels off
  somewhat (cf. Fig.~\ref{etemp.1} below).
  The total range of $\dot{R}_{\rm out}$ at the end of our simulations 
\changed{($\simeq$15\,000 yr) is $\simeq$55--80 \kms\ for the metallicities used here.
  In the $3Z_{\rm GD}$ sequence with its powerful stellar wind we see the
  rim shock overtaking the more slowly expanding (38~\kms) outer shock.}  This occurs at
  ${t\simeq 13\,000}$ yr when $\dot{R}_{\rm out}$ jumps from 38 \kms\
  to ${\dot{R}_{\rm out}=\dot{R}_{\rm rim}\simeq60}$ \kms\ (cf.
  Figs.~\ref{evol.3}--\ref{evol.4}, leftmost panels).

  The maxima of the gas velocities are always achieved right behind the outer shock,
  and are depicted in the right panel of Fig.~\ref{vmax}.  The trend with metallicity is, 
  of course, the same as seen for $\dot{R}_{\rm out}$, but the absolute values are 
  lower: starting from the shared
  AGB wind velocity of 10 \kms, the gas becomes rapidly accelerated to maximum values
  of 32--62 \kms, depending on metallicity.
  We see again the velocity jump in the $3Z_{\rm GD}$
  sequence when the shell becomes swallowed by the faster rim (cf.
  Figs.~\ref{evol.3}--\ref{evol.4}, leftmost panels).

\changed{Our simulations demonstrate clearly that the expansion of a planetary nebula
         is ruled by the electron temperature of the shell gas 
         \citepalias[cf.][]{schoenetal.05a}, and not by the wind from the central star 
         as is predicted by the favourite theory of interacting winds 
         put forward by \citet{kwoketal.78}.           
         Wind interaction is only responsible for the shape and acceleration of
         the rim, i.e. the inner, bright parts of a PN, and under most conditions 
         of our simulations the rim expands \emph{slower} than the shell. 
         With the reasonable assumption made here that the stellar wind strength 
         decreases with metallicity, it follows that the nebular expansion 
         of the outer shell is fastest if the wind is weakest!}

  The question arises whether the large variation of the expansion velocity with
  nebular age and metallicity can be measured.  For spatially resolved objects there
  is the possibility to measure the fast moving matter behind the outer shock front
  using high-resolution line profiles taken for the central line-of-sight
  \citep{corradi.07}.  In the case of distant objects which cannot be spatially
  resolved, only the half width of the (integrated) line profile can be used.
  We computed therefore spatially integrated line profiles for H$\beta$ and
  \oiii\ 5007 \AA\ \changed{and determined $V_{\rm HWHM}$ as a measure for the
  nebular expansion.}
  The results are summarised in Fig.~\ref{HWHM}.

  First of all, the \changed{spread in $V_{\rm HWHM}$} due to the metal content is, 
  if compared with the situation in Fig.~\ref{vmax}, surprisingly small and partly 
  irregular, and the increase \changed{of $V_{\rm HWHM}$} with time appears to 
  be more gentle in both lines, especially 
  during the early phase of evolution.  This can be understood as a consequence of 
  integrating over the whole object which gives much more weight to gas elements 
  with low velocity components along the line-of-sight if compared to the central 
  line-of-sight case.  This effect is enhanced by the density structure of the
  models: the gas velocity is lowest where the gas density is highest.

  The HWHM velocities based on H$\beta$ are higher than those derived from 5007 \AA\
  during the early evolution, which is due to both the different thermal line widths,
  indicated in Fig. \ref{HWHM} by the shadowed areas,
  \emph{and} the ionisation structure: if the central star is still not very hot, 
  the O$^{2+}$ zone is confined to the inner part of the ionised region
  only, which is strongly
  decelerated by the thermal pressure (cf. Fig.~\ref{evol.1}). Thus,
  $V_{\rm HWHM}$ as measured from the (integrated) \oiii\  line starts at very low 
  values, which are significantly \emph{below} the original AGB-wind velocity of 10 \kms\ 
  assumed here.
  Later on, at larger ages when the Doppler width of \hb\ becomes larger than the 
  thermal one and the O$^{2+}$ zone is extending to the outer nebular edge (i.e. to
  the outer shock), both lines behave very similarly.   In the following all 
  velocities are from \oiii\ only if not specified otherwise.

  In any case, the HWHM method severely
  underestimates the true expansion of the nebula which is given by the propagation
  of the outer shock front, as seen in Fig.~\ref{vmax} (left panel).
  For instance, at age 5000 years the HWHM velocities are between 10 and 20 \kms\
  while the true expansion velocities are already significantly higher, viz.
  between 30 and 60 \kms!  Thus, the rather large sensitivity of the nebular expansion
  on metallicity is almost lost by using spatially integrated line
  profiles.  An observational test of the predicted dependence of the expansion
  velocities on metallicity by using an appropriate sample of PNe drawn from
  galaxies with different metal contents as done by \citet{richer.06} appears to be
  difficult if not impossible.  We will come back to this point in more detail
  in Sect. \ref{distant}.

\begin{figure*}[t]
\sidecaption
\includegraphics*[bb= 0.7cm 0.5cm 15.4cm 15.1cm, width=0.35\linewidth]
                {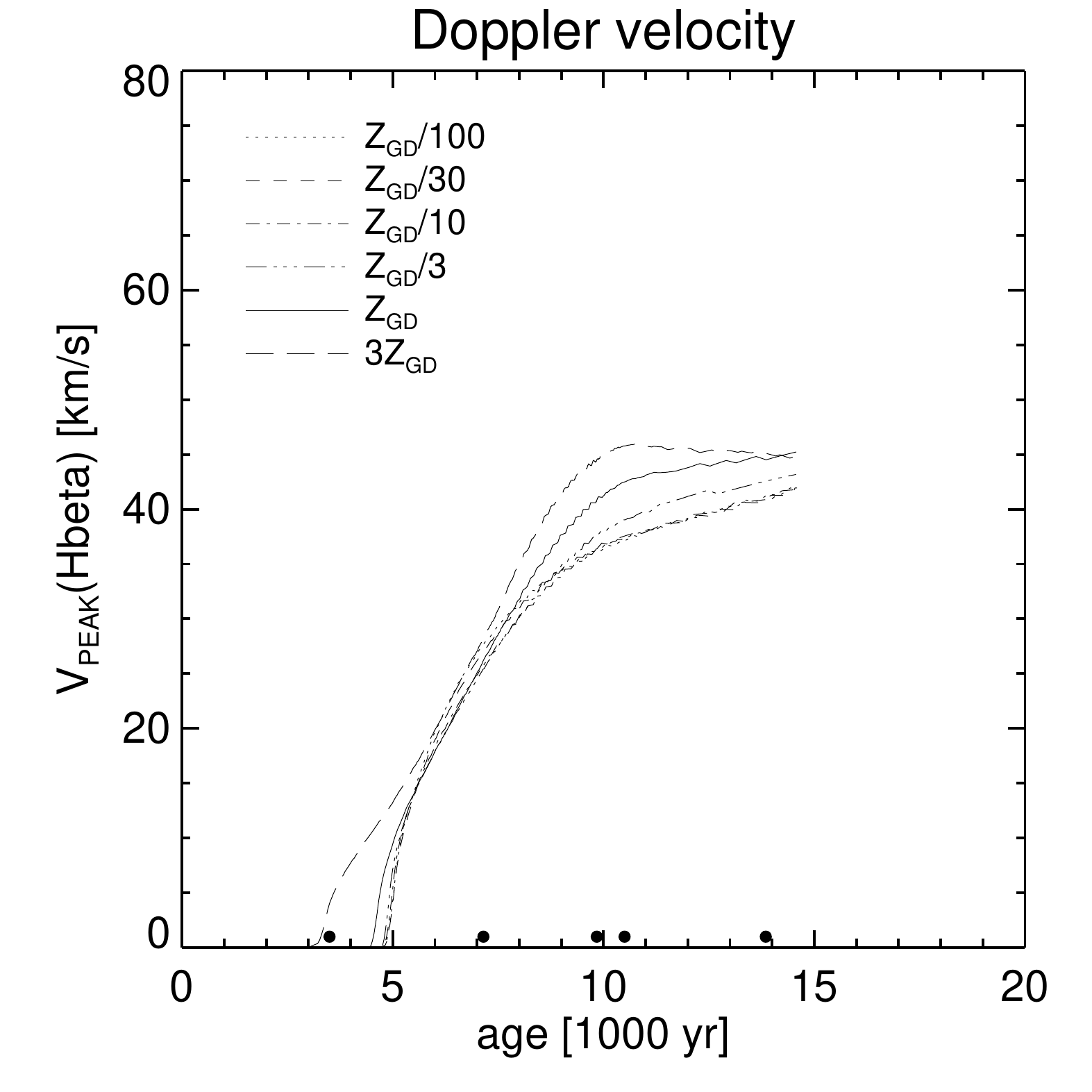}
\includegraphics*[bb= 0.7cm 0.5cm 15.4cm 15.1cm, width=0.35\linewidth]
                {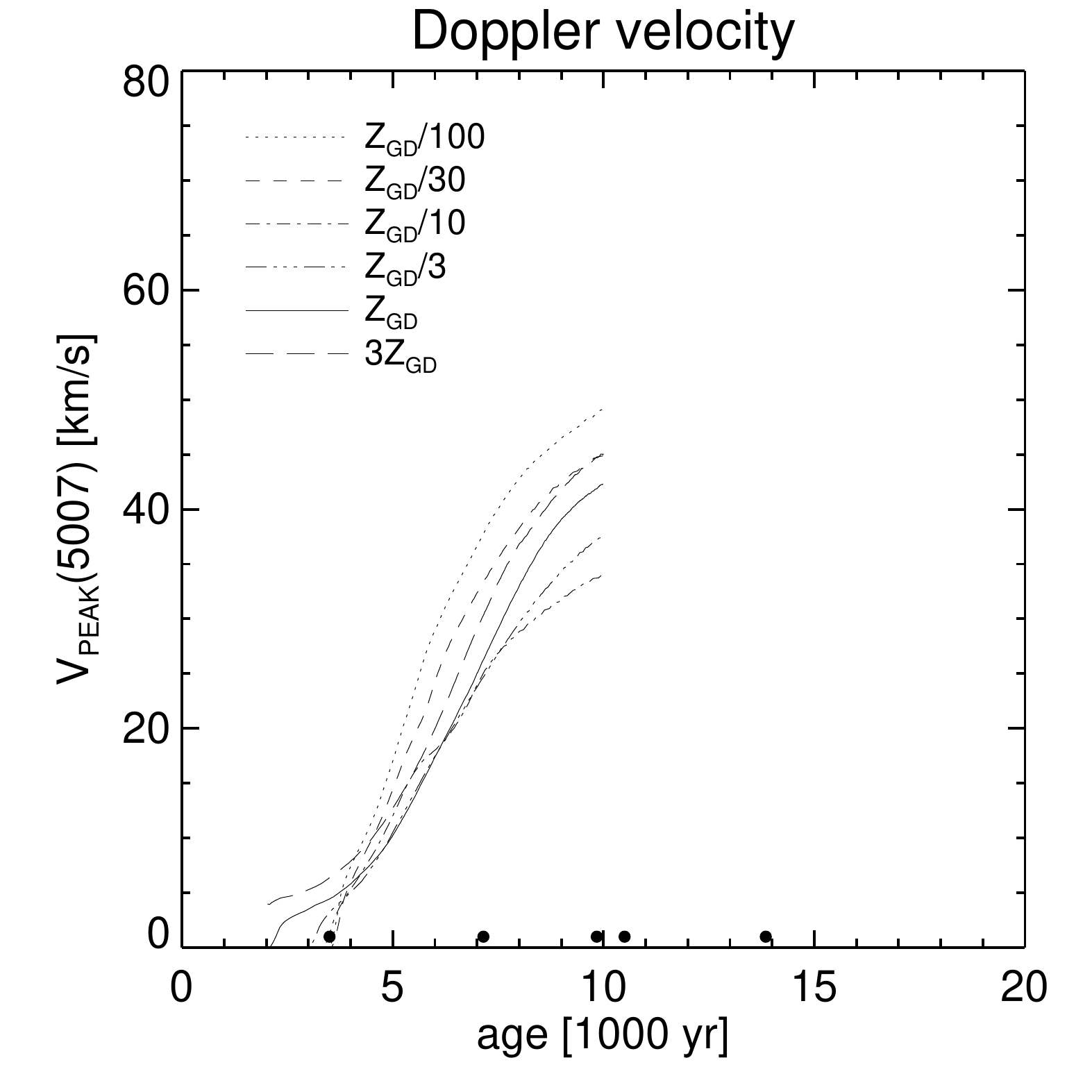}
\caption{\label{peak}
\changedIII{Expansion {properties as} derived from the line peak separation of 
         spatially resolved line profiles for the same six  ${\alpha=3}$ sequences 
         shown in Figs.\,\ref{vmax} and \ref{HWHM} vs. the post-AGB age, for H$\beta$ 
         (\emph{left}) and \oiii\ 5007~\AA\ (\emph{right}). The line profiles are
         simulated using a central numerical aperture of ${1\times10^{16}}$~cm
         (or 0\farcs67 at a distance of 1~kpc).
         At low ages, the profiles are still singly peaked, and thus ${V_{\rm peak}=0}$. 
         The \oiii\ profiles are
         only plotted up to model ages of about 10\,000~yr because beyond this age
         the profiles develop a complex structure due to recombination. The five 
         black dots along the abscissas correspond again to the snapshots depicted in 
         Figs. \ref{evol.1}--\ref{evol.4}. }
        }
\end{figure*}

\paragraph{Applications}
\label{app}

\changedIII{It is tempting to apply our model sequences to metal-poor objects of
            the Milky way. The two well-known objects \object{NGC 4361} and 
            \object{NGC 1360} appear especially suited for this purpose because
            structure and velocity informations are available from the literature. 
            \object{NGC 4361} belongs to the Galactic halo, while \object{NGC 1360} 
            is not known as an halo object but is metal-poor as well (see
            below).  According to \citet{mendetal.92}, both objects consist of a
            very highly excited nebula surrounding a very hot and luminous central star.}

\changedIII{Since both objects are spatially resolved, we computed the line
            profiles for an aperture centred on the position of the central 
            star (Fig.~\ref{peak}).}
\changedIII{At some time the lines become double-peaked, and this figure demonstrates that 
            the expansion rates as derived from the
            line peak separations, $V_{\rm peak}$, are very similar to those derived
            from the integrated profiles: The dependence on metallicity is weak or
            even partially absent (hydrogen lines only), and the true expansion
            velocity is strongly underestimated as well, especially for the metal-poor
            cases.}

\begin{figure}[t]
\includegraphics[width=\linewidth]
                {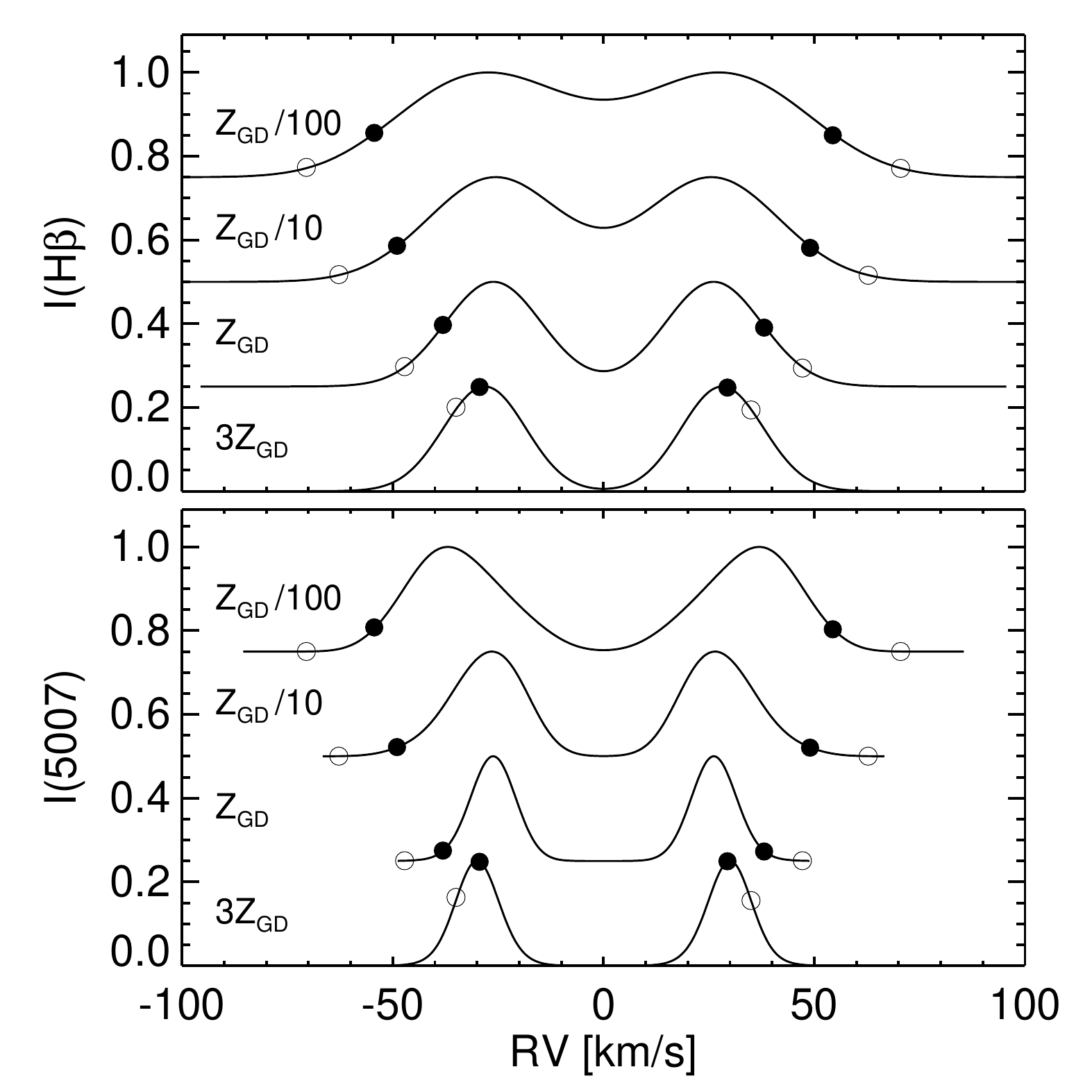}
\caption{\label{detail}
\changedIII{Line profiles of H$\beta$\ (\emph{top}) and \oiii\ 5007~\AA\ (\emph{bottom})
         computed for the models displayed in Fig.~\ref{evol.2} for a central aperture 
         of ${1\times10^{16}}$ cm and broadened by a Gaussian of 10 \kms\ FWHM.
         The profiles are normalised to $I=0.25$ and shifted by $\Delta I = 0.25$ each
         for more clarity. Different symbols mark shock (open circles) and post-shock 
         velocities (filled circles) of the models. }
        }
\end{figure}

\changedIII{For illustration, the (resolved) line profiles of hydrogen and \oiii\ for
            model ages of about 7100 years (cf. Fig.~\ref{evol.2}) are rendered in 
            detail in Fig.~\ref{detail}. At these ages, our models have very high nebular
            excitations and still luminous central stars with about 100\,000~K effective 
            temperature. The peak line emission corresponds to the denser
            inner regions where the emission measure is high but the gas velocity still
            rather modest (cf. upper panels of Fig.~\ref{evol.2}). The fast moving
            nebular layers immediately behind the outer shock contribute very little to 
            the total profile and are masked by the strong emission from the denser 
            nebular regions. This figure emphasises vividly how the real expansion
            of a PN is underestimated by employing the peak separation of resolved
            line profiles.\footnote
            {For the $3Z_{\rm GD}$ case the post-shock velocity coincides just by chance 
            with the peak velocity because the rim which is resposible for the peak 
            emission is being accelerated by the strong wind/bubble pressure 
            (see upper left-most panels of Figs.~\ref{evol.2} to \ref{evol.4}).}}

\changedIII{Figure~\ref{detail} displays also an interesting line width behaviour: 
            The widths increase with decreasing metallicity. This is due to the different
            ranges of gas velocities encountered in the models (cf. Fig.~\ref{evol.2}). 
            From the highest to the lowest metal content shown in Fig.~\ref{evol.2}, 
            this velocity range varies from about 10 to about 44 \kms. Also, we see for
            the lower metallicities a trend that the \oiii\ lines have larger peak
            separations than the hydrogen lines. The reason is the ionisation profile of 
            oxygen: At the evolutionary stage considered here, the O$^{2+}$ concentration
            increases towards the outer nebular region where the velocities are higher
            (cf. 6th left panel in Fig.~\ref{ionisation}).  } 
  
  Guided by our simulations, we expect metal-poor PNe to show a
  more smooth or diffuse morphology 
  without a clear-cut signature of a central cavity (or hole) and a sharp outer boundary. 
  The surface brightness profile of \object{NGC 4361} shows indeed neither a pronounced 
  central cavity nor a bright, distinct rim \citep{mon-iber.05}. 
  The metal content of this object is, on the average, $\simeq\!0.1Z_{\rm GD}$ 
  \citep[][Table 4 therein]{howard.97}. It has also one of the highest electron temperatures,
  19\,300 K, of the whole \citeauthor{howard.97} sample (see Fig.\,\ref{roiii}). 
  {The central star has an effective temperature of $\simeq\! 82\,000$~K \citep{mendetal.92}.}  

\begin{figure}[t]
\includegraphics[width=\linewidth]
                {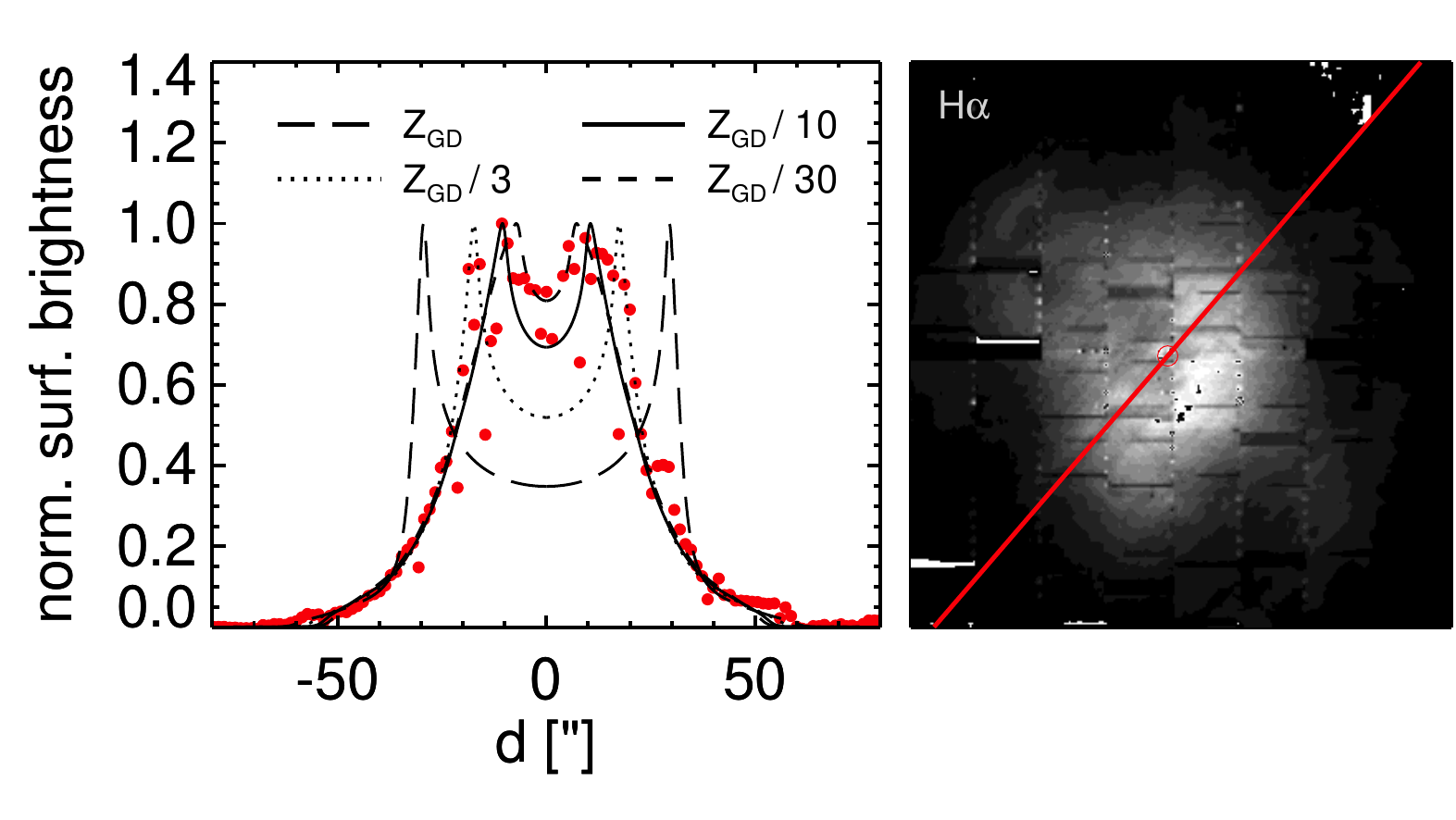}
\vskip-3mm
\caption{\label{ngc4361}
         H$\alpha$ surface brightness of \object{NGC 4361} (dots, \emph{left panel}) 
         derived from a cut along the semi-minor axis of the H$\alpha$ image as
         indicated (\emph{right panel}). We extracted this image from 
         observations reported in \citet{mon-iber.05}. 
         The dark horizontal bars are artefacts of the VIMOS spectrograph.
         Also shown are the predicted H$\alpha$ surface brightness distributions 
         of models with different metal contents (see figure legend), taken 
         from the ${\alpha=3}$ sequences.
         These models are chosen such that they match the observed size of 
         \object{NGC 4361}, assuming a distance of 1\,kpc. 
\changedIII{The model ages vary from about 6700 years ($Z_{\rm GD}/30$, fast 
            expansion) to 9100 years ($Z_{\rm GD}$, slow expansion). }
        }
\end{figure}

  In Fig.\,\ref{ngc4361} we compare the H$\alpha$ surface brightness of \object{NGC 4361} 
  with the predictions of our ${\alpha=3}$ models discussed here.  The surface brightness
  is taken from a cut along the semi-minor axis in order to avoid complications
  due to the weak ansae seen at the poles.  We conclude from the figure that 
  \emph{only} the models with a reduced metal content, ${Z \la Z_{\rm GD}/10}$,
  provide a quite reasonable agreement with the observation in terms of central dip 
  and intensity profile.  

\changedIII{The kinematics of our models can be tested as well: 
            The latest velocity measurements of \object{NGC 4361} are 
            ${V_{\rm peak}(5007)={27\pm2}}$~\kms\ \citep[][]{muan.01} and 
            ${V_{\rm peak}(\mbox{H}\beta)={22\pm5}}$~\kms, 
            ${V_{\rm peak}(5007)={26\pm5}}$~\kms\ \citep{medal.06}.
            \citet{muan.01} provide \oiii\ line profiles, and the one taken from the 
            central region of {NGC 4361} (their Fig. 4c) is compared with theoretical 
            profiles generated from the models of Fig.~\ref{ngc4361}.  Our metal-poorer
            models predict the right observed apparent expansion velocity, broader line
            profiles, and also ${V_{\rm peak}(5007)>V_{\rm peak}(\mbox{H}\beta)}$.  }

  The other case, \object{NGC 1360}, consists also of a hot 
  (97\,000 K), luminous (${\approx\!5000}$ \Lsun) central star \citep{trauletal.05} 
  and a rather smooth, extended nebula without a clear cavity/rim/shell signature 
  \citep[][Fig. 5 therein]{goldmanetal.04}. It is reported to
  have a very high electron temperature \citep[16\,500 K,][]{kaleretal.90}, indicative 
  of a correspondingly low metal content (see Sect. \ref{T.electron}).  Abundance 
  determinations for this object are rare: \citeauthor{kaleretal.90} provide only 
  ${\epsilon(\rm O^{2+}) = 7.8}$, while \citet{manchado.89} found 
  ${\epsilon(\rm O) > 8.2}$, 
  ${\epsilon(\rm Ne) > 7.6}$,
  ${\epsilon(\rm Ar) = 5.8}$, which are, however,
  based on an electron temperature of only 10\,000~K. A higher electron
  temperature as mentioned above would result in correspondingly lower abundances. 
\changed{\cite{trauletal.05} provide also photospheric abundances of the central star
         which, converted to the notation used here, are: $\epsilon(\rm C)=8.3$,
         $\epsilon(\rm N)= 7.7$, and $\epsilon(\rm O)=8.3$. 
         Thus, \object{NGC 1360} is certainly also a metal-poor object
        (compare with the Galactic disk abundances in Table \ref{tab.element}). }

\changedIII{\citet{garcetal.08} measured recently the apparent expansion of 
            \object{NGC 1360} by means of high-resolution echelle spectrograms: 
            ${V_{\rm peak}(\mbox{H}\alpha)= 26}$~\kms, consistent with older
            measurements of 24~\kms\ by \citet{goldmanetal.04}.
            Again, the agreement with the predictions of our models for this particular 
            stage of evolution ($\simeq$7100 yr, Fig.~\ref{peak}) is satisfying. 
            We repeat that the real expansion rate of both objects is expected to be 
            considerably higher, viz. up to a factor of 
            about two (cf. Fig.~\ref{vmax}, left panel, and Fig.~\ref{detail}). }

\changedIII{\citet{goldmanetal.04} determined also an elliptical spatio-kinematic 
            model of the nebula structure of \object{NGC 1360}, based on the 
            position-velocity ellipse and the surface-brightness distribution. 
            The basic result is that the gas density falls off outwards, while the 
            gas velocity increases outwards, i.e. the density (or surface brightness) 
            is highest close to the central star where the expansion velocity is 
            lowest. The fastest expanding shell has $\simeq$35 \kms\ at the equator
            and $\simeq$70 \kms\ at the pole. The density, velocity, and surface
            brightness profiles predicted by our metal-poorer models 
            (Fig.~\ref{evol.2}, top panels, and Fig.~\ref{sb}) are in  
            agreement with the mean properties of the \citeauthor{goldmanetal.04} 
            model of NGC 1360.  }    

\changedIII{\citet{garcetal.08} constructed a dynamical (2D) model including
            magnetic fields, but their main point was the following: They assumed 
            that the stellar wind has stopped shortly after the formation of the PN (i.e.
            after 1000 years) and followed the dynamical evolution with a collapsing
            hot bubble for additional 10\,000 years. Despite the authors' claim that their
            model ``successfully reproduce many of the key features of NGC 1360'', we
            note following inconsistencies: }
\begin{enumerate}
\item
\changedIII{The final nebular model at $t= 11\,000$~yr in Fig. 6 of \citet{garcetal.08}
            shows a positive gradient for both the velocity \emph{and} density: the outer
            fastest layers are also the densest, i.e. the model is limb-brightened
            (except at the poles), which is in clear contradiction with the observations.}
\item
\changedIII{Despite the long simulation time of 11\,000 years, an evolution of the star 
            is obviously not considered. 
            The stellar model envisaged is rather massive ($\simeq$0.75~\Msun,
            \citealt{vawo.94}) and crosses the HR diagram within
            only about 1000 years and fades thereafter, accompanied by a drop of
            wind power (cf. Figs.~\ref{schoen.star} and \ref{schoen.windmodel}). 
            After 11\,000 years, the model's luminosity is then only $\approx$100~\Lsun. }
\changedIII{But presently the central star of NGC 1360 is still a high-luminosity 
            object with several 1000~\Lsun, luminous enough to sustain a radiation-driven
            wind which, because of the low metallicity,
            may be too weak for developing spectroscopic signatures. }   
\end{enumerate}

\begin{figure}[t]
\vskip-2mm
\includegraphics[width=\linewidth]
                {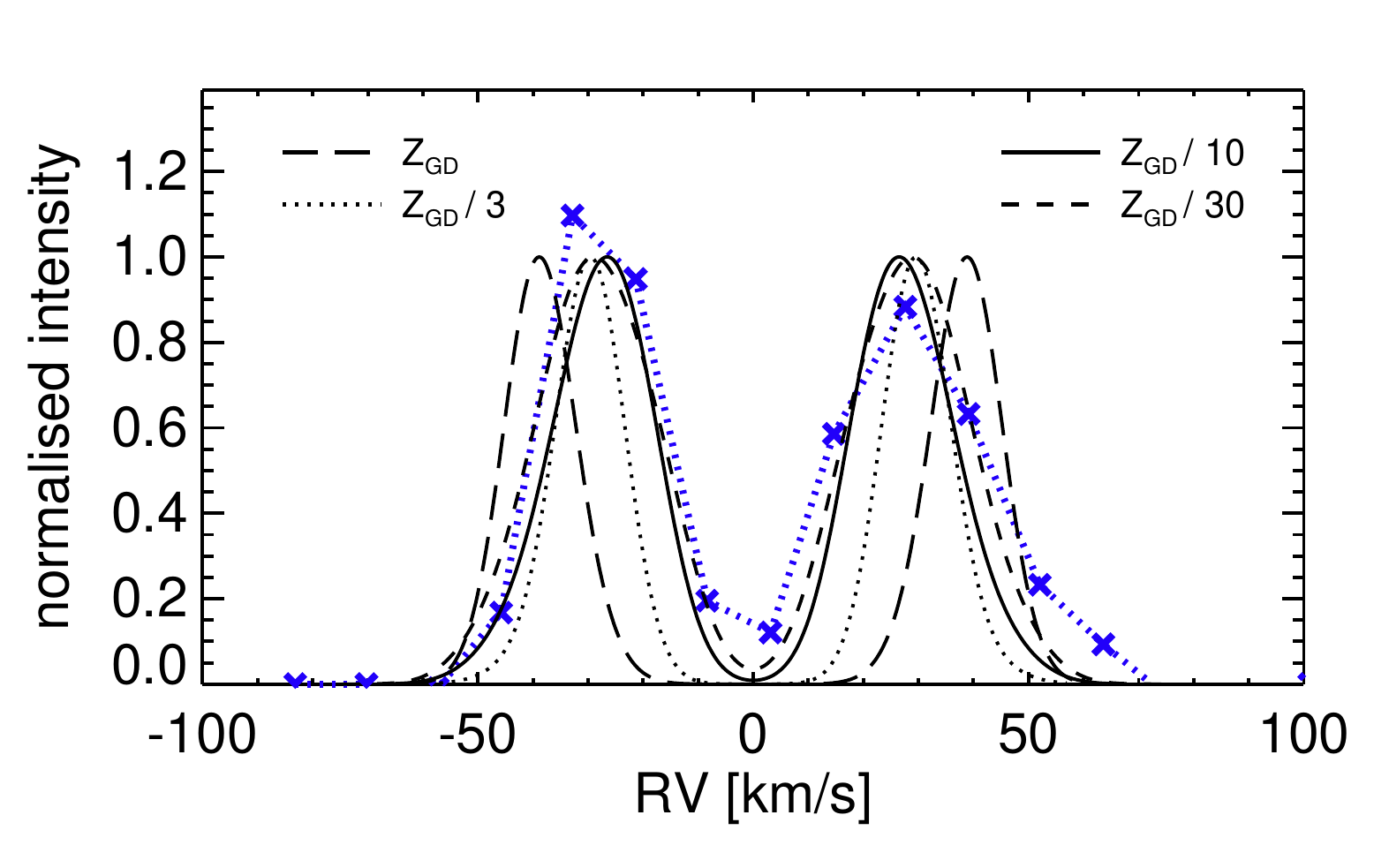}
\vskip-2mm
\caption{\label{ngc4361.muhtu}
\changedIII{Observed central 5007~\AA\ line profile of NGC 4361 (connected crosses) as 
         measured by \citet{muan.01} compared with the profiles generated from the 
         models used in Fig.~\ref{ngc4361}. As in Fig.~\ref{detail}, the theoretical 
         profiles are computed for a central aperture of of ${1\times10^{16}}$ cm and 
         then broadened by a Gaussian of, now, 12 \kms\ FWHM.       }
        }
\end{figure}

\changedIII{We emphasise at the end of this discussion that models taken from the ${\alpha=3}$ 
            sequences displayed in Figs.~\ref{evol.1} to \ref{evol.4} can explain the
            (mean) properties of two rather well-known metal-poor objects very successfully,
            although they were not designed to fit any particular object! Furthermore,}
  according to our models, it is not justified to conclude, just from the non-existence of
  an apparent central hole in the surface brightness distribution of a PN, that the 
  central-star wind has already stopped and that the nebular gas is now falling back 
  onto the star. 
\changedIII{Rather, a detailed analysis of the density and velocity structure} 
  of the object in question appears to be mandatory in order to decide whether a
  deceased central-star wind is responsible for a more smooth or diffuse appearance of a 
  PN, or whether we observe a low-metallicity system with a weak central-star wind and a 
  more extended nebular structure.

\changedIII{We acknowledge, however, that our spherical models are certainly too simple
            in order to provide detailed models of the objects discussed in this section. 
            Important further ingredients are, e.g., inhomogeneities, jets, and other
            means that impose departures from sphericity.  We note also that magnetic fields
            have been detected in NGC 1360 by \citet{jordetal.05}, but whether they are
            really important in shaping NGC 1360, as believed by \citet{garcetal.08},
            must be seen in the future when more realistic simulations become possible. }

\begin{figure*}[t]
\sidecaption
\includegraphics*[bb= 0.7cm 0.5cm 15.4cm 15.07cm, width=0.35\linewidth]
                {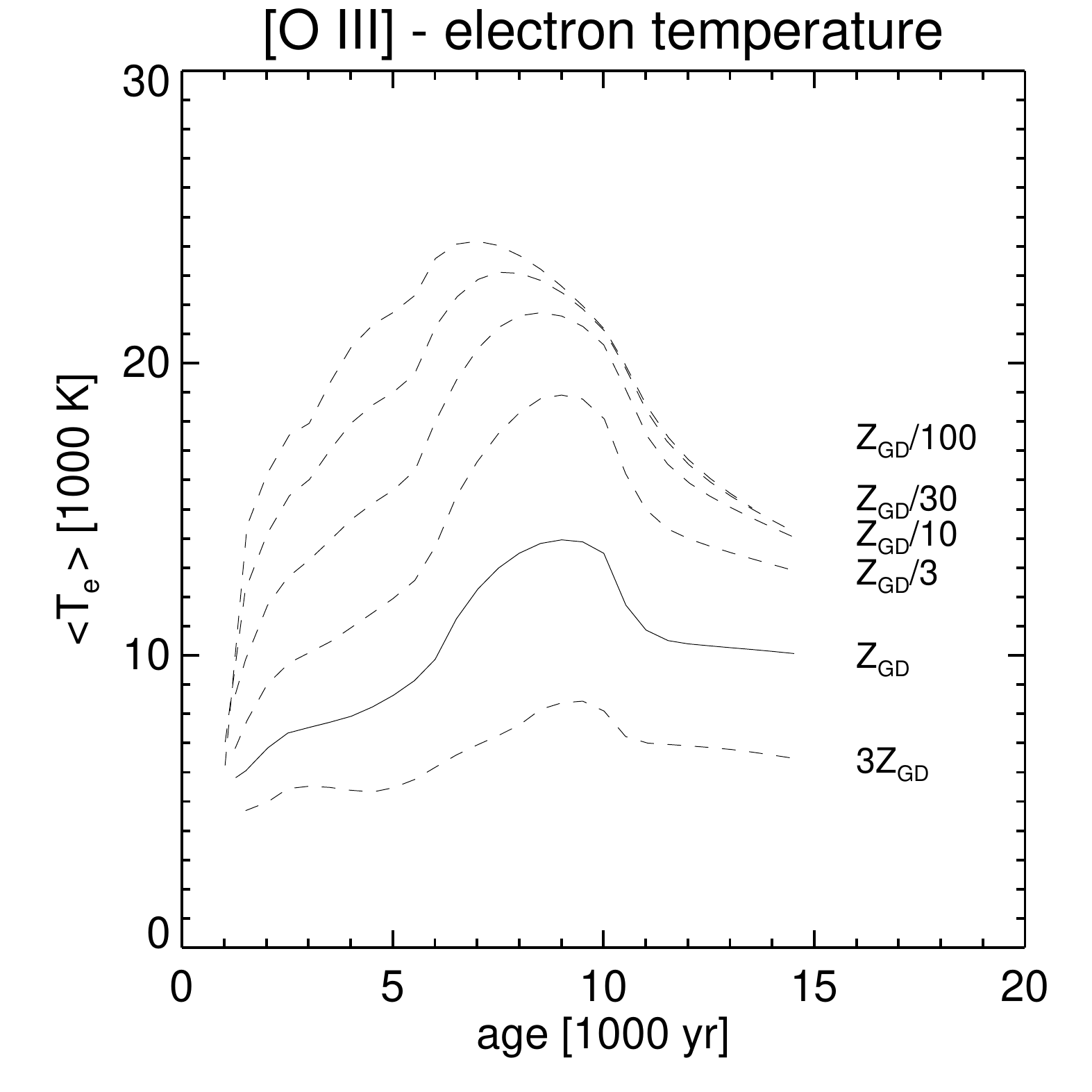}
\includegraphics*[bb= 0.7cm 0.5cm 15.4cm 15.07cm, width=0.35\linewidth]
                {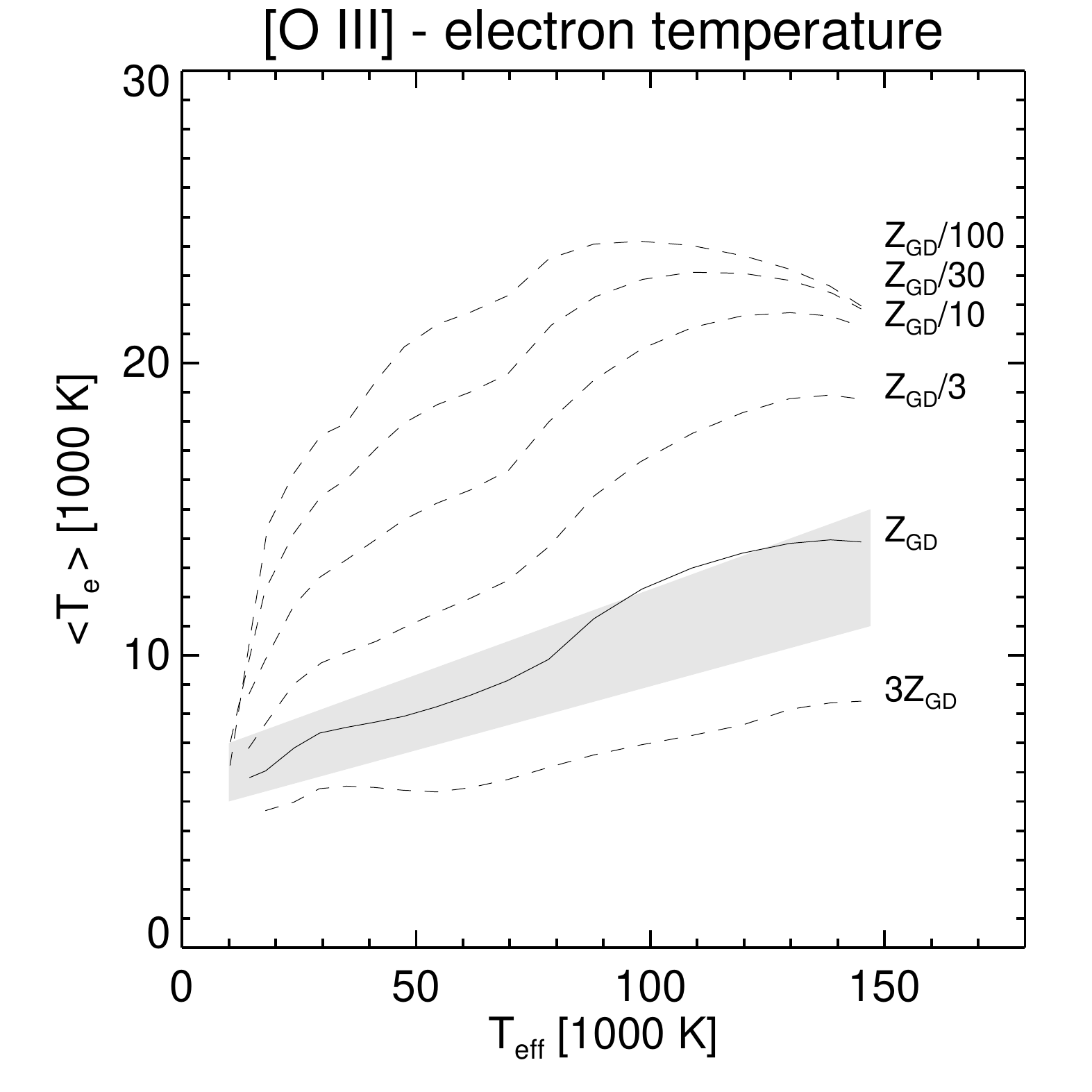}
\caption{Mean [\ion{O}{iii}] electron temperatures, computed according to
         Eq.~(\ref{1}), for all ${\alpha=3}$ nebular models
         as function of the post-AGB age (\emph{left}) and stellar effective temperature
         (\emph{right}).  In the right panel, the models are only plotted until maximum 
         central star
	 temperature (corresponding to a post-AGB age of $\simeq$9800 years in the
         left panel) in order to avoid confusion due to overlapping lines.
         The grey area depicts approximately the region occupied by the hydrodynamic
	 models presented in \citetalias{perinotto.04} and \citetalias{schoenetal.07}
         which have all ${Z=Z_{\rm GD}}$, but various initial conditions and central 
         stars.
        }
\label{etemp.1}
\end{figure*}

\subsubsection{Electron temperatures}
\label{T.electron}

  We computed the mean electron temperatures of the models according to
\begin{equation}
   \langle T_{\rm e} \rangle=\frac{\int{T_\mathrm{e}(r)N_\mathrm{e}(r)N_\mathrm{i}(r)\,
    \mathrm{d}V}}{\int{N_\mathrm{e}(r)N_\mathrm{i}(r)\,{\rm d}V}},       \label{1}
\end{equation}
  with $N_\mathrm{e}(r)$ being the electron number density, $N_\mathrm{i}(r)$ the O$^{+2}$
  number density, $T_{\rm e}(r)$ the electron temperature,
  and ${\rm d}V$ the volume element.  The upper boundary for the volume
  integration is set at the outer shock front; the 
  halo (= ionised former AGB wind) is thus excluded.  In passing we remark that 
  Eq. (\ref{1}) corresponds in practice to an \emph{emission weighted} mean
  of the electron temperature, in contrast to the usual meaning of a mean value.
  Differences are only expected for cases with radial temperature gradients and/or
  strong shocks as seen in the late, metal-poor models of Figs. \ref{evol.1}--\ref{evol.4}.

  The determination of the mean electron temperature from the model structure by using 
  Eq. (\ref{1}) is convenient but different from the one used in practice if the
  nebular structure is not known.
  The alternative is to compute first the (total) line strengths from the model and
  to apply then, e.g., Eq. (5.4) of \citet{ost.89} to get a volume averaged
  $T_{\rm O\,III}$ from the ratio $R_{\rm O\,III}$.  Provided the collision strengths 
  are correct, both methods should give the same result.  We performed a test using
  two sequences with rather extreme metallicities, viz. with $Z_{\rm GD}$ and
  $Z_{\rm GD}/100$.  In the first case we found always
  $|\langle T_{\rm e}\rangle -  T_{\rm O\,III}|\la 60$ K, in the second
  $|\langle T_{\rm e}\rangle -  T_{\rm O\,III}|\la 170$ K.  Thus the 
  (relative) differences between both methods are less than about 1\%,
  and the temperature determination according to Eq. (\ref{1}) provides an excellent 
  value for $T_{\rm O\,III}$.

  The run of the mean electron temperature, $\langle T_{\rm e} \rangle$, 
  is displayed for all ${\alpha=3}$ sequences
  in Fig.~\ref{etemp.1}, and we see, as expected, a strong temperature dependence on
  metallicity.  The first rapid increase of   
  the mean electron temperature is, of course, due to hydrogen ionisation.
  Already at this stage one sees clearly how rapidly the electron temperature
  increases with decreasing cooling efficiency.  Then the temperatures increase further
  as the central star becomes hotter and its photons more energetic.
  The second temperature ``jump'' after about 5000 years is due
  the second ionisation of helium, providing additional energy input to the gas.

  Because the effective temperature decreases once the central star begins to 
  fade rapidly at an age of about 10\,000 years (cf. Fig. \ref{schoen.star}), the 
  electron temperature in the nebula must drop accordingly.  This is clearly seen
  in Fig.\,\ref{etemp.1} (left) for the more metal-rich models in which expansion cooling is
  not very important.  However, the relation $\langle T_{\rm e}\rangle$-$\teff$\ 
  differs between the stellar high- and low-luminosity branch because of differences in 
  the nebular ionisation: At a given stellar temperature, the nebular ionisation is
  higher and line cooling lower at the high-luminosity branch. 
  Hence, the electron temperature along 
  the low-luminosity branch of evolution is significantly \emph{below} that along 
  the corresponding high-luminosity branch, independently of non-equilibrium effects.

  An interesting effect is seen for the more metal-poor sequences (${Z < Z_{\rm GD}/3}$)
  \changed{which expand faster and become hence more dilute}:
  Here expansion cooling 
\changed{becomes relevant during more advanced evolutionary stages} 
  and forces the nebular 
  electron temperature to drop at progressively lower stellar temperatures (cf.
  Fig. \ref{etemp.1}, right panel).
\changed{A more detailed discussion about expansion cooling
  is given in Sect. \ref{line.exp.cool}. }

  Note that the interplay between line and expansion cooling, and their relative importance, 
  depend sensitively on gas density and expansion velocity.  It is, however, clear from our
  simulations that a low metal content results in high  electron temperatures because of 
  less line cooling, leading consequently to faster expanding and more diluted nebulae 
  which are then prone to departures from thermal equilibrium.
\changed{The grey region seen in the right panel of Fig. \ref{etemp.1} delineates the
         approximate area occupied by $\langle T_{\rm e} \rangle$ of all the 
         hydrodynamical sequences previously used in \citetalias{perinotto.04} and
         \citetalias{schoenetal.07}, which have very different 
         combinations of initial AGB-envelope configuration and central star,
         but all with the $Z_{\rm GD}$ composition. These models sequences demonstrate 
         that only a low metal content can be responsible
         for unusually high electron temperatures in nebular envelopes.} 

  We note also that the conspicuous shock regions seen in Figs. 
  \ref{evol.1}--\ref{evol.4} do not contribute to the electron temperatures, although 
  the post-shock temperatures become very high and are included in the integration 
  according to Eq. (\ref{1}).  The reason are the low ion (i.e. O$^{2+}$) densities, 
  giving these outer shock regions only a minute weight in the integral.

\subsubsection{Comparisons with equilibrium models}
\label{equi.model}

  With the computation of equilibrium models (with respect to ionisation and
  thermal energy) at selected positions along the evolutionary path we have a unique
  tool to investigate under which circumstances non-equilibrium effects may become
  important.

\paragraph{Electron temperatures}
\label{el.temp}

  In equilibrium, the electron temperature is determined by the balance between radiative 
  heating and cooling processes:  
\changed{Because heating is mainly due to ionisation of hydrogen and helium, and because 
         generally a significant contribution to cooling comes from line radiation of 
         heavier ions,} 
  the models must become hotter with lower metallicities.
  During the course of evolution, the electron temperatures generally increase 
  because the stellar photons become more energetic (see discussion in 
  Sect. \ref{T.electron} above).   At low metallicity the 
\changed{contribution of radiation from collisionally excited metal ions is reduced, 
         and line cooling is limited mainly to free-bound emission of hydrogen and 
         helium. Hence, cooling by expansion becomes more important for the  
         energy balance,} thereby limiting the electron temperature increase.  
         This process is favoured by the larger expansion rates of the low-metallicity 
         models. \changed{See Sect. \ref{line.exp.cool} for more details.}

\begin{figure}[t]
\vskip-5mm
\includegraphics[width=\linewidth]{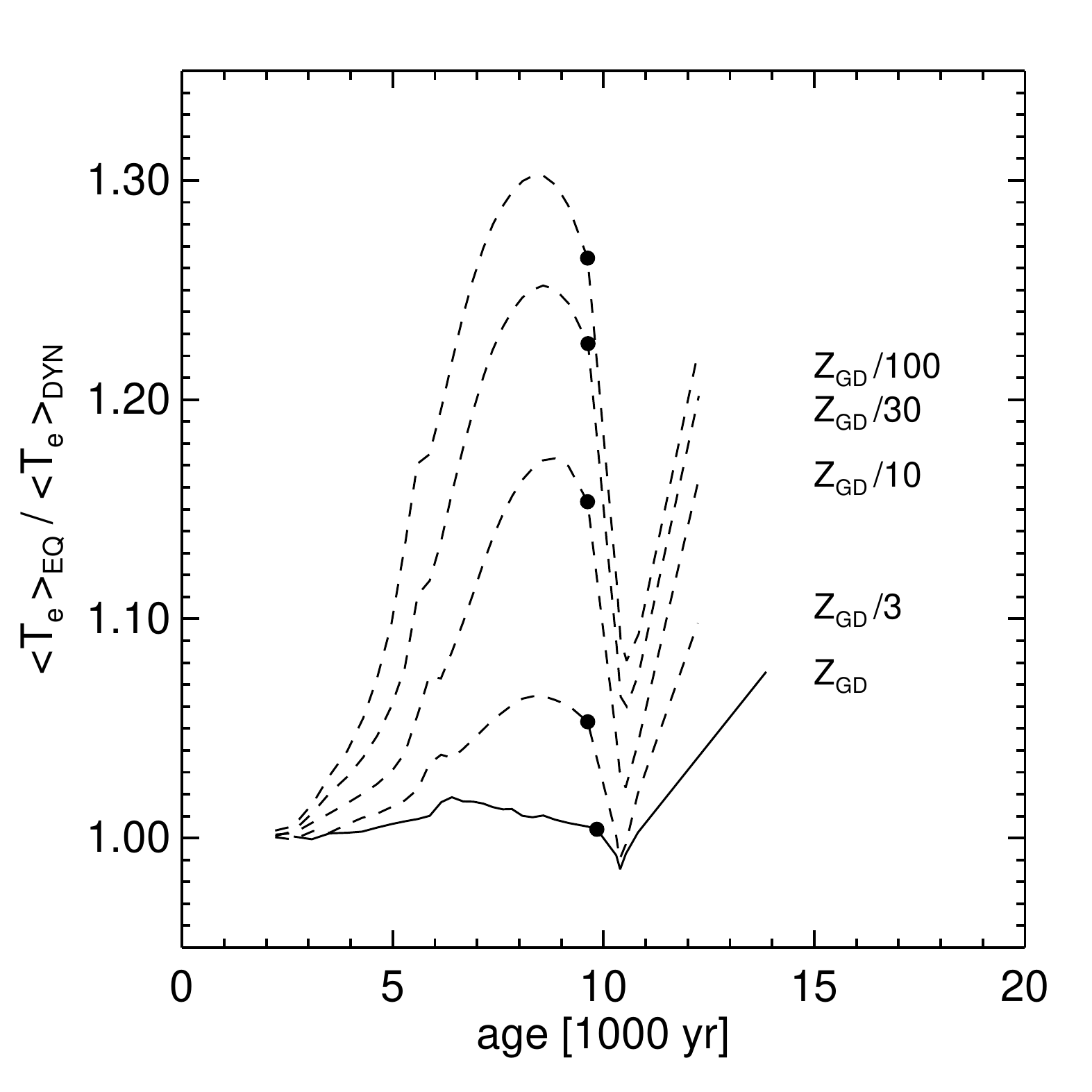}
\vskip-2mm
\caption{Ratios between electron temperatures of equilibrium (eq) and dynamical (dyn)
         ${\alpha=3}$ models with different metal contents are plotted versus post-AGB age.
         The filled dot on each curve indicates the position of maximum stellar temperature.
\changed{The sharp local minima of the temperature ratios occur during nebular 
         recombination while the central star fades rapidly (cf. Figs. \ref{schoen.star}
         and \ref{evol.3a}). See text for details.}   
}
\label{comp.temp}
\end{figure}

  The bottom panels of Figs. \ref{evol.1}--\ref{evol.4} display both
  the dynamic and the equilibrium electron temperatures. 
\changed{The close correspondence between the dynamic and equilibrium electron 
         temperatures seen in the metal-rich models during the horizontal evolution
         across the HR diagram indicates that the gas heating time scales are much
         smaller than the stellar time scales for heating due to the ionisation of
         hydrogen and helium \citep[see also][]{marten.95}.  This statement holds 
         independently of the metal content
         of the gas, and any deviations seen at later stages and/or for lower 
         metallicities must be due to the dynamics.}
  As expected, the
  temperature differences become largest for the models with the lowest metallicity, 
\changed{and hence lowest line cooling,}  
  and can reach up to about 10\,000 K in extreme cases.

  Figure \ref{comp.temp} gives a more detailed illustration of how the deviations 
  between dynamical and
  equilibrium models, measured by their mean O$^{2+}$ electron temperatures, 
  develop with time (or evolutionary stage of the central star).  
  One sees that, in general, the differences 
  increase with evolution, and they become large for the metal-poor models: 
  Under equilibrium conditions the electron temperatures can be higher by up to 30\,\% 
  in the metal-poorest case of Fig.\,\ref{comp.temp}.
  
\changed{During the fast luminosity drop of the central star between 10\,000 and 
         11\,000 years (cf. Figs. \ref{schoen.star} and \ref{evol.3a}) one sees 
         in Fig. \ref{comp.temp} that the mean temperature ratio eq/dyn becomes 
         equal to or even slightly lower than unity.
         Line cooling increased by recombination is not sufficient to explain 
         this; rather we see a typical time-scale effect: The fading time scale of the star, 
         $-L/\dot{L}$, becomes for about 300 years as short as 400 years, which is quite
         comparable to the cooling time scale of the diluted nebular gas. Thus, it may well
         happen that the electron temperature in thermal equilibrium comes close to or
         even falls \emph{below} non-equilibrium value
         because the latter cannot keep pace with the fading central star.
         After stellar fading is completed (at ${\approx\!\! 11\,000}$ years in 
         Fig. \ref{comp.temp}), the previous differences of electron temperatures are 
         restored quickly. }

\begin{figure}[t]
\includegraphics[width=\linewidth]{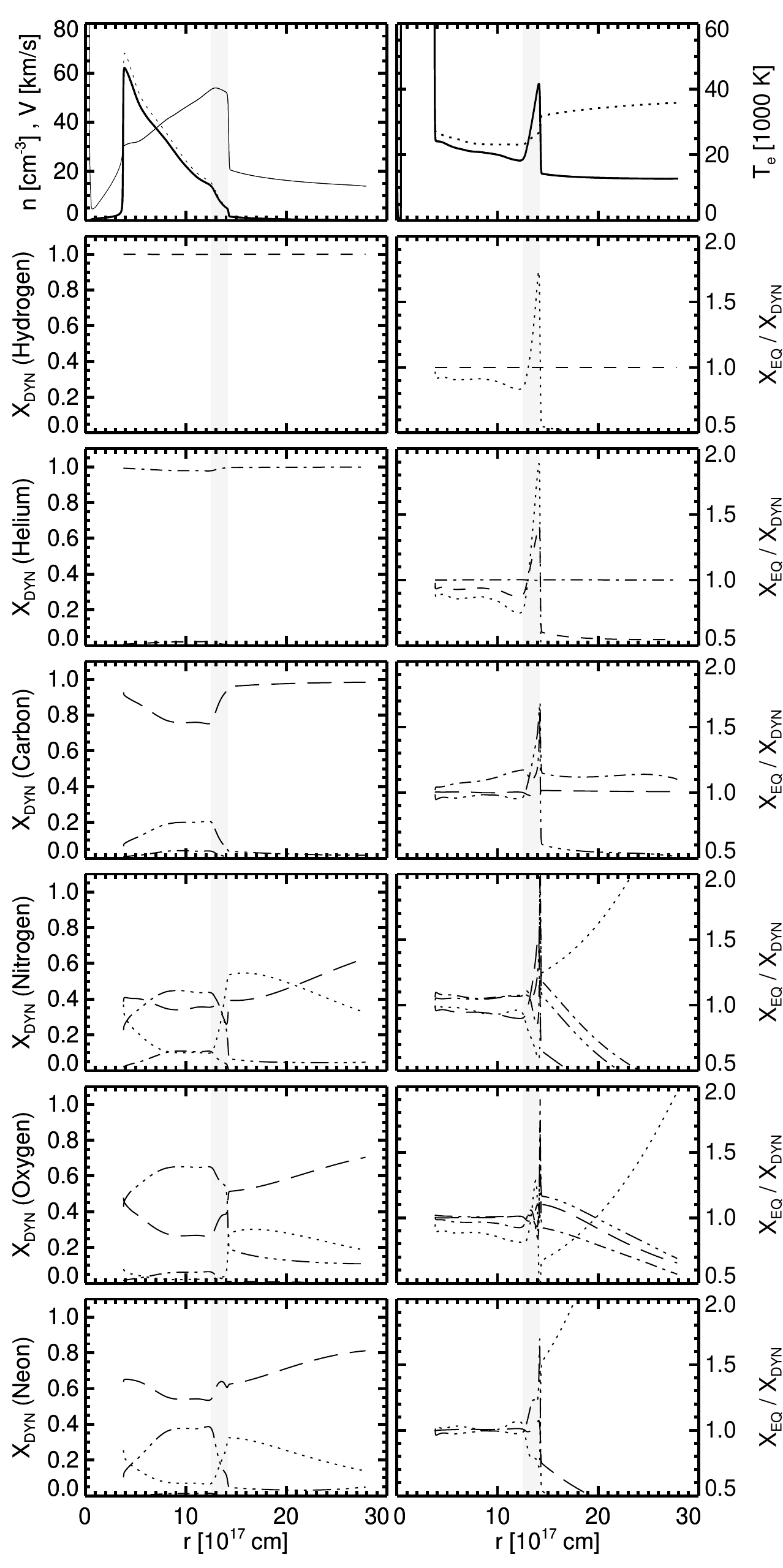}
\vskip-1mm
\caption{\changed{Radial ionisation structures for hydrogen, helium, carbon, nitrogen, oxygen, 
         and neon of the ${Z=Z_{\rm GD}/10}$ model presented in Fig.\,\ref{evol.3}. 
         The two \emph{top panels} show again the underlying nebular structures: 
         densities 
         (ions = thick, electrons = dotted), gas velocity (thin), both \emph{left},
         and electron temperatures (dynamical = solid, equilibrium = dotted, 
         \emph{right}).  The shadowed vertical strip marks the region occupied by 
         the shell's leading shock in all panels.  
         The ionisation fractions, $X$, are labelled as follows: dotted = neutral
         (hydrogen and helium only), short-dashed = 1st ionisation, dash-dotted = 
         2nd ionisation, dash-dot-dot-dot = 3nd ionisation, long-dashed = 
         4th ionisation, and dotted = 5th ionisation (except for hydrogen and helium), 
         and are displayed in the \emph{left panels} for the dynamical model 
         ($X_{\rm dyn}$). The ionisation fractions are omitted inside the 
         contact discontinuity (${r < 3.5\times 10^{17}}$ cm) in order to avoid 
         confusion. On the \emph{right}, the ratios between the equilibrium (eq) and 
         dynamical case (dyn) are shown. Only the ratios of major ionisation 
         fractions, which are seen also in the left panels, are rendered for clarity.}
}
\label{ionisation}
\end{figure}

  A few comments concerning the halo, i.e. the still rather undisturbed but ionised
  AGB material ahead of the outer shock front, are in order here.
  \citet{marten.93a, marten.95} showed that, because of its rather low density,
  the halo gas is especially prone to non-equilibrium conditions, even with solar
  metal content.  
\changed{Generally one can say that the energy balance of the inner, denser halo regions
         of young PNe are controlled by line cooling, while in the outer, less dense 
         regions expansion cooling prevails. }
  
\changed{Right after the passage of the ionisation front, the halo becomes quite hot
         and cools then slowly down, with a (local) time scale ruled by the local density. 
         Such temperature profiles can be seen for the metal-rich 
         models in Figs.~\ref{evol.2}--\ref{evol.4} (lower leftmost panels) which 
         are very compact and become optically thin quite late when the central star is
         already quite hot ($\teff \simeq 40\,000$--$42\,200$ K).}

\changed{The situation is different for the other models. They become optically thin
         earlier with cooler central stars ($\teff \simeq 33\,000$--$35\,000$ K), 
         and the halo is not so much heated in the first place.  
         According to \citet{marten.95} the mean heating time scale of the halo gas can be 
         comparable to or even larger than the stellar heating time scale due to
         ionisation of hydrogen and helium.
         Under such condition the dynamical haloes remain always quite cool during the
         whole evolution, in contrast to the equilibrium condition.  In general, the halo 
         temperatures follow the trend of the main nebulae and become hotter with 
         decreasing metal content.  
         In the extreme cases the equilibrium electron temperatures exceed
         locally 40\,000~K (cf. Figs. \ref{evol.2}--\ref{evol.3}).

\paragraph{Ionisation structure}
\label{ion.eq}

  Although a PN might well be out of thermal equilibrium because of dynamics, ionisation
  equilibrium is generally rather well established.
{\changed{During the high-luminosity phase as the object crosses the HR diagram, ionisation
          dominates over recombination in the ionisation equations, and the 
          nebular ionisation quickly adjusts to the stellar ionising photon flux.
          Our hydrodynamical models are thus always 
          close to ionisation equilibrium. An exception is only possible for a brief period 
          during the rapid stellar fading when the ionisation time scale may become larger 
          than the {recombination} time scale (e.g., of hydrogen), and provided the latter 
          is then comparable to or larger than the fading time scale of the star. 
          Ionisation equilibrium is quickly re-installed later when the stellar luminosity
          evolution slows down on the white-dwarf cooling track.}

\changed{As an example we compare in Fig.\,\ref{ionisation} the ionisation properties} 
  of the dynamical and the thermally relaxed $Z_{\rm GD}/10$ model 
  from Fig.\,\ref{evol.3}.
\changed{We selected this model because it is far evolved and has  
  a very high degree of excitation because of the very hot and luminous central star. 
  The metallicity chosen, $Z_{\rm GD}/10$, is typical 
  for very metal-poor
  populations, e.g., in the Galactic halo and in some distant stellar systems.}

  One sees in Fig. \ref{ionisation} that behind the 
  shock (${r\la 12\times10^{17}}$ cm), i.e. in the nebula, the ionisation 
  structures of both model types are very similar \changed{but not identical.}  
  The deviations seen in some cases are up to  
  10 to 15\,\% \changed{for main ionisation stages.
  These obvious deviations from a pure \emph{photo}ionisation equilibrium are due to
  \emph{collisional} ionisation whose contribution, albeit quite small, increases 
  with electron temperature. Different electron temperatures lead then 
  consequentially also to small differences of the ionisation.}

\changed{The situation is demonstrated for hydrogen and helium (2nd and
         3rd right panels in Fig. \ref{ionisation}). In the dynamical case, 
         hydrogen and helium are \emph{less} ionised both in the halo and the
         nebula proper. The  departures of the ionisation fraction ratios
         from unity follows closely the difference between, or ratio of, the
         electron temperatures. Only during the passage through the of the
         high-temperature shock region the ionisation of hydrogen and helium
         becomes temporarily \emph{larger} than in equilibrium because of 
         enhanced collisional ionisation. }

  Note also that for the particular, very diluted model shown in 
  Fig.\,\ref{ionisation} where the central star is very hot and still quite
  luminous, doubly-ionised oxygen which is so important for analysing nebular
  spectra is only a minority species. 

\changed{Because of its lower density, the halo, i.e, the matter ahead of the
         shock at ${r\ga 14\times10^{17}}$ cm, shows generally a larger degree of 
         ionisation than the main nebula. For instance, the fraction of neutral
         hydrogen is about $10^{-4}$ in the halo, but about $10^{-3}$ in the 
         nebula. Similar differences occur for the fraction of singly ionised helium: 
         about $4\times10^{-3}$ in the halo and about ${4\times10^{-2}}$ in the 
         nebula. The heavier elements show the same behaviour (cf. left panels of
         Fig. \ref{ionisation}, for N, O, and Ne).}

  It is also seen in Fig.\,\ref{ionisation} (right panels) that the halo region
\changed{is also prone to deviations from the}
  ionisation/recombination equilibrium:  the haloes of the dynamical models have  
  lower degrees of ionisation than in the corresponding equilibrium cases
  (cf. nitrogen, oxygen, and neon in
  Fig.\,\ref{ionisation}), indicating that the gas ionisation time scale exceeds 
  the stellar ionisation time scale. 
  The reason is the small fraction of high-energy photons 
  which are able to sustain such a high ionisation. 
\changed{Thus, the ionisation time scale becomes the longer the higher the gas ionisation.}  
\changed{Additionally, ionisation by electron collisions is more important
         in the cases of the very high halo temperatures. In equilibrium, the
         5th ionisation (N, O, and Ne) appears to be the preferred one. After the 
         halo gas becomes swallowed by the shock recombination brings the gas 
         immediately back to a lower degree of ionisation very close to equilibrium 
         conditions.}

\changed{The generally higher degree of ionisation within the haloes leads to a 
         lower line cooling efficiency which in turn is responsible for 
         the quite substantial temperature jumps across the nebula/halo boundaries, 
         i.e. the shock, seen in the equilibrium models (top right panel 
         of Fig. \ref{ionisation}). }

\subsubsection{Line vs. expansion cooling}
\label{line.exp.cool}

\changed{Here we want to discuss in more detail the relevance of the different
         cooling processes encountered in our models.  The nebular gas is heated 
         by ionisation and cooled by line radiation from free-free, bound-free 
         (recombination) transitions, and 
         from collisionally excited atomic or ionic levels.\footnote
         {\changed{Our time-dependent code treats heating by ionisation and 
                   cooling by line radiation separately, without considering 
                   \emph{net} heating rates which hold only
                   under the condition of strict ionisation equilibrium.} }
         Additionally, the 
         gas is subject to dynamical processes leading to cooling by expansion or heating
         by compression.} 

\changed{In standard photoionisation modelling it is implicitly assumed that dynamics is 
         unimportant, hence the thermal balance of the heated nebular gas is controlled by 
         line radiation only. Since the most important coolants are metal ions, the cooling
         function decreases with metallicity until radiation from hydrogen and helium 
         prevails. Furthermore, radiation cooling depends on
         gas density squared, leading to a reduction of the cooling efficiency also because
         the nebula expands with time. In contrast, cooling by expansion depends only on
         the gas velocity.}
  
\changed{Under adiabatic conditions, the total change of thermal energy content of 
         a volume element, $U$ (in erg\,cm$^{-3}$), with time is given by
        \begin{equation}  \label{A11}
        \frac{{\rm D}U}{{\rm D}t} = - p\,(\nabla\cdot\vec{v}) - U\,(\nabla\cdot\vec{v}) , 
        \end{equation} 
        where \vec{v} is the flow velocity and $p$ the thermal pressure 
        \citep[see Eq. A11 in][]{marten.97}.  
        The term ${p\,(\nabla\cdot\vec{v})}$ is the usual local source (sink) term 
        due to compression (expansion).}
\changed{The second term of the left-hand side of Eq. (\ref{A11}) accounts for 
         the change of the volume of a gas parcel with time, i.e. a gas parcel becomes 
         expanded or compressed while streaming.} 

\changed{It is more convenient to follow a mass element and
         rewrite Eq. (\ref{A11}) using the thermal energy per mass, $e=U/\rho$ 
         (in erg\,g$^{-1}$), as
         \begin{equation}  \label{A11a}   
          \rho\, \frac{{\rm D}e}{{\rm D}t} = - p\,(\nabla\cdot\vec{v})\, .
         \end{equation} 
         Note that only in case of constant density, 
         $\rho\, {\rm D}e/{\rm D}t = {\rm D}U/{\rm D}t$ follows.}

\changed{Introducing ${p=n\, k_{\rm B} T_{\rm e}}$, 
         with $n$ being the total particle density
         (ions + electrons), $k_{\rm B}$ the Boltzmann constant, and $T_{\rm e}$ the
         electron temperature, we get from Eq. (\ref{A11a}) in spherical coordinates
         \begin{equation}  \label{E.exp}
         \rho\,\frac{{\rm D}e}{{\rm D}t} =  - n\,k_{\rm B} T_{\rm e}
         \left \{\frac{\partial v}{\partial r} + \frac{2v}{r} \right \}
         \quad \rm [erg\,cm^{-3}\,s^{-1}].
         \end{equation}  }

\changed{Estimating possible dynamical effects demands knowledge of the density 
         and the velocity structure of the object in question, both of which are 
         difficult -- if not impossible -- to get observationally. One may use the 
         measured line width $V_{\rm HWHM}$ which, however, is an unsuitable choice: 
         This velocity value is (i) much lower than the real expansion speed 
         and (ii) is constant, (i.e. the velocity gradient disappears), leading to a
         substantial underestimate of expansion cooling.  The assumption of a 
         constant electron density may also introduce an uncertainty.
         It remains the likewise difficult choice of a -- distance dependent --
         radius range.
         One could determine the radial extent of an object from a model, or
         select the outer edge of the nebula.  In the latter case, an additional 
         systematic underestimate of the expansion cooling would result.}

\begin{figure}[t]
\includegraphics[width=\linewidth]{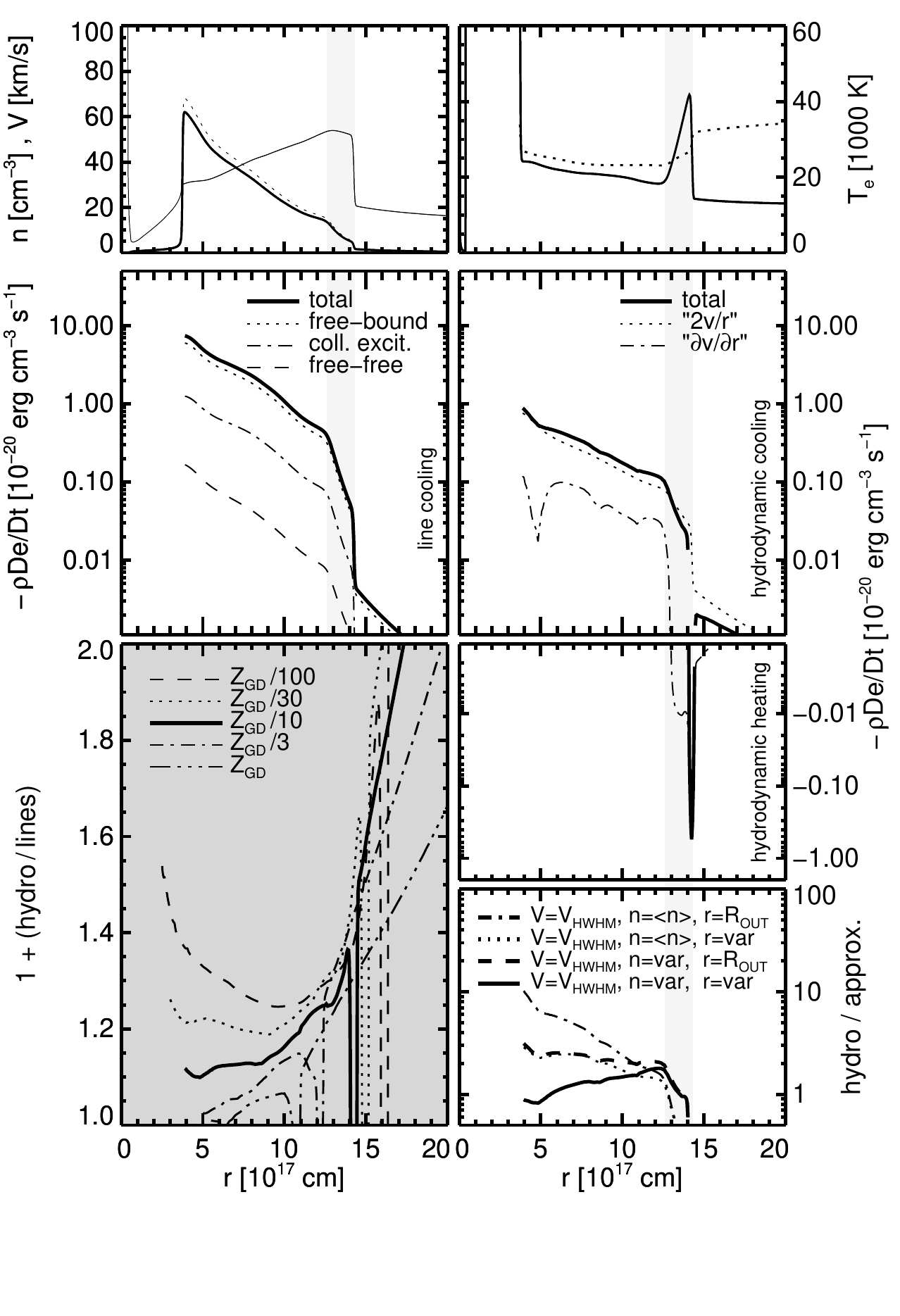}
\vskip-1.3cm
\caption{
\changed{The effect of dynamical/expansion cooling as described by the source/sink term
         ${p\, (\nabla\cdot\vec v)}$ for the  $Z_{\rm GD}/10$, ${\alpha=3}$ 
         model shown in Fig. \ref{evol.3} where the central star has gained its maximum
         effective temperature (see the track in Fig. \ref{schoen.star}).  
         \emph{Top row}: model structure with particle densities and flow velocities 
         (\emph{left}, ions = thick solid, electrons = dotted, velocity = thin solid) 
         and temperatures (\emph{right}, dynamical = solid, equilibrium = dotted).  
         The shadowed vertical strip indicates the thickness of the outer shock region. 
         \emph{Second row}: different contributions of line cooling (\emph{left}) and 
         dynamical cooling (\emph{right}), as indicated in the legends, both 
         based on the equilibrium temperature. 
         The right panel with the two hydrodynamical terms 
         ($\partial v/\partial r$ and $2v/r$) is ``mirrored'' at the horizontal axis 
         in order to show also heating due to compressions if $\partial v/\partial r$
         becomes negative (shock front and halo/upstream region). Note that no cooling 
         terms are plotted for the hot bubble, i.e. inside the contact discontinuity at 
         ${\simeq\! 3.5\times10^{17}}$ cm.
         \emph{Bottom left (shadowed)}:  total cooling (line\,+\,hydro) to line cooling, 
         taken from the middle panels (thick solid) and for 
         models with the same evolutionary time but
         with other metallicities (see legend). Note the different
         radius ranges covered by the models: the inner boundaries 
         are set by the respective contact discontinuities, and the vertical `double'
         lines mark the outer shock regions where heating by compression prevails.
         \emph{Bottom right}: ratio between adiabatic cooling using Eq. (\ref{E.exp}),
         with the model data displayed in the top panels (hydro) and using various 
         approximations (approx), always based on the equilibrium electron temperature:
         mean of total particle density of 40 cm$^{-3}$ and fixed outer radius 
         ${R_{\rm out} = {1.4\times10^{18}}}$ cm (dash-dotted), the (same) mean 
         density but a variable radius (dotted),  variable density and fixed outer 
         radius (dashed), and finally variable density and radial coordinate (solid). 
         In all cases, 
         the model velocity $V_{\rm HWHM}$ = 40 \kms\ is used.
         }
        }
\label{cooling}
\end{figure}

\changed{The whole situation is illustrated in Fig. \ref{cooling} for an evolved 
         nebular model of the ${\alpha=3}$ sequence with $Z_{\rm GD}/10$ where the
         central star is at its maximum effective temperature (cf. Figs.~\ref{evol.3} 
         and \ref{ionisation}). We selected this
         metallicity because we have previously seen that at this value 
         hydrodynamical effects become significant for more evolved and diluted
         models. This figure allows both a detailed insight into the behaviour of 
         the different cooling processes and an assessment of dynamical cooling.}

\changed{We employed the equilibrium temperature profile for estimating hydrodynamical 
         effects since this is the case if one uses a photoionisation code.
         For this metallicity and ionisation stage line cooling is dominated 
         by free-bound transitions throughout the entire nebula (Fig. \ref{cooling},
         middle left).  The largest contribution to cooling from collisional excitation 
         comes from neutral hydrogen, albeit its fraction is very small 
         (${\approx\!10^{-3}}$).  
         Expansion contributes to the total cooling everywhere,
         except at the shock where the gas is heated
         considerably by shock compression (Fig. \ref{cooling}, middle right). 
         Expansion cooling makes up for about 35\% of the line
         cooling in the outer regions of the model immediately behind the
         shock where the expansion is the fastest and the density the lowest.
         This extra cooling is responsible for the temperature difference of a few 
         thousand 
         degrees seen between the dynamical and equilibrium model  
         in the right top panel of Fig. \ref{cooling}.}\footnote 
{\changed{It should be remarked that for the dynamical case all the cooling contributions
          seen in Fig. \ref{cooling} are slightly changed because of the different
          electron temperatures.} }

\changed{The bottom left panel of Fig. \ref{cooling} compares the relative importance 
         of hydrodynamical cooling for additional models of the ${\alpha=3}$
         sequences. All these models are of the same age, which corresponds to
         maximum stellar temperature (cf. Fig. \ref{evol.3}). 
         For these models, the contribution 
         from hydrodynamics must be considered if the metallicity drops below 1/3 
         to 1/10 of the Galactic disk (or solar) value. Furthermore, we also
         see that the hydrodynamic cooling in the metal-rich models is restricted 
         to outer regions. In their inner regions, i.e. the rims, a negative
         velocity gradient indicates compression which, at least partly, compensates 
         the second term within the curly brackets of Eq. (\ref{E.exp}).  This 
         behaviour disappears with decreasing metallicity. One can also see that the shapes 
         of the curves representing the ratios of total cooling to line cooling
         correspond directly to the electron temperature differences between the dynamical
         and equilibrium models in Fig. \ref{evol.3}. }

\changed{The bottom left panel of Fig. \ref{cooling} also shows that for all 
         these very advanced models the dynamics is important for the 
         thermal balance of the halo regions: 
         Expansion cooling is larger than line cooling for all shown models,
         although the former is quite small because of the low flow 
         velocities and a slight compressional contribution from a tiny negative 
         velocity gradient (see top left panel and right panel of the second row, both
         for $r\ga 14\times10^{17}$ cm). }

\changed{The right bottom panel of Fig.\,\ref{cooling} illustrates the errors made 
         when Eq. (\ref{E.exp}) is used with various simplifications.
         Generally, the 
         dynamical cooling is always underestimated, in one case locally up to 
         a factor of ten! The best result is achieved by using the correct density
         and radius variables of a model (provided they are available) together 
         with the $V_{\rm HWHM}$ velocity. 
         Since the electron temperature is fairly constant through the whole nebular 
         structure, an empirical mean value is a reasonable choice for all cases. 
         The error figures found here are model-dependent and
         cannot be generalised.}

\changed{We conclude from the discussion in this section that any method to estimate 
         empirically the contribution of dynamics to the total cooling function from an 
         otherwise static model, as done by \citet{stasetal.10} in their study of the
         metal-poor PN G135.9+55.9, will not provide a sound result.
         These authors used Eq. (\ref{E.exp}) with the assumption
         $\rho\, {\rm D}e/{\rm D}t = {\rm D}U/{\rm D}t$,
         but omitted the velocity gradient 
         by setting $v(r)\equiv V_{\rm HWHM}$ and $r\equiv R_{\rm out}$ 
         (cf. right bottom panel of Fig. \ref{cooling}). --
         Only from full dynamical simulations one can get hold of the
         correct contribution of dynamics to the heating/cooling balance of a nebular
         model. }

\begin{figure*}[ht]
\includegraphics[width=0.346\textwidth]{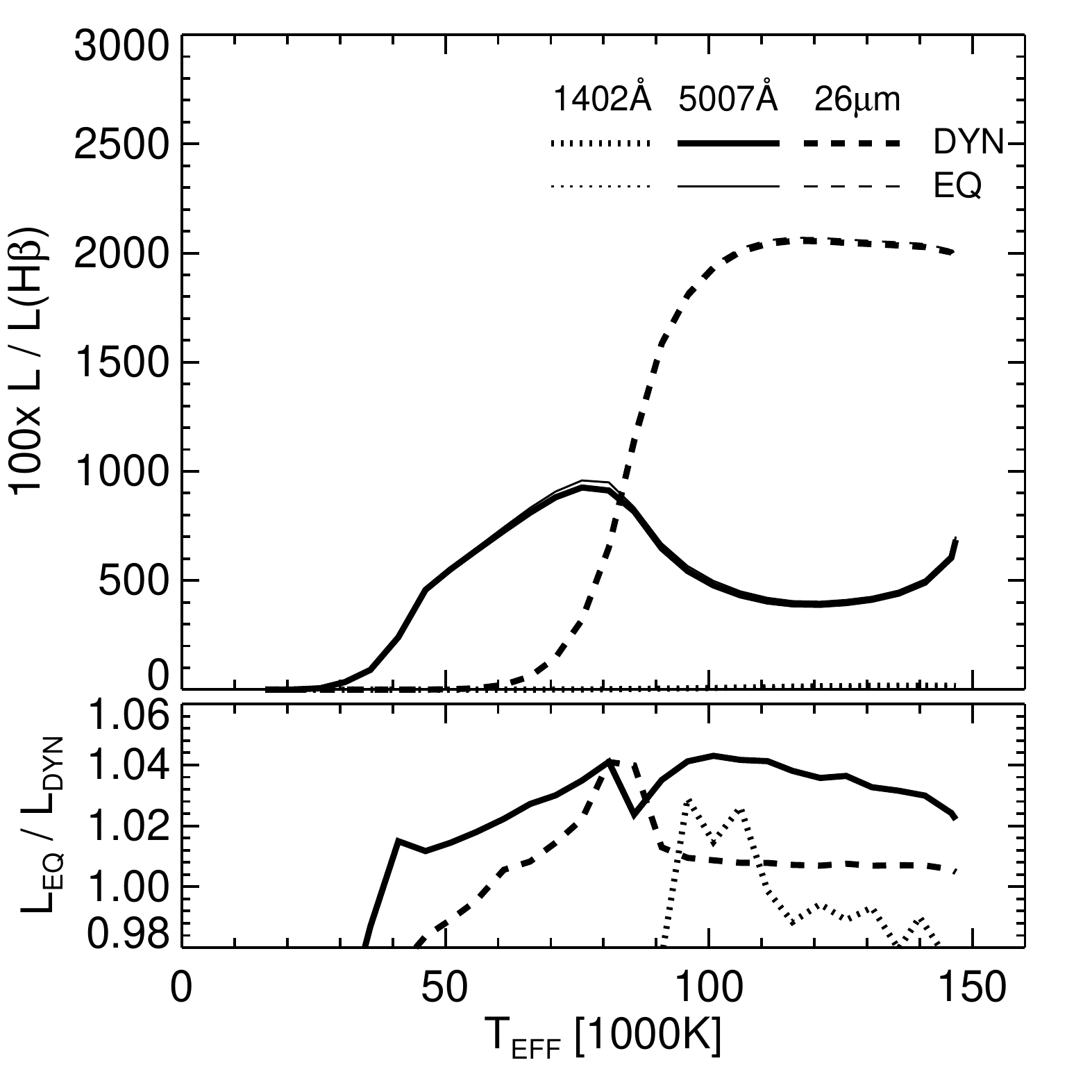}\hskip-3mm
\includegraphics[width=0.346\textwidth]{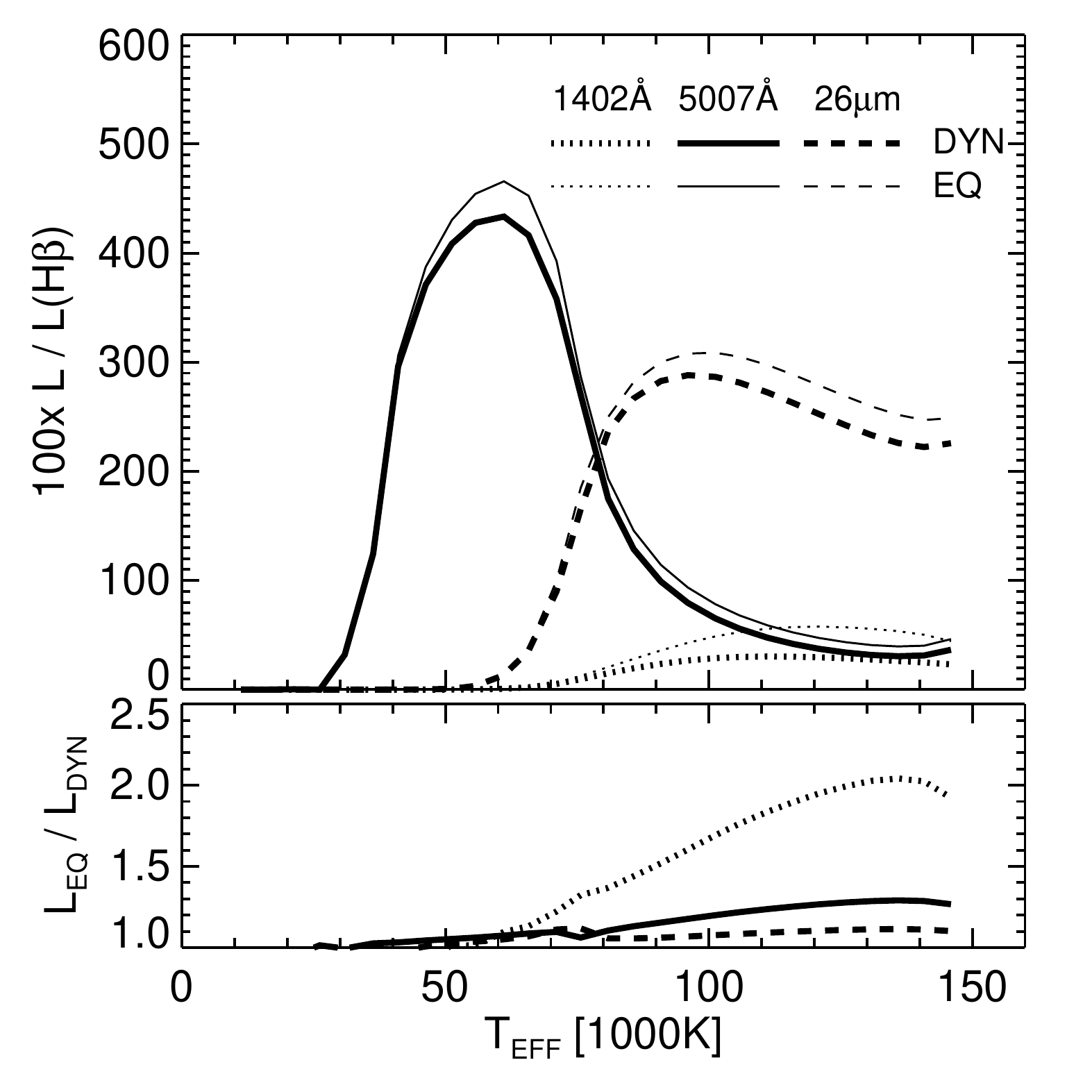}\hskip-3mm
\includegraphics[width=0.346\textwidth]{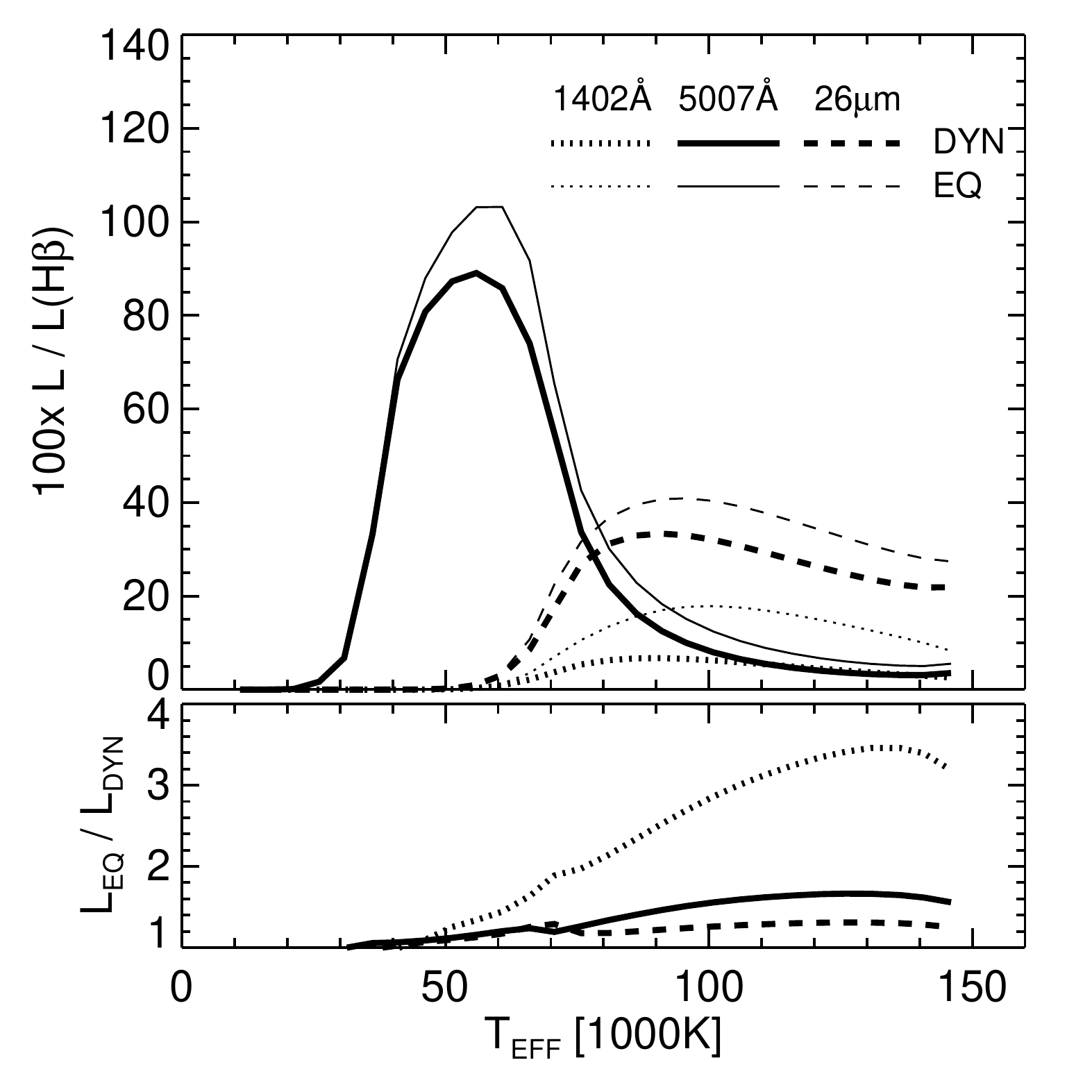}
\vskip-2mm
\caption{\emph{Top panels}:
         run of line strengths (relative to H${\beta=100}$) of three
         collisionally excited  oxygen
         lines, \ion{O}{iv}] $\lambda$\,1402 \AA, [\ion{O}{iii}]\,$\lambda$\,5007 \AA, 
         and [\ion{O}{iv}] $\lambda$\,26 $\mu$m, with stellar temperature 
         for three ${\alpha=3}$ model sequences, viz. ${Z=Z_{\rm GD}}$ (\emph{left}), 
         ${Z=Z_{\rm GD}/10}$ (\emph{middle}), and ${Z=Z_{\rm GD}/100}$ (\emph{right}).
         The evolution is only shown until maximum $\teff$ in order to
         avoid confusion.  Note the different ordinate ranges.
\changed{\emph{Bottom panels}: Line strength ratios $L_{\rm eq}/L_{\rm dyn}$.
          The small scatter seen for the 
         $\lambda$\,1402 \AA\ ratio in the metal-rich model  is due to numerics.}
        }
\label{CEL}
\end{figure*}

\subsubsection{Line strengths}
\label{lines}

  The temperature differences between dynamical and equilibrium models have, of course,
  consequences for the line emission, especially for that of the collisionally excited 
  lines, as has been already demonstrated by \citet[][]{schoenetal.05c}. 
  The effect is again illustrated in Fig.\,\ref{CEL} for three lines of oxygen, 
  in the UV (\ion{O}{iv}] $\lambda$\,1402 \AA), the optical  ([\ion{O}{iii}] 
  $\lambda$\,5007 \AA), and the IR region ([\ion{O}{iv}] $\lambda$\,26 $\mu$m). 
  The trend is as expected:  
\begin{itemize}
\item  \changed{In equilibrium} 
       the models show a tendency to have always higher line strengths,
\changed{in line with the temperature difference to the dynamical case.}
\item  The differences increase with evolution (or stellar temperature)
       and decreasing metallicity; they can, however, be totally neglected for more normal 
       or metal-rich compositions (cf. Fig. \ref{CEL}, left bottom panel). 
\item  The relative differences of line strengths between both types of models are the
       largest for UV lines and the smallest in the infrared. The optical lines behave 
       intermediately.  At the lowest metallicities considered here ($Z_{\rm GD}/100$), 
       \ion{O}{iv}] $\lambda$\,1402 \AA\ can be off by about 250\,\%  if thermal
       equilibrium is assumed.  The corresponding value for [\ion{O}{iv}] 
       $\lambda$\,26 $\mu$m is only about 30\,\% (cf. Fig. \ref{CEL}, right bottom panel).
\end{itemize}  
  Note that the main reason for the differences of line strengths between equilibrium 
  and dynamical models is the electron temperature because we found only very small 
  deviations from ionisation equilibrium (cf. Sect. \ref{equi.model} above). 
\changed{We repeat that the discrepancies found here are highly dependent on structure 
         and expansion properties of the model and should be regarded as indicative only.}

  We see also in Fig.\,\ref{CEL} that the expected changes of line strengths with 
  the corresponding abundances is
  partly compensated by the electron temperatures.  For instance, at medium
  stellar temperatures when O$^{2+}$ is the main ionisation stage 
  and the electron temperatures increase with decreasing $Z$ (cf. Fig.\,\ref{etemp.1}), 
  the line strength of 5007\,\AA\ decreases roughly with $0.5\epsilon({\rm O})$ only.
  At high excitations (${\teff \ga 100\,000}$ K) when most oxygen is O$^{3+}$, the
  electron temperatures of very metal-poor (dynamical) models become nearly independent 
  of  metallicity (Fig.\,\ref{etemp.1}), and hence the strength of, e.g., 26 $\mu$m
  varies proportional to $\epsilon({\rm O})$.  

  Our metal-poor models are well suited to address the question of how one can determine
  the chemical composition of metal-poor PNe if conventional plasma diagnostics is not
  feasible because of lack of suitable lines, as is the case for \object{PN G135.9+55.9}.
  One is then left with photoionisation modelling for constraining, e.g., the oxygen
  abundance, as has been shown extensively and illuminatingly by \citet{PT.05}. 
  According to these authors, the oxygen abundance of \object{PN G135.9+55.9} is 
  likely not below 1/30 of the solar value, assuming a stellar temperature
  of $\simeq\!130\,000$ K.\footnote{The analysis of \citet{PT.05} is
  entirely based on optical lines!}

  However, for a highly-excited PN like \object{PN G135.9+55.9}
  the oxygen content can best be estimated by using, e.g., the 26 $\mu$m line whose 
  strength depends only weakly on the electron temperature and its uncertainties
  (cf. Fig.\,\ref{CEL}).  The optical oxygen line at 5007~\AA\ cannot be recommended
  because its higher sensitivity to the electron temperature, especially if the stellar 
  temperature is above 100\,000~K.
  Ultraviolet lines of highly ionised elements, such as \ion{O}{iv}] $\lambda$\,1402 \AA, 
  should not be used at all for abundance studies because these lines depend strongly 
  on electron temperatures at nearly all stellar temperatures.

  We conclude from \changed{this discussion}
  that the use of photoionisation models assuming thermal equilibrium 
\changed {is not adequate 
         for objects with a diluted nebula and a very hot central star (i.e. for highly 
         excited objects as represented by our models) and with reduced metallicity 
         because of incorrect physical assumptions. Note that departures from thermal
         equilibrium depend only on the gas properties regardless of the nature of
         the central ionising object.}
  Of course, this will become evident only if the restriction to 
  one wavelength region (e.g. the optical) is relaxed. 
\changed{Our models show that, for UV lines, departures from equilibrium may become a
         problem already for ${Z\la Z_{\rm GD}/3}$.  They suggest also that the
         neglect, or a too simple consideration, of dynamical effects in the modelling 
         of a metal-poor and highly
         excited PN, as done in the recent study of \object{PN G135.9+55.9} by
         \citet{stasetal.10}, becomes problematic.}

   Based partly on new spectroscopic material, \citet{sandetal.10} performed an 
   abundance study of PN G135.9+55.9 \changed{taking full advantage of} the dynamical models 
  discussed here. 
\changedIII{For the details, the reader is referred to the cited paper.}

\subsubsection{The Milky Way halo PNe}
\label{halo.pne}

  In this context 
  it is interesting to see how the \changed{known} Galactic halo PNe fit into our grids
\changed{of mean electron temperatures vs. stellar effective temperatures or ages}
  of the simple hydrodynamical ${\alpha=3}$ models as shown in Fig.\,\ref{etemp.1}
  and whether hydrodynamical effects are discernable.
\changed{As ages of the halo PNe are not known, and their stellar temperatures are too 
         uncertain, we used the nebular excitation as given by the relative strength
         of \ion{He}{ii} $\lambda$ 4686 \AA\ for a proxy of the evolutionary 
         status of both the objects and models.}

  The result is seen in Fig.\,\ref{etemp.2} where the mean \oiii\ temperatures are
  now plotted against \changed{$L(4686)/L({\rm H}\beta)$} instead.  We compared our
  models \changed{(also in equilibrium)} with both the electron temperatures from 
  detailed nebular analyses  
  by means of photoionisation models as provided by \citet{howard.97}
  and with the corresponding temperatures derived directly from the observed temperature 
  sensitive \oiii\ line ratios.%
\footnote{The electron temperatures which we derived from the \emph{observed} \oiii\
  line ratios listed in Table 2 of \citet{howard.97} differ in some cases,
  for unknown reasons, from the temperatures given in the same table.}
\changed{\object{K\,648} and \object{DdDm-1} are omitted because they do not show 
         a \ion{He}{ii} $\lambda$ 4686 \AA\ line.}

\changed{Neglecting for the moment \object{M2-29}, the following emerges from Fig.  
         \ref{etemp.2}:
\begin{itemize}
\item  Compared with the simulations, the observed mean electron temperatures suggest 
       nebular abundances between $Z_{\rm GD}/3 < Z < Z_{\rm GD}/30$, which is
       in reasonable agreement with the spectroscopic analyses in the literature
       \citep[cf. Table 4 in][]{howard.97}.
\item  On average, discrepancies between observed and modelled temperatures increase
       with excitation: they are less than 1000 K at $L(4686)/L({\rm H}\beta)\simeq 0.1$,
       but up to 3500~K at the highest excitation, ${L(4686)/L({\rm H}\beta)\simeq 1}$.%
       \footnote{\changedIII{The trend seen for the three highly excited objects is 
                             most likely not real. Firstly, the relative importance of
                             expansion cooling depends on the object's density and
                             velocity structure, and secondly, the electron temperature
                             based on a photoionisation model is subject to the
                             imposed fit criteria.} } 
       This is consistent with the predictions of our model sequences if one neglects 
       the (unexplainable) fact that for the 3 objects with medium 4686 \AA\ line strengths
       the \citet{howard.97} temperatures are below the temperatures 
       derived from the observed line strenghts, while they are above  for the 3 objects 
       with the highest excitations.
\end{itemize}}
\noindent
\changed{Because abundances derived from collisionally excited lines
       depend strongly on the electron temperature, one is tempted to
       say that the use of standard photoionisation models cannot be recommended for cases
       where non-equilibrium (i.e. dynamical) effects become important.
       This statement may become applicable already at metallicities below about one
       third solar, for very diluted and rather fast expanding nebulae. }

\changed{For \object{M2-29} with its only  moderate excitation the discrepancy between 
         observed (from $R_{\rm OIII}$) and modelled temperatures is the largest, 
         $\simeq$8000 K, and at disturbing variance with the model predictions.}
  First of all, the photoionisation models of \citeauthor{howard.97} are based on 
  the incorrect \oiii\
  $\lambda$4363 \AA\ line strength of 15 (H${\beta = 100}$) which is the unde\-reddened
  value \citep[cf.][Table 2A]{pena.91}.  We used in Fig.~\ref{etemp.2} the dereddened value
  of 19.5 instead, leading to an electron temperature of 24\,000 K which is then the
  highest in the \citet{howard.97} sample of Galactic halo PNe. 
\changed{The photoionisation model of \citet{howard.97} predicts a $\lambda$4363~\AA\ 
         line strength of 11 only instead of the assumed value 15, which would
         correspond to an electron temperature of about 20\,000~K. }

\changed{Because of the high electron temperature, the plasma diagnostics of 
         \citet{pena.91} leads to a lower metallicity for \object{M2-29} than claimed
         by \citet{howard.97}: a metal depletion by factors between 20 and 30.
         Still, the problem of \object{M2-29}'s too high
         electron temperature remains a mystery: At its moderate excitation this high
         electron temperature would suggests a metallicity well below $Z_{\rm GD}/100$,
         which is not observed.
         In this context it is interesting
         to note that \citet{haijduk.08} claimed recently
         that the nucleus of \object{M2-29} consists of a binary system with an eclipsing disk.}

\begin{figure}[t]
\includegraphics*[bb= 0.7cm 0.4cm 15.1cm 15.0cm, width=0.99\linewidth]
                {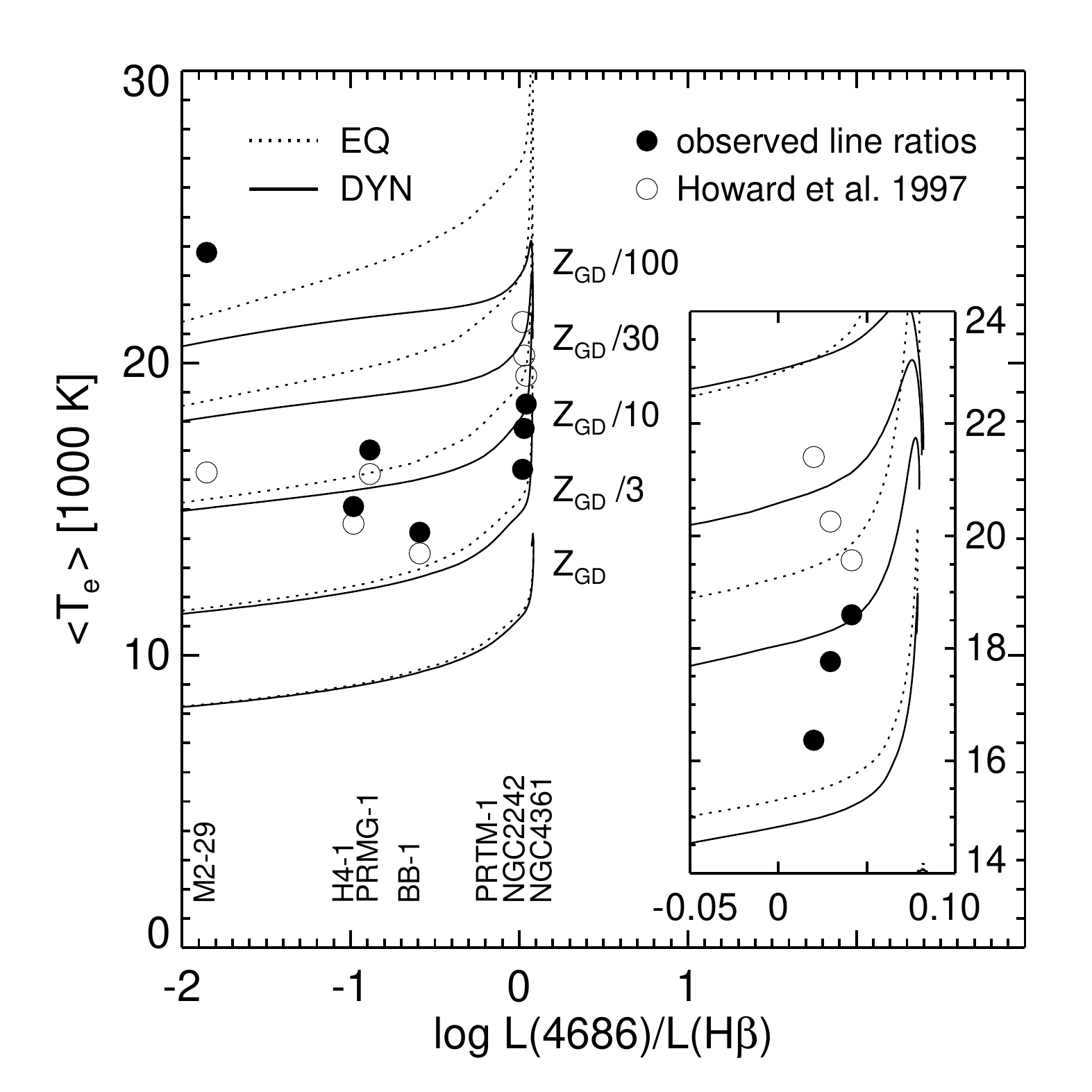}
\caption{\changed{Mean nebular \oiii\ electron temperatures vs. nebular line ratio
                  $L(4686)/L({\rm H}\beta)$ for the dynamical ${\alpha = 3}$ sequences
                  from Fig. \ref{etemp.1} (dotted in thermal equilibrium).
                  Again, the sequences are only plotted until maximum stellar 
                  temperature in order to avoid confusion.  The circles give the
                  corresponding positions of the Halo PNe: filled circles are based on 
                  the {observed} \oiii\ lines, open circles are results of 
                  detailed analyses using photoionisation models \citep{howard.97}.
                  The inset renders an enlarged view of the region around 
                  $L(4686)/L({\rm H}\beta) \simeq 1$.
}        }
\label{etemp.2}
\end{figure}

\subsection{The $\alpha=2$ sequences}
\label{alpha.2}

\begin{figure*}
\includegraphics*[width= \textwidth, height= 13.5cm]
      {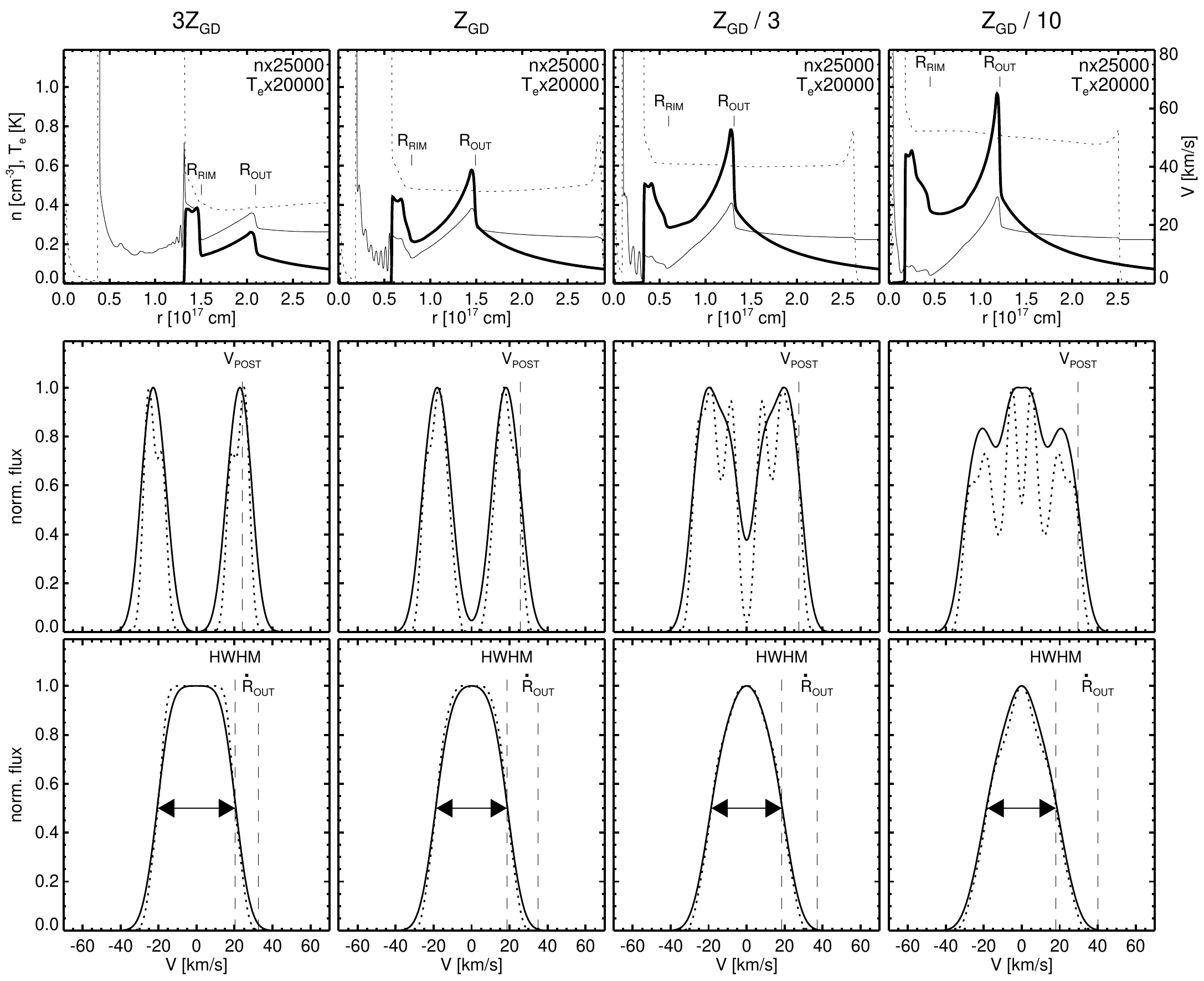}
\caption{Composite figure showing one snapshot taken at maximum 5007 \AA\ brightness
         along each of the four 0.625~\Msun\
         sequences (${\alpha =2}$), with the metallicities indicated over the panels. 
         Details of the models are listed in Table~\ref{tab.625}.
	 \emph{Top panels}: structures of the models: run of ion density
	 (thick solid line), flow velocity (thin solid line), and electron temperature 
         (dotted line) with radius.  Densities and temperatures are normalised
	 and must be multiplied with the factors given in each panel.
	 The radial positions, $R_{\rm out}$, of the nebular outer shock and of the rim,
         $R_{\rm rim}$ are also marked in each panel.  
         The central star is at ${r=0}$ cm. The electron temperature spikes 
         seen at larger radii are caused by the ionisation fronts, and
	 the jumps at the inner nebular edges mark the positions of the contact
	 discontinuity/conduction fronts.
	 \emph{Middle panels}: normalised spatially
	 resolved \oiii\ 5007 \AA\ and thermally broadened line profiles (dotted line)
	 as seen through a central numerical aperture with a diameter of $1\times10^{16}$ cm
         (corresponding to 0\farcs67 at 1 kpc), and
	 additionally broadened by a Gaussian of 10 \kms\ FWHM (solid line) in order to
         simulate a typical instrumental broadening.
	 The maximum velocity behind the outer shock at $R_{\rm out}$,
	 $V_{\rm post}$, is indicated by vertical dashed lines.
	 \emph{Bottom panels}:  \oiii\ 5007 \AA\ line profiles as integrated over
	 the whole nebula (dotted line), and additionally broadened by a Gaussian of
         10 \kms\ FWHM (solid line).  Two vertical dashed lines in each panel indicate
	 the HWHM and the shock velocity, $\dot{R}_{\rm out}$, respectively.
        }
\label{0.625}
\end{figure*}

\begin{table*}
\caption{Relevant parameters of the four ${\alpha =2}$ models with central-star masses
         0.625 \Msun, as depicted in 
         Fig.\,\ref{0.625}. Columns 2--4 describe the central star (post-AGB age, 
         luminosity, and effective temperature), and Cols. 5--12 the nebular properties
	 (\changed{maximum} \oiii-brightness, rim/shock positions, shock propagation, 
         post-shock velocity, 
         velocities at half-width-half-maximum (HWHM) and half-width-10\%-maximum 
         (HW10\%M) of the spatially integrated \oiii\ 5007\,\AA\ line profile, 
         the mean O$^{2+}$ electron temperature, 
         the mass of the rim, and the total ionised nebular mass).  
         The initial parameters of the circumstellar envelopes are 
         ${\dot{M}_{\rm agb} = 10^{-4}}$ \Mdot\ and ${v_{\rm agb} = 15}$ \kms\ in all 
         cases.  
        }
\label{tab.625}
\tabcolsep=4.9pt
\begin{tabular}{lccccccccccc}
\hline\hline\noalign{\smallskip}
Metall. $Z$  & $t  $ & $L$ & $T_{\rm eff}$ & $M(5007)$& $R_{\rm rim}/{R}_{\rm out}$ & 
$\dot{R}_{\rm out}$
       & $V_{\rm post}$    & $V_{\rm HWHM}^{\rm [OIII]}/V_{\rm HW10\%M}^{\rm [OIII]}$ 
       &   $\langle T_{\rm e}\rangle$ 
       & $M_{\rm ion}(R_{\rm rim})\,^{\rm a}$     & $M_{\rm ion}(R_{\rm out})\,^{\rm a}$ \\[2.5pt]
       & [yr]  &  [\Lsun]    & [K]           & [mag]  &[$10^{17}$cm] & [\kms]
       & [\kms]             &  [\kms]       &  [K]       & [\Msun]& [\Msun]       \\[2pt]
  (1) & (2) & (3) & (4) & (5) & (6) & (7) & (8) & (9)  & (10) & (11) & (12)       \\[2pt]
\hline\noalign{\smallskip}
$3Z_{\rm GD}$  & 2392& 6643&        117\,643& --4.66& 1.5\,/\,2.1 & 33& 24.1& 20.3\,/\,26.3\rlap{\,$^{\rm b}$} & \enspace8\,070 &0.043\ (0.017) & 0.174\ (0.109)\\
$Z_{\rm GD}$   & 1694& 7507& \enspace73\,338& --4.25& 0.8\,/\,1.5 & 35& 25.7& 18.6\,/\,25.1 & \enspace9\,553 &0.012\hspace{1cm}  & 0.131\hspace{1cm}  \\
$Z_{\rm GD}/3$ & 1460& 7667& \enspace60\,917& --3.80& 0.6\,/\,1.3 & 37& 27.6& 18.5\,/\,26.7 &      12\,080   &0.007\ (0.011) & 0.119\ (0.153)\\
$Z_{\rm GD}/10$& 1324& 7736& \enspace54\,296& --3.08& 0.4\,/\,1.2 & 40& 29.6& 18.0\,/\,27.9 &      15\,198    
               &0.012\ (0.013) & 0.115\ (0.169)\\[1.5pt]
\hline
\end{tabular}
\\[3pt]
$^{\rm a}$ Mass values in parentheses refer to the age of the $Z_{\rm GD}$ model, $t= 1694$ yr. 
\\
$^{\rm b}$ $V_{\rm HW10\%M} > V_{\rm post}$ because at this phase of advanced evolution the 
           rim matter expands faster than the shell (cf. top left panel of Fig. \ref{0.625}).
\end{table*}

  Recently we showed in \citetalias{schoenetal.07} that the bright cutoff of
  the PNLF can be explained by nearly optically thick nebulae around central stars
  with masses slightly above 0.6~\Msun.  However, in \citetalias{schoenetal.07} we 
  only assumed a metal content that is typical for Galactic disk objects. 
  We thus continue our work with investigating the properties of model nebulae 
  around more massive (and much faster evolving) central stars, 0.696~\Msun\ 
  and 0.625~\Msun, and with different
  metal contents, $3Z_{\rm GD}$, $Z_{\rm GD}$, $Z_{\rm GD}/3$, and $Z_{\rm GD}/10$.
  This is also the mass range of interest in observations of very  
  distant stellar populations because there only the most luminous nebulae are
  accessible for more detailed studies \citep[see, e.g.,][]{arnaboldi.08}.
  All initial models have the same constant-outflow conditions, viz. 
  ${\rho \propto r^{-2}}$, ${\dot{M}_{\rm agb} = 10^{-4}}$ \Mdot, and 
  ${v_{\rm agb} = 15}$ \kms\ (cf. Sect. \ref{models}).  The metallicity dependent 
  central-star wind model used is the same as introduced in Sect. \ref{models}.
\changed{For these always rather compact and dense models the dynamic contribution
         to the cooling function remains always negligible, also at lower metallicities.
        }

\subsubsection{Expansion properties}
\label{exp.vel}

  We start with a discussion of the expansion properties of models around a 
  0.625~\Msun\ central-star.
  For this purpose, Fig.\,\ref{0.625} contains four snapshots,
  each taken at the moment of maximum \oiii\ 5007 \AA\ luminosity of the respective
  sequence, showing the model structures and the corresponding profiles of the
  5007\,\AA\ line, both spatially resolved and volume integrated.  The four sequences
  started from exactly the same initial configuration (see above)
  but with different metallicities: $3Z_{\rm GD}$, $Z_{\rm GD}$, $Z_{\rm GD}/3$, and
  $Z_{\rm GD}/10$. Relevant model parameters are listed in Table\,\ref{tab.625}. 

  The model structures of all sequences shown are very similar, with a pronounced 
  shell/rim morphology.  The mass fractions contained in the rims are, however,
  quite different (Table \ref{tab.625}, Cols. 11 and 12) and reflect the dependences of
  stellar wind power and shock speed on metallicity:
\begin{enumerate}
\item  For a given time, the total nebular mass, $M_{\rm ion}(R_{\rm out})$, increases 
       strongly with decreasing metal content because of the increasing shock speed, 
       $\dot{R}_{\rm out}$ (Col. 7 of Table \ref{tab.625}).
\item  The case for the rim is more complicated:  In general, the decreasing wind power
       leads also to a decreasing bubble pressure, and hence also to smaller rim masses.
       This trend is inverted for the metal-poorest sequence because here the bubble
       becomes so small that its pressure increases again somewhat, despite of the 
       very weak stellar wind.
\item  In all cases, however, the nebular mass fraction contained in the rim remains 
       rather small as long as the shell is not recombining (see below)!
\end{enumerate}

\changed{A remark concerning the terminology ``young'' planetary nebula is in order
         here. The bright models shown in Fig. \ref{0.625} and whose parameters are listed in 
        Table \ref{tab.625} are all very compact (${R_{\rm out} < 0.07}$ pc) and less than 
        2500 years old. Objects with such properties are usually considered to be young PNe.
        However, because of the higher mass (0.625 \Msun) of the central stars these systems 
        are already quite evolved with hot stars of effective temperatures ranging between 
        55\,000 and 120\,000 Kelvin! The presence of a rim is the result of this advanced
        stage of evolution. Thus, in terms of evolution such objects must be considered
        at least as being middle-aged, but not ``young''. 
        The Milky Way object \object{NGC 7027} is an even more extreme example of this kind:
        Because its central star is already close to its maximum temperature,
        the term ``old'' is a better designation for this PN 
        \citep[for more details on this object, see][Paper III hereafter]{schoenetal.05b}. }   

  Concerning their general expansion properties the models with ${\alpha=2 }$ behave 
  (qualitatively) in the same way as the ${\alpha=3}$
  models discussed extensively in Sect.~\ref{alpha}: A lower metallicity provokes
  higher electron temperatures and faster expansion, but a lower pressure exerted
  by the shocked stellar wind (top panels of Fig.~\ref{0.625}).
  Consequently, the velocity difference between the outer and the inner nebular
  boundaries increases with decreasing metallicity as well.  This effect is
  clearly seen in the middle panels of Fig.~\ref{0.625} where the computed, spatially
  resolved \oiii\ 5007 \AA\ line profiles of the models are shown:  At lower $Z$,
  the fast shell and the slow rim matter are spectroscopically resolved, and the 
  entire profile becomes wider.  In the $Z=Z_{\rm GD}$ case both velocities are 
  already too similar and cannot be separated.
  Note that in all cases shown (except for ${Z=3Z_{\rm GD}}$ where the rim matter 
  expands faster than the shell matter) the extreme values of the profile derivative, 
  measured at the outer flanks of the profile, underestimates only slightly 
  the post-shock velocity, $V_{\rm post}$, i.e. the gas velocity immediately
  behind $R_{\rm out}$ \citep[see also the discussion in][]{corradi.07}. 

  Spatially resolved line profiles are not available for PNe in extragalactic
  systems, and we thus show the corresponding spatially integrated profiles in the 
  bottom panels of Fig.~\ref{0.625} as well.  Despite of the different velocity 
  and density profiles of the models caused by their different metallicities and which
  result in quite different spatially resolved
  line profiles (as is discussed above and shown in the middle panels of 
  Fig.~\ref{0.625}), all four spatially integrated profiles 
  are very similar, but with much \emph{smaller} line widths than found for their 
  spatially resolved counterparts.

  \changed{The expansion rates as they follow} from the line widths}, $V_{\rm HWHM}$,
  vary only between 18 and 20 \kms and are well below $V_{\rm post}$
  (${\simeq\!24\ldots30}$ \kms), and only about half of $\dot{R}_{\rm out}$
  (${\simeq\! 33\ldots40}$ \kms)  -- the true expansion rates (cf. Table \ref{tab.625},
  Cols. 7, 8, and 9)!

  The \changed{surprisingly low $V_{\rm HWHM}$ can be understood as follows:}  
  By integrating over the nebular volume the fraction of nebular gas with the highest 
  \changed{line-of-sight}
  velocities is small and \changed{is only responsible for the faint wings}
  of the total line profile.
  The high-speed matter immediately behind the outer shock is only
  detectable either by means of spatially resolved high-resolution spectroscopy
  \citep{corradi.07}, which is not possible for distant objects, or by measuring the
  width of the integrated profile close to the bottom of the profile. This, however
  demands a sufficiently high signal-to-noise ratio.

\begin{figure}[t]
\includegraphics[width=\linewidth]{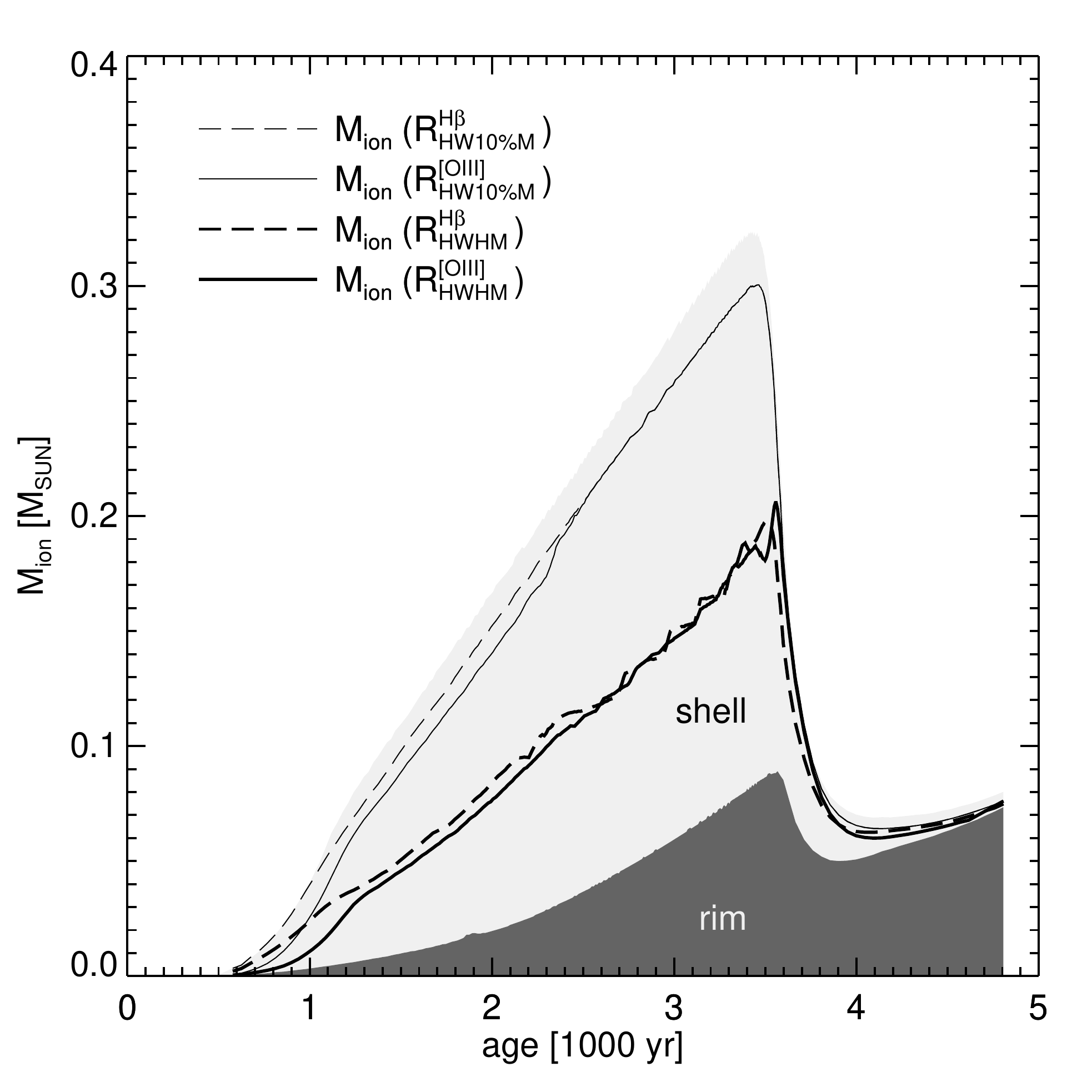}
\caption{Evolution of nebular mass fractions vs. time for 
         the 0.625 \Msun\ sequence from Fig. \ref{0.625} with ${Z=Z_{\rm GD}}$. 
         Shown are the total (ionised) mass enclosed by the outer nebular boundary,
          $M_{\rm ion}(R_{\rm out})$, with the different contributions of the shell 
         (light grey) and rim (dark grey) indicated.
         The different lines indicate mass fractions contained within certain distances 
         $R$ from the star for which the \changedIII{real} 
         gas velocities reach distinct limits: 
         $M_{\rm ion}(R_{\rm HWHM}^{\rm H\beta})$ contains all mass shells with
         ${v\le V_{\rm HWHM}^{\rm H\beta}}$ (thick dashed line), and  
         $M_{\rm ion}(R_{\rm HWHM}^{\rm [O\,III]})$ is the corresponding mass value for 
         \oiii\ 5007 \AA\ (thick solid line).  HWHM refers to the total, volume integrated
         line profile. The thin lines (solid and dashed) indicate the masses enclosed 
         by the radii $R_{\rm HW10\%M}$ where the \changedIII{(real)} 
         gas velocity $v$ corresponds to the half width of the 10\% intensity level of the  
         integrated line profile, i.e. ${v= V_{\rm HW10\%M}}$.
        }
\label{fract.mass}
\end{figure}

  The case is illustrated in more detail in Fig. \ref{fract.mass} where the
  time evolution of various mass fractions of the model nebula with ${Z=Z_{\rm GD}}$ around 
  the 0.625 \Msun\ central star is provided.  One sees a large temporal
  variation of the total ionised nebular mass: a mass increase due to ionisation and the 
  expanding shock 
\changed{which swallows the upstream AGB matter and adds it to the nebula}, 
  followed then by a rapid decline when recombination destroys nearly the
  entire shell after about 3500 years.  For the density and velocity profiles of this
  model at an age of 1694 years, see Fig. \ref{0.625} (second top panel). 

\changedIII{Figure \ref{fract.mass} can be considered as a tool for interpreting the
            kinematics of a PN from spatially-integrated line profiles:}
  Thick lines (solid for \oiii\ 5007\,\AA\ and dashed for H$\beta$) 
  indicate the mass fractions where \changedIII{the \emph{real}} gas velocities are lower 
  than the value deduced from the corresponding half width of the total (volume integrated) 
  line profile.
\changed{This matter 
         contributes fully to the line emission within the halfwidth value, 
         \changedIII{next to matter with \emph{projected} velocities $\le V_{\rm HWHM}$.}
         The important velocity 
         information from the faster moving gas, the mass fraction of which may 
         even exceed half of the total ionised mass, can only be retrieved from 
         the outer wings of the line profile.}

  Obviously the only method to get hold of the \changed{signature of the} 
  fastest expanding nebular parts is to
  measure the line width at a very low (ideally zero) intensity, as proposed by
  \citet{dop.85, dop.88} \changed{who suggested to measure line widths at the
  10\% level from maximum (see also discussion in Sect. \ref{MC}).}
  One sees indeed from Fig. \ref{fract.mass} that 
\changed{such a choice, ${V_{\rm HW10\%M}}$, ensures that the mass contained in 
         the outermost shell in which the \changedIII{(real)}
         gas velocities are higher than ${V_{\rm HW10\%M}}$ becomes very small, i.e.} 
  ${V_{\rm HW10\%M}}$ is close to the post-shock velocity $V_{\rm post}$
  (Table \ref{tab.625}, Cols. 8 and 9).  

  The situation becomes easier \changedIII{for this particular model}
  when recombination sets in after 3400 years of evolution:
  The shell disappears nearly completely, and because the velocity within the rim
  is roughly constant, $V_{\rm HWHM}$ represents, in the particular case shown here, 
  the overall PN kinematics well (Fig. \ref{fract.mass}).  The now rather low PN mass
  (${<\! 0.1}$\,\Msun) corresponds roughly to that of the rim formed by the stellar
  wind action during the previous high-luminosity part of evolution.
  
\begin{figure}[t]
\includegraphics*[bb= 0.6cm 0.5cm 15.2cm 15.2cm, width= \linewidth]
      {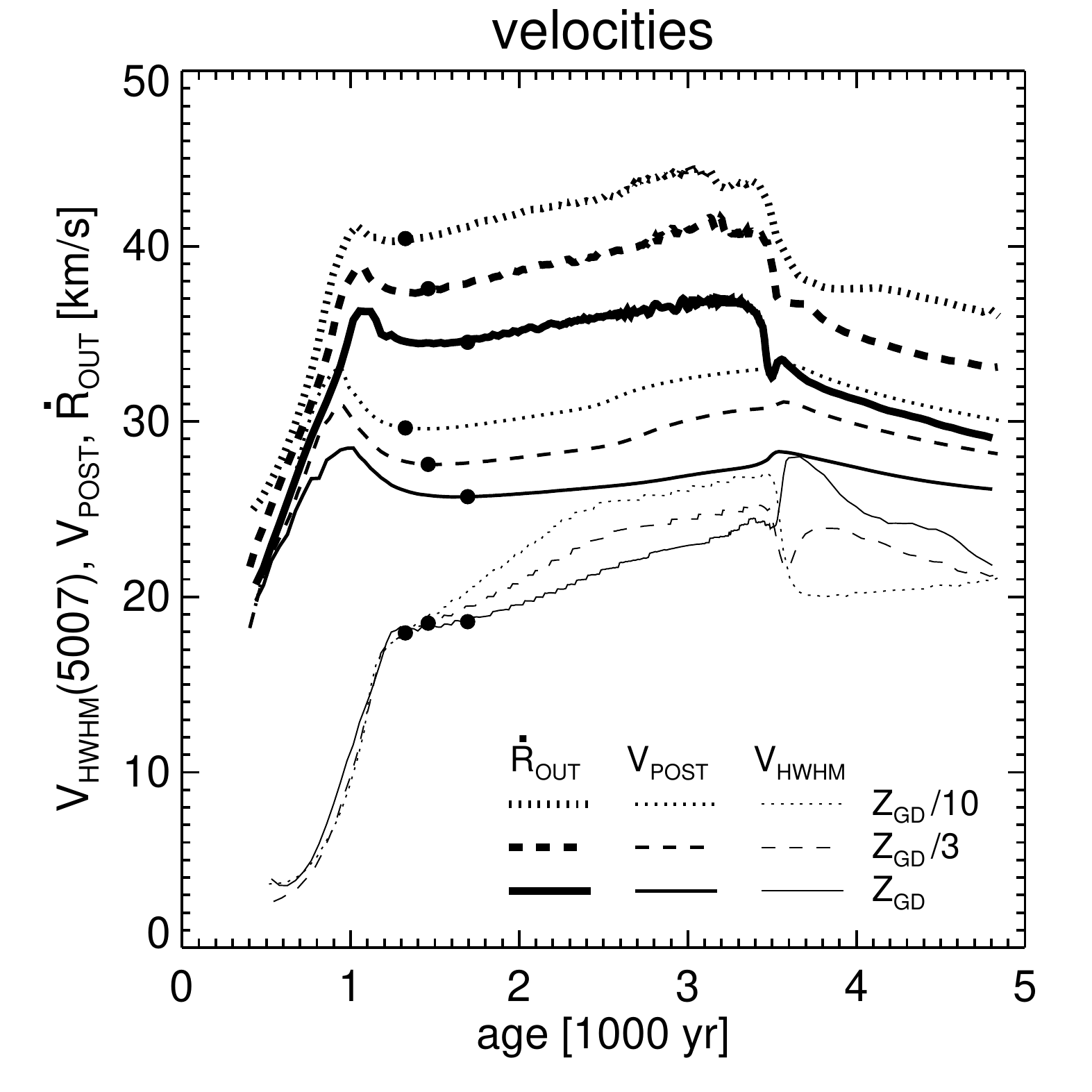}
\caption{Three different expansion velocities of a PN:  $V_{\rm HWHM}$ of
         the spatially integrated \oiii\ 5007 \AA\ line profile,
         the post-shock velocity, $V_{\rm post}$,  behind the outer shock at
	 $R_{\rm out}$, and the shock velocity, $\dot{R}_{\rm out}$, itself.
	 Shown is the complete time evolution of these velocities for three
	 sequences with the 0.625 \Msun\ central-star model
	 and different metallicities (see figure legend).
	 The filled circles mark the snapshots shown in Fig.~\ref{0.625},
	 which correspond also to the moment of maximum \oiii\ emission. The 
         relevant model parameters are listed in Table \ref{tab.625}
 \changed{where also the values of $V_{\rm HWHM}$, $V_{\rm post}$, and 
         $\dot{R}_{\rm out}$ for the $3Z_{\rm GD}$ sequence, not plotted for clarity,
         are given.}     
         The bumps at
         age ${\simeq\!1000}$ years are due to the optically thick/thin transition. 
	 }
\label{0.625.vel}
\end{figure}

  The complete velocity evolution of three 0.625 \Msun\ sequences is rendered in
  Fig. \ref{0.625.vel}.  Plotted are three velocities, $\dot{R}_{\rm out}$, $V_{\rm post}$,
  and $V_{\rm HWHM}$ as a function of the post-AGB age.  We see that in general
  the difference between the measurable $V_{\rm HWHM}$ and the real expansion speed,
  $\dot{R}_{\rm out}$, is quite large:  about a factor of two during the \oiii\ bright
  phase of evolution (e.g., filled circles in the figure).  During the early, optically
  thick stage, the velocities increase rapidly with time in all cases.
  The $V_{\rm HWHM}$ velocities start at very low values, well \emph{below} the
  AGB wind velocity of 15 \kms, because early in the PN evolution when the
  central star is not very hot, O$^{2+}$ is restricted to the inner, slowly expanding
  matter in or around the rim only (cf. Fig.~\ref{0.625}, top panels).

  As we have already seen for the $\alpha=3$ sequences, $V_{\rm HWHM}$ is quite insensitive 
  to the metal content for the whole evolution, in comparison to the shock speed 
  $\dot{R}_{\rm out}$ (cf. Fig. \ref{HWHM}).  However, the post-shock gas velocity,
  $V_{\rm post}$ reflects the metal dependence of the shock speed quite well 
  (Fig. \ref{0.625.vel}), but it requires accurate measurements of the faint wings of 
  the integrated line profile, as discussed above. 

  The velocity bumps ($V_{\rm post}$ and $\dot{R}_{\rm out}$ only) seen close to
  ${t\simeq1000}$ yr indicate that at this time the models become optically thin for
  Lyman continuum photons and enter the `champagne' phase of expansion during which
  the shock speed depends only on electron temperature (sound speed) and radial
  density gradient \citepalias[see][]{schoenetal.05a}.
  This transition is not seen in $V_{\rm HWHM}$
  because the outer nebular region behind the shock contains only little O$^{2+}$
  at this time.  Later, at ${t\simeq1200}$ yr, also the outer nebular regions start to
  contribute significantly to \oiii\ line profiles, and the velocity increase flattens
  out.   The deceleration of the shock speed, $\dot{R}_{\rm out}$, at higher ages,
  ${t>3000}$ yr, is due to recombination behind the shock, leading there to much
  lower electron temperatures and hence lower pressure.

   Of course, the maximum amount of \oiii\ emission decreases with the oxygen content (see
   Table \ref{tab.625}). This table also shows that this maximum emission value occurs the
   earlier (i.e. at lower stellar temperatures) the lower the oxygen abundance is 
   (cf. filled circles in Fig. \ref{0.625.vel}). The reason for this behaviour is
   twofold: (i) the $\rm O^+$ ionising continuum becomes optically thin earlier for 
   lower abundances, and (ii) the number of photons ionising $\rm O^+$ 
   ($\lambda < 353 $ \AA) decreases rapidly with $\teff$.\footnote{%
   The stellar radiation field is assumed to be independent of the chosen metallicity.}
    
   We note also that the expansion speeds do not reach the very high levels of the
   models shown in Figs.~\ref{vmax} and \ref{HWHM}.  It has nothing to do with the
   different central-star masses (0.625 vs. 0.595 \Msun).  Rather it is caused by
   the different density profiles of the circumstellar matter.  We recall that
   the nebular models shown in the Figs.~\ref{vmax} and \ref{HWHM} have initial
   models with power-law density profiles, ${\rho(r)\propto r^{-\alpha}}$, with
   ${\alpha=3}$,
   while the sequences discussed in this section have ${\alpha=2}$, \changed{which
   is the signature of an AGB wind with fixed speed and mass-loss rate. 
   For ${\alpha=2}$, the shock speed relative to the upstream flow is considerably
   lower than for the ${\alpha=3}$ case \citepalias[see Fig. 7 in][]{schoenetal.05a}.}
   The modest shock acceleration 
   during the (optically thin) `champagne' phase seen in Fig.~\ref{0.625.vel} is due 
   (i) to the rise of the electron temperature as the star becomes hotter, and 
   (ii) to the weak acceleration of the AGB wind (= upstream flow) after the passage 
   of the ionisation front.

\subsubsection{The \oiii\ brightness}
\label{M.5007}

   The \oiii\ emissivity should depend directly on the oxygen content itself.
   \citet{jacoby.89} showed, however, that the
   5007 \AA\ luminosity changes with roughly the square root of the abundance only.
   This more modest dependence is explained by two effects which compensate
   each other partly:  The lower oxygen abundance leads to a higher electron
   temperature, which in turn enhances the emission of collisionally excited lines.
\changed{Based on optically thick photoionisation models, \citet{dop.92} found that
         there exists a maximum efficiency of converting stellar radiation into \oiii\ 
         5007\,\AA\ line emission at about twice the solar oxygen abundance.}\footnote
         {This statement refers to the old solar oxygen abundance of
         $\epsilon(\rm O)_{\sun}=8.84$.}

\begin{figure}[t]
\includegraphics[width= \linewidth]{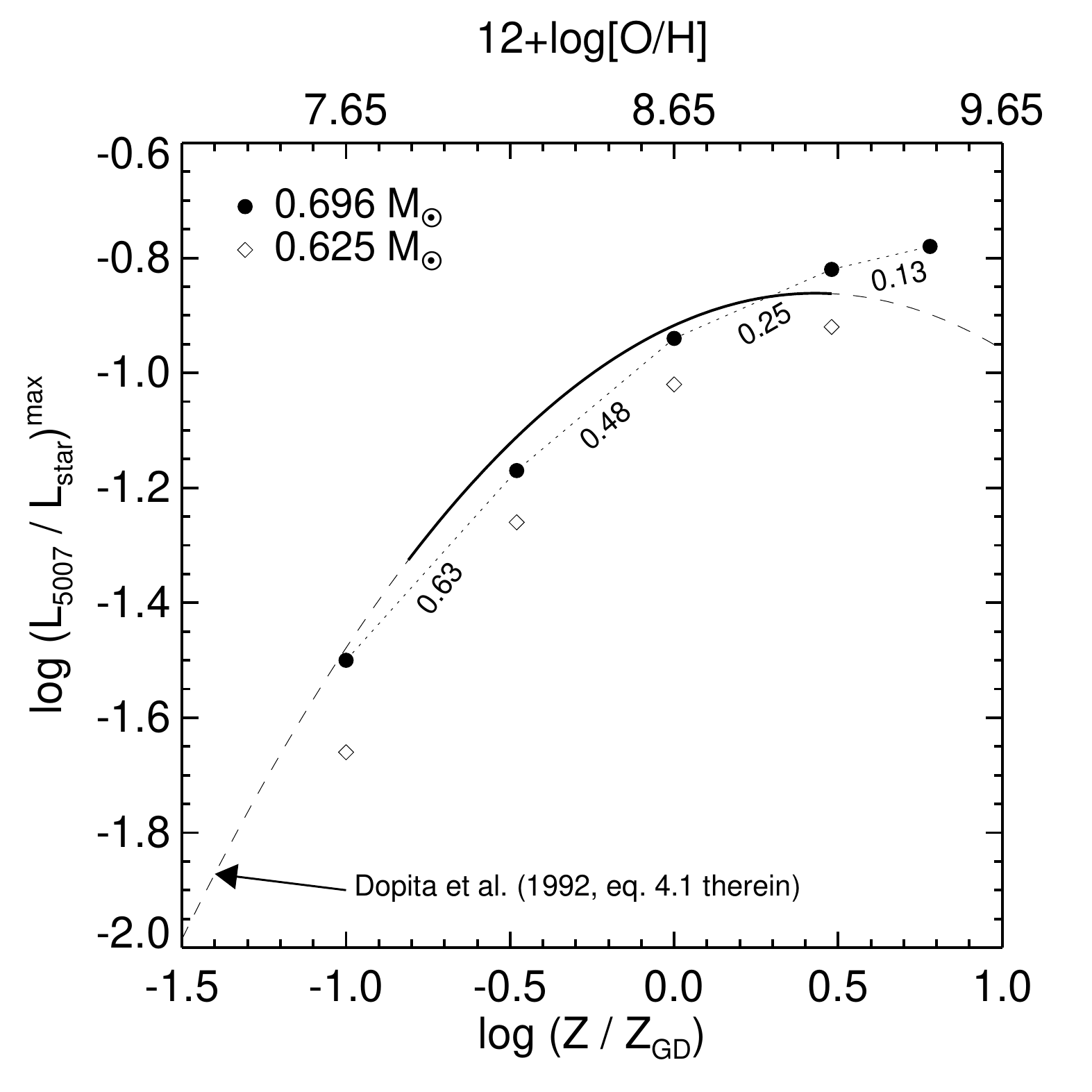}
\vskip-2mm
\caption{The maximum efficiency of converting stellar radiation into \oiii\ 5007 \AA\
         line radiation, $(L_{5007}/L_{\rm star})^{\rm max}$, of 0.696 \Msun\ (filled 
         circles) and 0.625 \Msun\ models (diamonds) against their metallicity,
         $Z/Z_{\rm GD}$. 
\changed{The upper abscissa gives the corresponding oxygen content relative to hydrogen,
         $\epsilon (\mbox{O})\ (= \log ({\rm O}/{\rm H}) + 12)$.}
         The numbers along the dotted line segments 
         connecting the 0.696 \Msun\ data points indicate the slope.  The inverted
         parabola (dashed and thick line) is a fit to the predictions of (optically thick)
         photoionisation models computed by \citet[][Eq. 4.1 therein]{dop.92}.  
         The thick portion corresponds to the metallicity range covered by the 
         Dopita et al. models. 
        }
\label{max.oiii}
\end{figure}

  Since our ${\alpha=2}$ sequences (0.696 \Msun\ and 0.625 \Msun) cover the range 
  about the bright 5007 \AA\ cut-off of the PNLF, we used them also to
  check the variation of the 5007 \AA\ luminosity with metallicity $Z$.%
\footnote{Note that $Z$ refers to an abundance pattern scaled to those
          of the Galactic disk PNe.  Thus, we always have $Z_{\rm oxygen}\propto Z$.}
  For this purpose
  we show in Fig.\,\ref{max.oiii} how the maximum 5007\,\AA\ luminosity,
  relative to the bolometric luminosity of the central star (i.e. the efficiency of 
  converting stellar UV radiation into \oiii\ 5007\,\AA\ line emission), varies with $Z$, 
  and compare the outcome of our hydrodynamical models with predictions of (optically
  thick) photoionisation models of \citet{dop.92}.  We used their Eq. (4.1) which is a 
  fit to their model predictions, corrected for the fact that \citeauthor{dop.92}
  used a higher solar oxygen abundance (8.84 on the usual scale).

  The maximum efficiency of converting
  stellar radiation into \oiii\ 5007 \AA\ line emission is 0.11 for
  the (optically thick) 0.696 \Msun\ models with the Galactic disk
  composition used here (i.e. $\epsilon(\rm O) = 8.65$).  
  The corresponding stellar effective temperature
  is about 110\,000 K.  The maximum efficiency decreases with $Z$
  and occurs at progressively lower stellar temperatures, i.e. 
  $\teff\simeq\! 136\,000$ K for $Z=3Z_{\rm GD}$,
  $\teff\simeq\! 101\,000$ K for $Z_{\rm GD}/3$ and ${\simeq\! 95\,000}$ K for
  $Z_{\rm GD}/10$.  Our models confirm that this efficiency decrease is weaker than
  expected because the reduced oxygen abundance is partly compensated for by an
  increase of the mean electron temperature: \changed{from 
  ${\langle T_{\rm e}\rangle=8170}$\,K ($6Z_{\rm GD}$) over} 9500 K ($3Z_{\rm GD}$),
  11\,440 K ($Z_{\rm GD}$), and 13\,980 K ($Z_{\rm GD}/3$) to 16\,650\,K ($Z_{\rm GD}/10$).
  According to our models, the general $ Z\,^{0.5}$ dependence of the 5007 \AA\ conversion
  efficiency quoted in the literature is inaccurate: it holds only for PNe with a
  metal content around $Z_{\rm GD}/3$ (see Fig.~\ref{max.oiii}).\footnote{%
  We recall that the metallicity discussion refers only to the circumstellar
  matter.  The central-star models (and the radiation fields) used are the same in 
  all four sequences and are based on evolutionary computations with
  about solar initial composition \citep{bloecker.95}.}

  Our models are, however, partly discrepant to the results of \citet{dop.92}: 
  For metal contents below the solar one the agreement between our hydrodynamic models 
  and the \citeauthor{dop.92} fit is reasonably good, but according to 
  their Eq. (4.1) the 5007 \AA\
  conversion efficiency reaches a maximum at ${Z\simeq 3\,Z_{\rm GD}}$ not found
  with our models (cf. Fig. \ref{max.oiii}). \changed{Our models are rather suggesting 
  a levelling off at extremely high oxygen abundances instead.} 
  We do not know the reason for this discrepancy,
  since our hydrodynamic models, being still quite dense 
  and \changed{optically thick}, are also in ionisation \emph{and} thermal equilibrium. 
\changed{However, our hydrodynamical sequences mimic consistently the  evolution 
         of bright nebulae systems as realistic as possible, whereas the parameter 
         space of the Dopita et al. models is limited and does not cover the high stellar
         luminosities necessary for PNe close to the luminosity function cut-off.}
  Therefore, we caution that \citeauthor{dop.92}'s fit formula for estimating the metallicity 
  dependence of the \oiii\ cut-off brightness might lead to erroneous conclusions if it is
  extrapolated into the metal-rich domain \citep[cf., e.g.,][Fig. 5 therein]{ciardetal.02}.

  The models around the 0.625 \Msun\ star become optically thin and therefore have always
  lower conversion efficiencies.  
\changed{A more detailed description how the \oiii\ brightness depends on model
         parameters and/or evolutionary stages can be found in \citetalias{schoenetal.07},
         although only for the Galactic disk chemical composition.}

\section{PNe in distant stellar systems}
\label{distant}

  We have already shown at different occasions, viz. \citetalias{schoenetal.05a},
  \citetalias{schoenetal.05b}, \citet{SS.06}, and recently in 
  \citetalias{schoenetal.07}, 
  that the basic physical model used in our simulation of the planetary nebula evolution 
  is very successful in explaining basic observed structures of PNe.
  A further important test of the quality of our hydrodynamical models is to answer 
  the question if and how well the models are also able to explain global properties 
  of PNe samples in stellar systems so distant that a PN cannot be spatially resolved.
  Very useful for testing are basic diagrams as those involving characteristic line
  ratios and expansion velocities measured from the Doppler width of bright emission lines.

\begin{figure}[t]
\vskip -3mm
\includegraphics[width=\columnwidth]{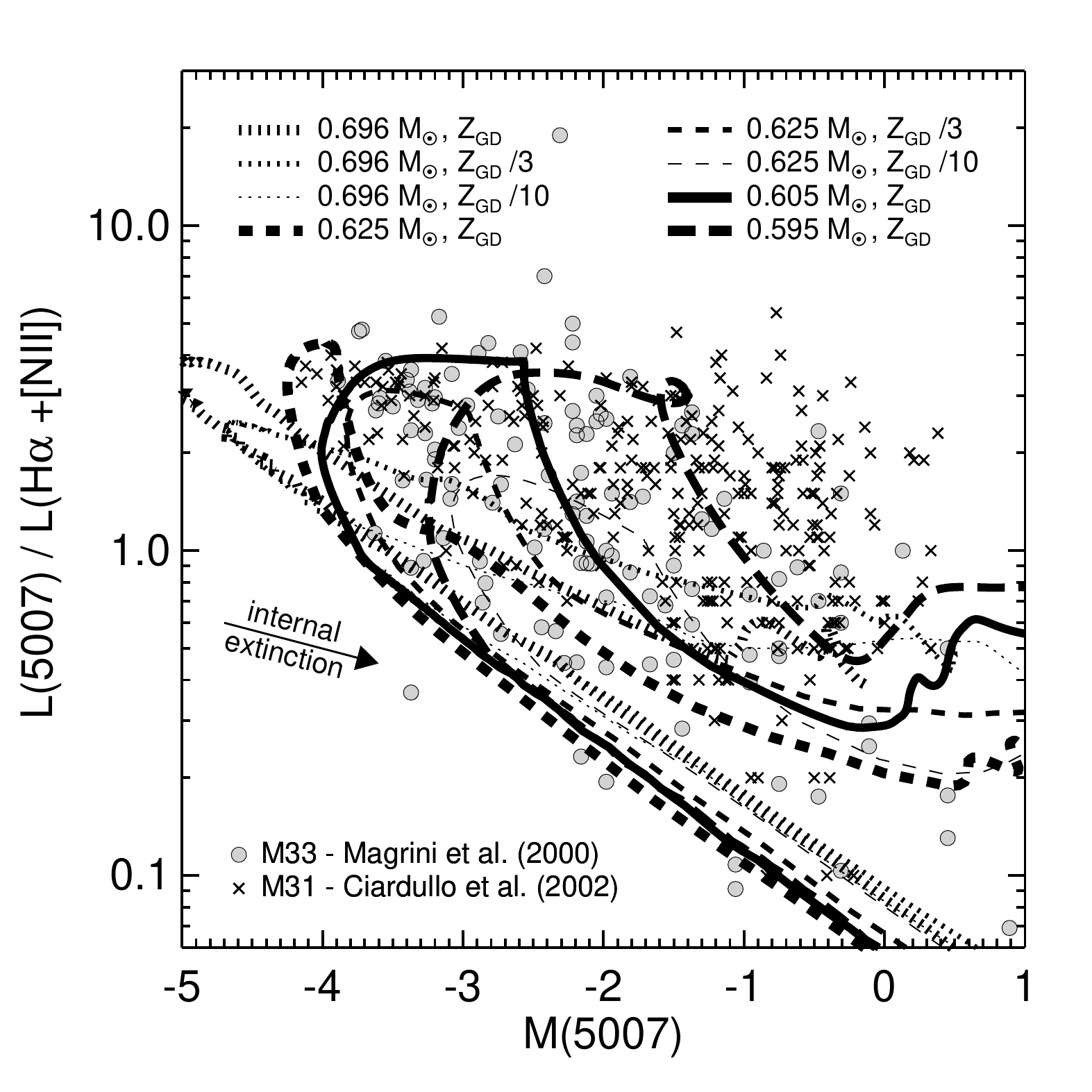}
\vskip-3mm
\caption{Observed line ratios $L(5007)/L(\rm H\alpha$\,+\,[\ion{N}{ii}]) for PNe of the 
         bulge of M\,31 \citep{ciardetal.02} and the disk of M\,33 \citep{magretal.00}, as 
         a function of $M(5007)$, compared with the predictions of various hydrodynamical
         sequences with different metallicities.  Evolution proceeds always from
         faint magnitudes and low line ratios up to maximum 5007 \AA\ emission and then
         back again, but at higher line ratios.  Foreground extinction is considered
         for the data, but not internal extinction.  The arrow indicates the direction 
         of shifts to be applied to the tracks if additional internal extinction occurs.
        }
\label{ciardullo}
\end{figure}

\subsection{Line-ratio diagrams}
\label{lineratio.distant}

  We begin with a diagram which combines the line ratio $L(5007)/L(\mbox{H}\alpha$\,+\,[\ion{N}{ii}])
  with the brightness in \oiii, $M(5007)$.  As an example, Fig. \ref{ciardullo} compares
  the data of two Local Group galaxies, M\,31 \citep{ciardetal.02} and M\,33 
  \citep{magretal.00}, with our model predictions. The observations occupy a wedge-like
  region in this diagram and are, according to \citet{herrman.08}, roughly bounded 
  from below by a line 
\\[4pt]
 $ \log\, \{L(5007)/L({\rm H}\alpha$\,+\,[{\rm N\,{\sc ii}]})$\}= - 0.37\,
    M(5007) - 1.16,$\\[4pt]
  and from above by a horizontal line at \\[4pt] 
 $ L(5007)/L({\rm H}\alpha$\,+\,[{\rm N\,{\sc ii}]$) \simeq 4 .$
\\[4pt]
\indent
  The lower boundary is nearly perfectly matched by all of our models during their 
  early, optically-thick evolution towards maximum 5007 \AA\ emission, virtually 
  independently of their parameters and chemical composition.  This boundary is not 
  very well populated because this optically-thick phase of evolution is
  rather short \citepalias[cf.][Fig. 15 therein]{schoenetal.07}.  The maximum reachable 
  brightness in $M(5007)$ depends, of course, on the central star mass, the metallicity,
  and the thick/thin transition.  Models which remain optically thick (i.e. the ones
  with the 0.696 \Msun\ central stars) evolve then backwards a little bit above this 
  lower boundary.  Models which become optically thin turn then further upwards until 
  $L(5007)/L(\rm H\alpha$\,+\,[\ion{N}{ii}]${) \simeq 4}$ is reached and stay there 
  until the central star fades: The line ratio decrease after the `kink' is the result 
  of the rapid luminosity drop of the central star as hydrogen burning ceases, with a
  corresponding increase of the [\ion{N}{ii}] lines because of a temporary recombination 
  to N$^+$.  A slow increase of this line ratio occurs only at fainter magnitudes 
  (${M(5007)\ga 0}$) because of re-ionisation during the very slow early white-dwarf phase.
\changed{Because of the fast model evolution between the `kink' and the 
         re-ionisation stage, it is very unlikely that  observed objects are in 
         this particular phase.} 

\begin{figure}[t]
\vskip -3mm
\includegraphics[width=\columnwidth]{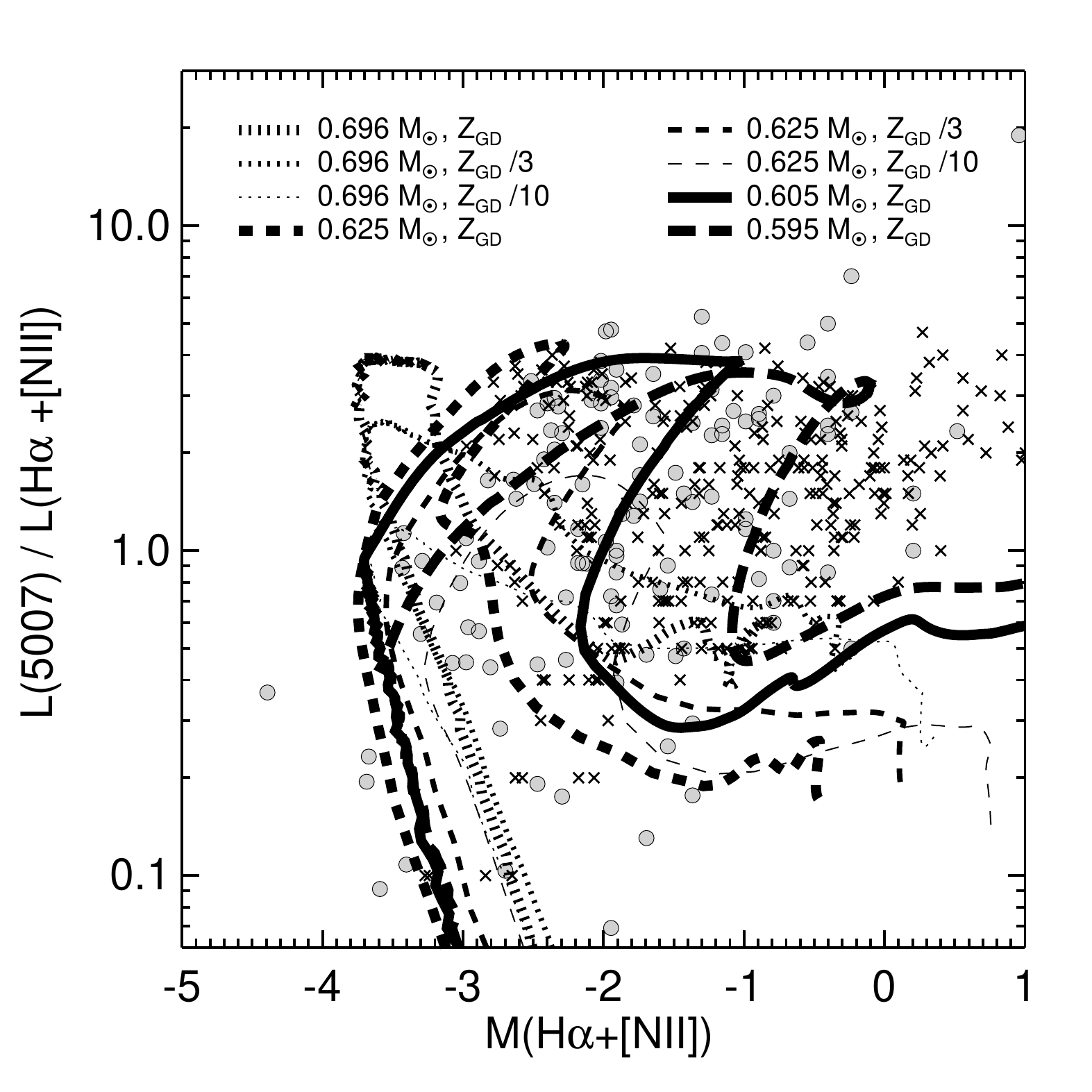}
\vskip-3mm
\caption{\label{ciardullo.2}
        The same data and theoretical sequences as presented in Fig. \ref{ciardullo}, 
        but now plotted over $M$(H$\alpha$\,+\,[\ion{N}{ii}]). Objects and sequences are
        only plotted down to $M(5007) = 1$, the brightness limit of the samples (cf. 
        Fig. \ref{ciardullo}).
        }
\end{figure}

  We can therefore state that the wedge-shaped region of Fig. \ref{ciardullo} is 
  populated by PNe which are evolving along the horizontal part of the HR diagram 
  through maximum \oiii\ brightness until the central star begins to fade.
  Nebulae around the most massive central stars, $\ga\!0.63$ \Msun, stay 
  always rather close to the lower boundary because they remain optically thick 
  or become only modestly optically thin.  Optically thin 
\changed{(or partly thin) models around central stars of $\simeq$0.63 \Msun\ to well 
         below 0.59 \Msun\ which are on their evolution across the HR diagram}
  fill the entire region between the lower and upper boundary and towards the right 
  until faint magnitudes.

  A similar diagram can be made for 5007/(H$\alpha$\,+\,[\ion{N}{ii}) vs.
  $M$(H$\alpha$\,+\,[\ion{N}{ii}]) and is shown in Fig. \ref{ciardullo.2}. 
  In this diagram the PNe remain bright in  (H$\alpha$\,+\,[\ion{N}{ii}]) during the 
  fading of the central stars towards white-dwarf configurations because of 
  recombination and a temporary dominance of the [\ion{N}{ii}] lines:
  The models are now in H$\alpha$\,+\,[\ion{N}{ii}] brighter than in \oiii\ 5007 \AA\
  by up to 1.5 mag.  Hence, our hydrodynamical models suggest that 
  the deficit of objects in the lower right corner of this diagram is 
  not due to sample incompleteness (as proposed by \citealt{ciardullo.09})
  but is rather a consequence of central star evolution: The early white-dwarf 
  phase where the nebula re-ionises limits the observability of PNe in distant 
  systems if observed in H$\alpha$\,+\,[\ion{N}{ii}].

  Another useful diagram for a comparison between data and the predictions of our
  simulations is that presented in \citet[][Fig.\,1 therein]{richer.06}.  
  It combines the nebular ionsation
  structure, as measured by [\ion{O}{iii}] $\lambda$5007 \AA\ and [\ion{O}{ii}] 
  $\lambda$3727 \AA,  with the strength of
  \ion{He}{ii} $\lambda$4686 \AA, which is essentially an indicator of the stellar
  surface (effective) temperature. 
  An adaptation of \citeauthor{richer.06}'s Fig.\,1
  is shown in Fig. \ref{richer.oii}, with the predictions of our evolutionary 
  models included.  This diagram is nearly independent of metallicity, as is 
  demonstrated by the three 0.625 \Msun\ sequences with different $Z$, but contains 
  only objects with rather hot and thus evolved central stars because 
  $\lambda$4686 \AA\ must be measurable.

\begin{figure}[t]
\vskip -4mm
\includegraphics[width=\columnwidth]{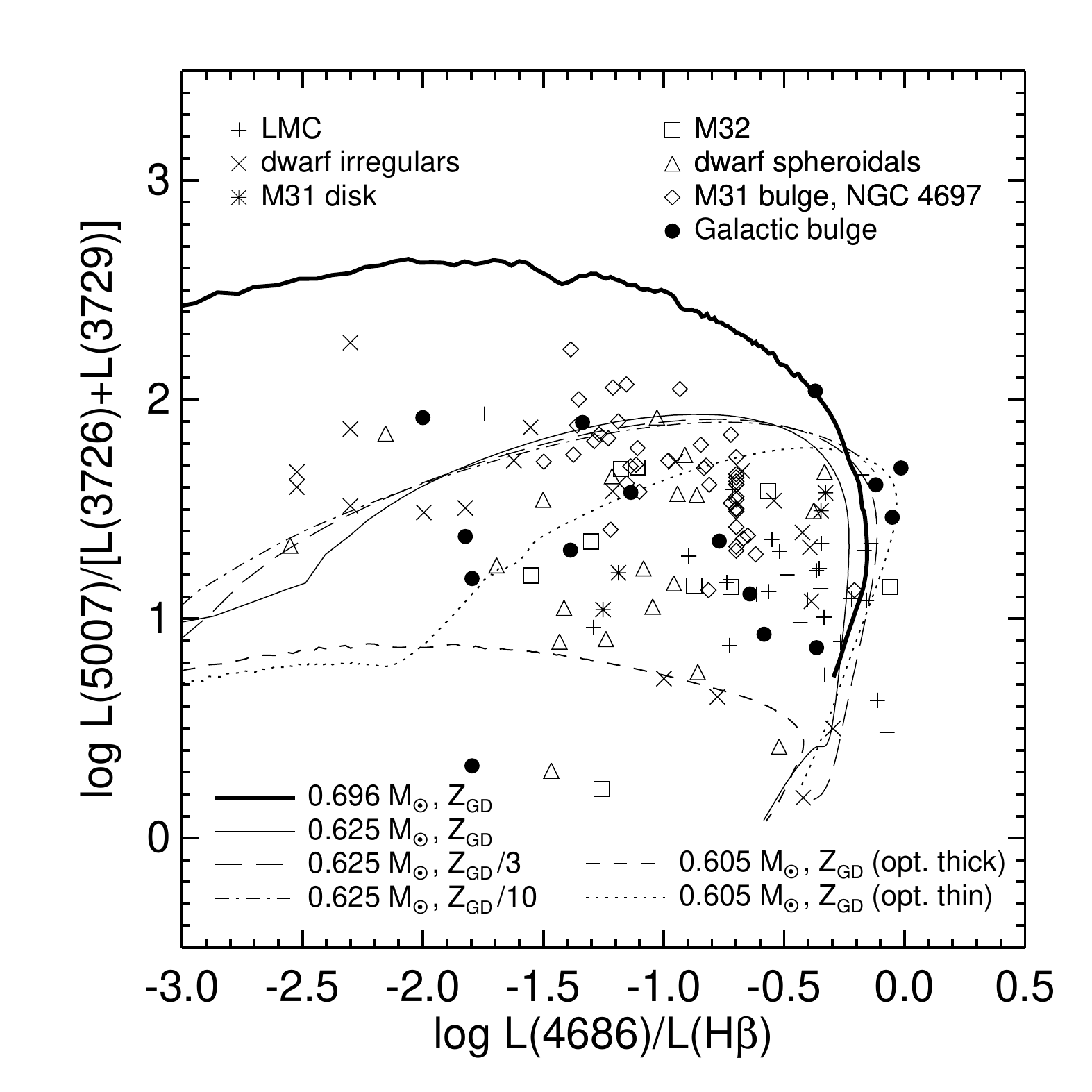}
\vskip-2mm
\caption{The ratios $L(5007)/[L(3726)\! +\! L(3729)]$ are shown against 
         $L(4686)/L(\rm H\beta)$.  The predictions from a selection of our simulations 
(\changed{the two 0.605 \Msun\ sequences are from \citetalias{schoenetal.07}})
         is compared with Local Group data taken from 
         \citet[][references therein and priv. comm.]{richer.06}.
         The Galactic bulge data are from \citet{stasetal.98}.
         Shown are only objects with well-determined line 
         strengths and which are not fainter than two magnitudes below the \oiii\ PNLF
         cut-off.  Evolution proceeds from left to right and then down 
\changed{because of recombination.}  
         In order to be consistent with the data, the tracks are truncated at two 
         magnitudes below the PNLF cut-off. The kinks seen in two sequences are due to the 
         optically thick/thin transition.  
        }
\label{richer.oii}
\end{figure}

  The interpretation of this diagram, based on our evolutionary models, is as follows:
\begin{itemize}
\item  The right border, at ${L(4686)/L(\rm H\beta)\simeq 1.0}$, is occupied by highly
       ionised objects for which the emitting volumes of H$\beta$ and \ion{He}{ii}
       $\lambda$4686 \AA\ are about equal.  When recombination occurs, this ratio becomes
       smaller, as is seen in our models.
\changed{At the same time, the strength of the \oii\ doublet increases considerably,
        more than the $\lambda$5007 \AA\ line, because the models become optically 
        thick. These objects close to the right border have very hot and still luminous
        central stars, producing high nebular excitation, but 
        at the same time also strong \oii\ emission, and \nii\ emission as well.}
\item  The upper limit ${L(5007)/[L(3726) + L(3729)]\simeq 100}$ is 
 \changed{explained either from above by optical thick sequences with massive
          central stars (${\approx\! 0.65}$ \Msun), or from below by sequences with less 
          massive central stars} whose models become optically thin during the course 
          of evolution. 
       Objects with massive
       central stars with $\ga\!0.65$ \Msun\ are obviously not existent (cf. also
       Fig. \ref{ciardullo}).
\item  A lower boundary for ${L(5007)/[L(3726) + L(3729)]}$ is provided by optically
       thick models around central stars of about 0.6 \Msun\ 
       \changed{(the three outliers neglected).}
\end{itemize}

\changed{The whole observed sample can be interpreted to consist of objects with
         rather highly ionised/excited optically thick or thin nebular envelopes, 
         as dependent on their evolutionary stage,
         with central stars with masses between about 0.6 and 0.65 \Msun.}

\subsection{Expansion properties}
\label{expansion.distant}

\begin{figure}[t]
\vskip 1mm
\includegraphics*[bb= 0.5cm 0.5cm 15.2cm 15.7cm, width= \linewidth]
         {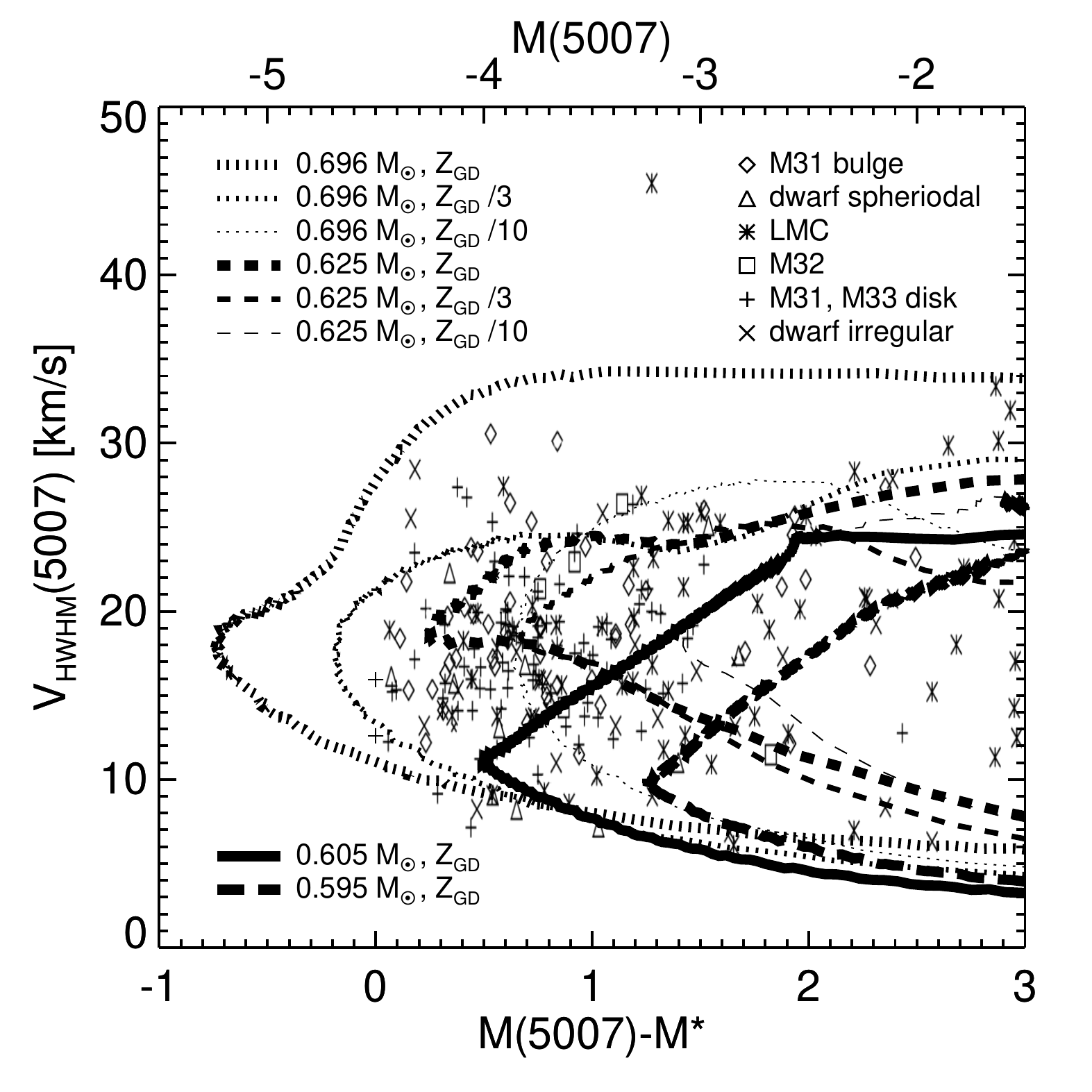}
\caption{Evolution of PN models in the $V_{\rm HWHM}$-$M(5007)$ plane.  Shown are
         nebular sequences around central star models of 0.696 and 0.625 \Msun,
	 each of which started with the same initial model but with three different
	 metal contents, as indicated in the figure legend.  
\changed{These sequences are supplemented by ones with less massive central stars of
         0.605 and 0.595 \Msun, both with hydrodynamically computed initial envelopes
         (cf. Table \ref{sequences}).}     The abscissa gives either
	 the \oiii\ magnitude $M(5007)$ as defined by \citet{jacoby.89} (\emph{top}) or
	 ${M(5007)-M^{\star}}$, with ${M^{\star}=-4.48}$ (\emph{bottom}).  The evolution
	 proceeds from low velocities and faint magnitudes over medium velocities
	 and bright magnitudes towards high velocities and again faint magnitudes.
         The models are compared with a large data set of bright PNe belonging to a
         variety of galaxies (or parts of them) as indicated in the legend and 
         extracted from \citet[][Fig. 4 therein]{richer.06}. 
}
\label{vel.richer}
\end{figure}

  In distant stellar systems one cannot spatially resolve the profiles of
  emission lines, and information on expansion properties can only be gained from
  measuring the halfwidth of the profile.  
  Although relevant studies are rare, the following picture emerged:
  \cite{dop.85, dop.88} showed for a large sample of Magellanic
  Cloud objects that there exists an evolution of \changed{nebular expansion rates,
  in terms of line widths,}
  from low to higher values with nebular excitation class or stellar effective temperature.
  \citet{richer.06, richer.07} reported, based on an extensive spectroscopic study of
  PNe in the Local Group galaxies, that there is virtually no dependence of the mean
\changed{expansion properties, as measured from the halfwidth of the 5007 \AA\ \oiii\
         line, $V_{\rm HWHM}(5007)$,}
  on the metallicity of the parent stellar population.
\changed{This halfwidth varies} between about 8 and 30 \kms, with a clear accumulation around
  18 \kms\ for the brightest objects \citep[][]{richer.06, richer.07}.  For the 
  Galactic bulge PNe, \citet{gesicki.00} found also a mean $V_{\rm HWHM}(5007)$ of 18 \kms.

  This obvious observed invariance of PNe \changed{emission line widths} with metallicity 
  seems to contradict strongly the prediction of our hydrodynamical models that expansion 
  speeds generally increase with decreasing metal content.
  We thus used the sequences discussed in the previous sections for clarifying the situation
  and interpreting existing observations.  

\begin{figure*}[t]
\vskip-1mm
\includegraphics[width=0.33\linewidth]{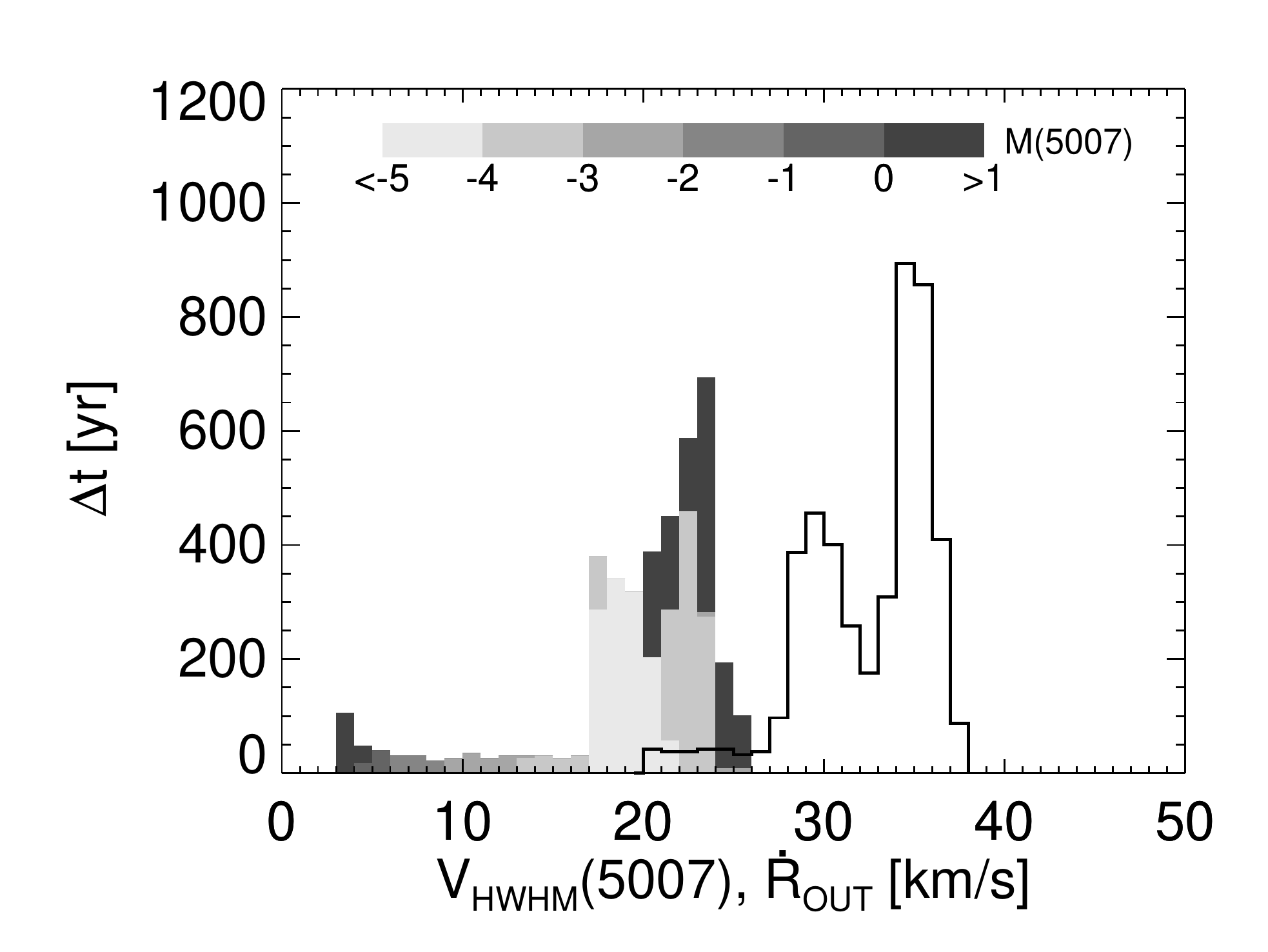}
\includegraphics[width=0.33\linewidth]{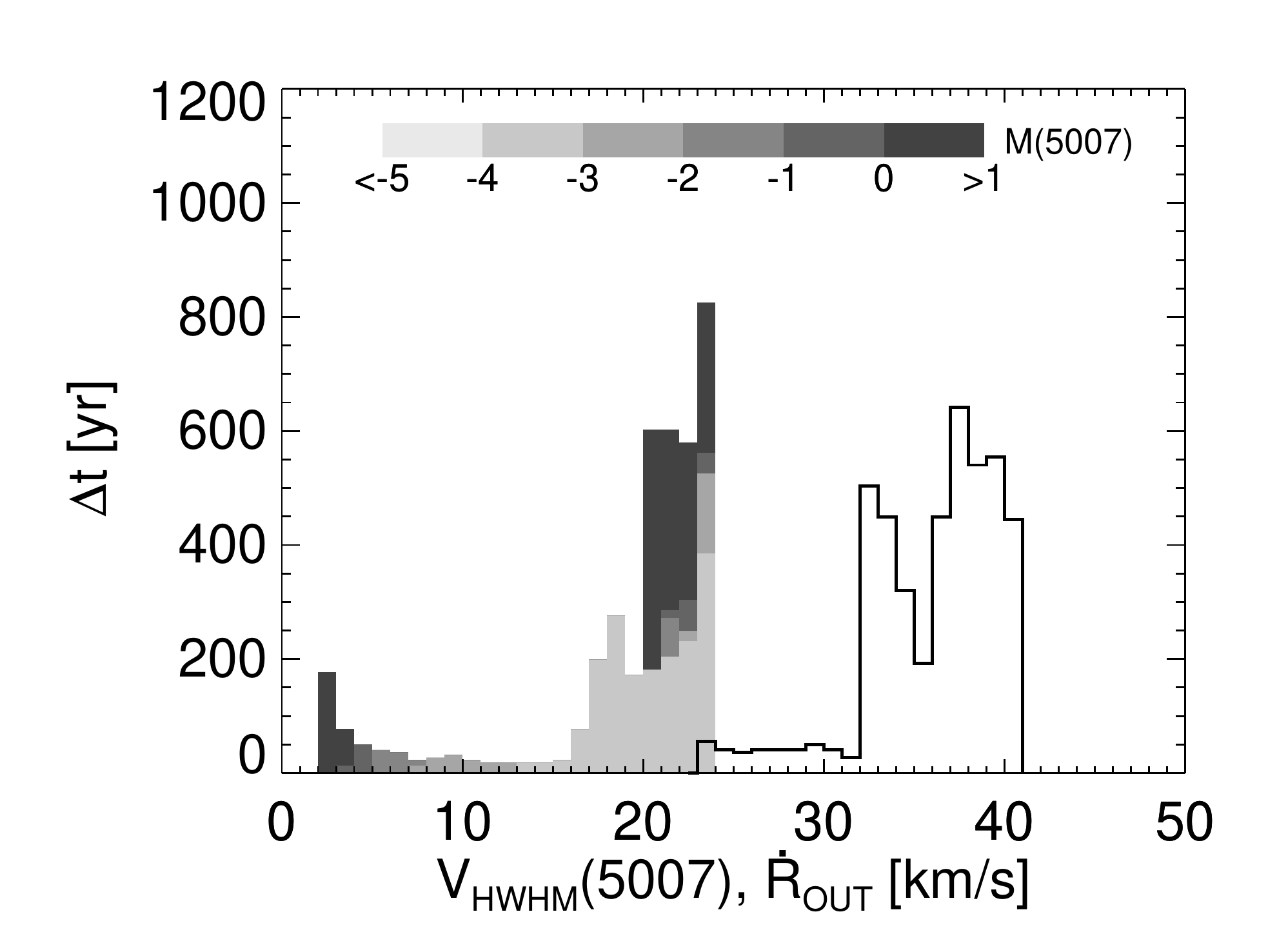}
\includegraphics[width=0.33\linewidth]{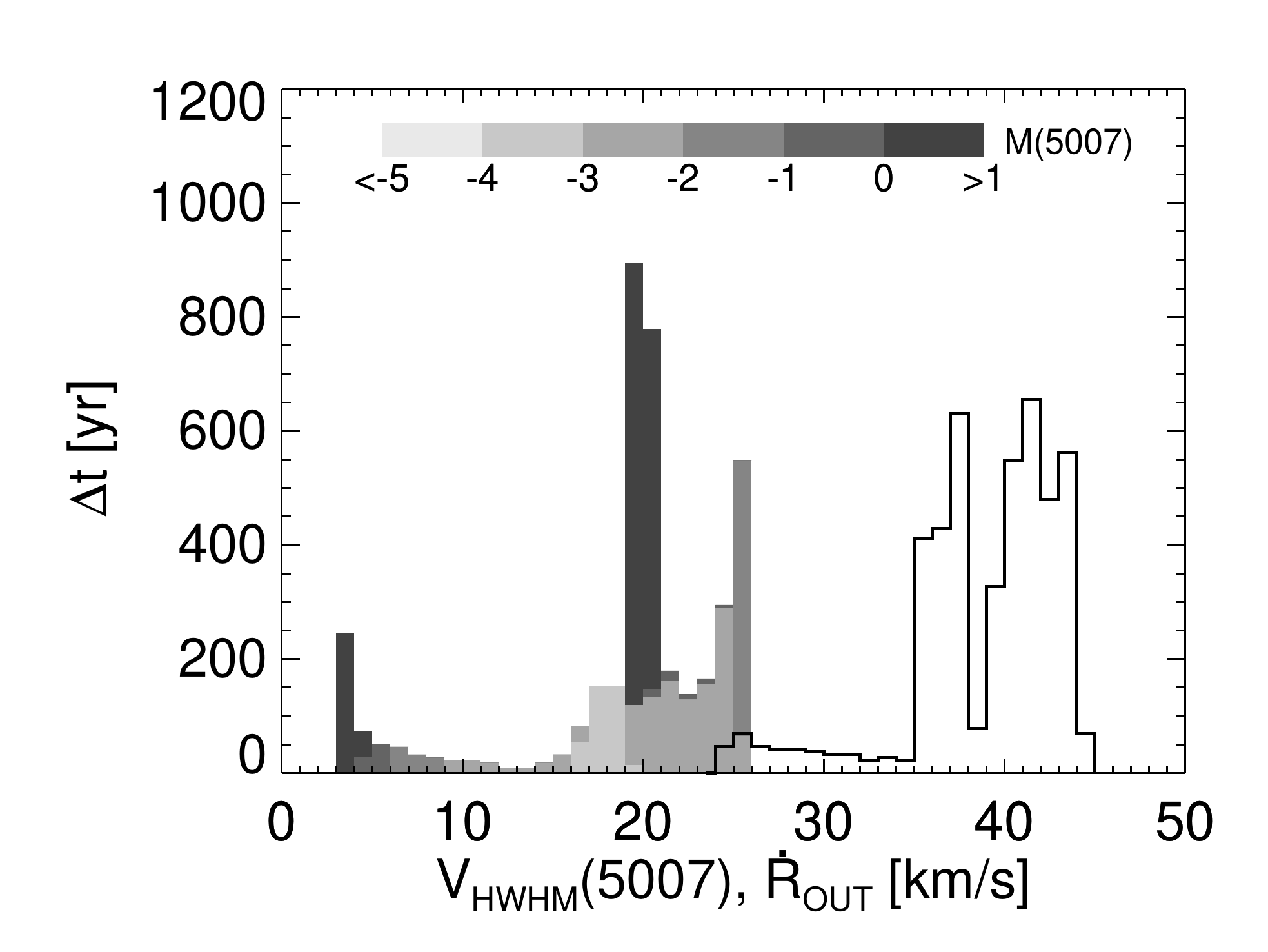}
\vskip-2mm
\caption{Times $\Delta t$ spent by the models within the velocity intervals
         ${\Delta V_{\rm HWHM} =1}$ \kms\ and ${\Delta \dot{R}_{\rm out}= 1}$ \kms.
	 The $V_{\rm HWHM}$ bins are broken down into $M(5007)$ intervals as
	 indicated by the different shades.  The $\Delta \dot{R}_{\rm out}$ histogram
	 serves only for comparison and is not broken down into magnitude intervals.
	 Shown are the three 0.625 \Msun\ sequences with
	 $Z=Z_{\rm GD}$ (\emph{left}), $Z=Z_{\rm GD}/3$ (\emph{middle}), and
         $Z=Z_{\rm GD}/10$ (\emph{right}) whose velocity and brightness evolution
	 are also seen in the Figs.~\ref{0.625.vel} and \ref{vel.richer}.
	 The maximum post-AGB age considered is 5000 years.  The $\Delta t$ for
	 the faintest magnitudes are lower limits only because of this time limit.
        }
\label{histo.0625}
\end{figure*}
\begin{figure*}[t]
\vskip-1mm
\includegraphics[width=0.33\linewidth]{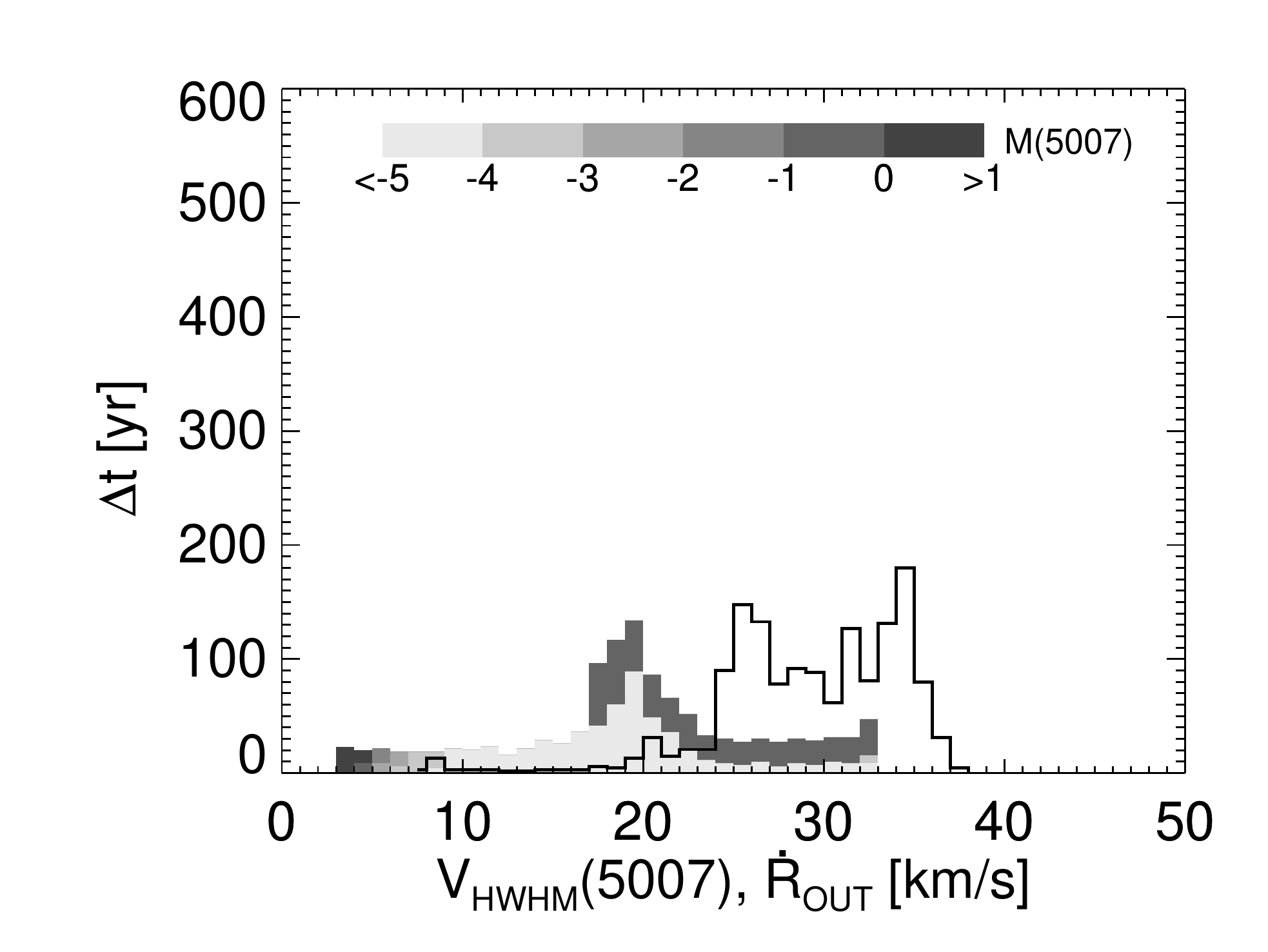}
\includegraphics[width=0.33\linewidth]{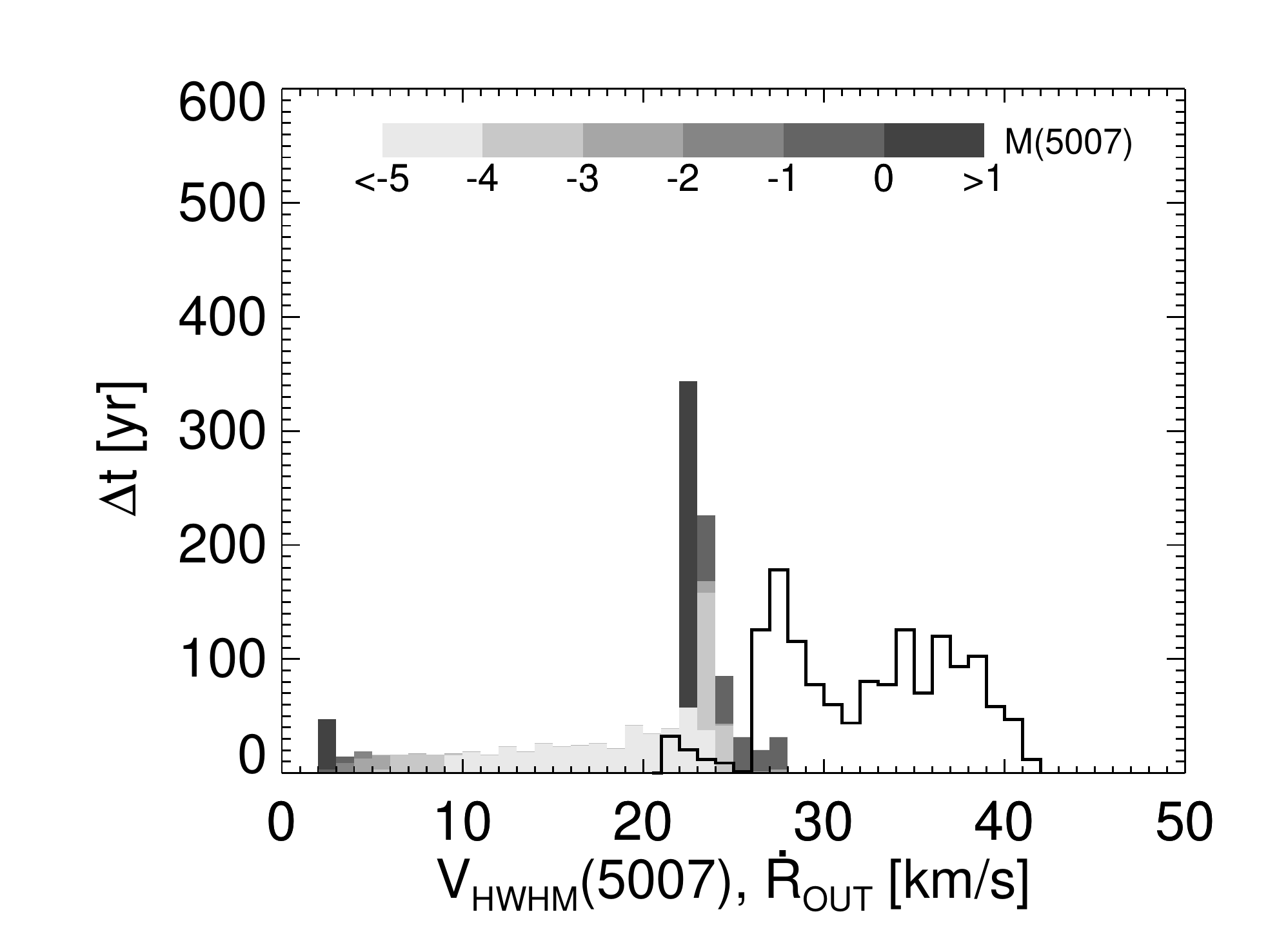}
\includegraphics[width=0.33\linewidth]{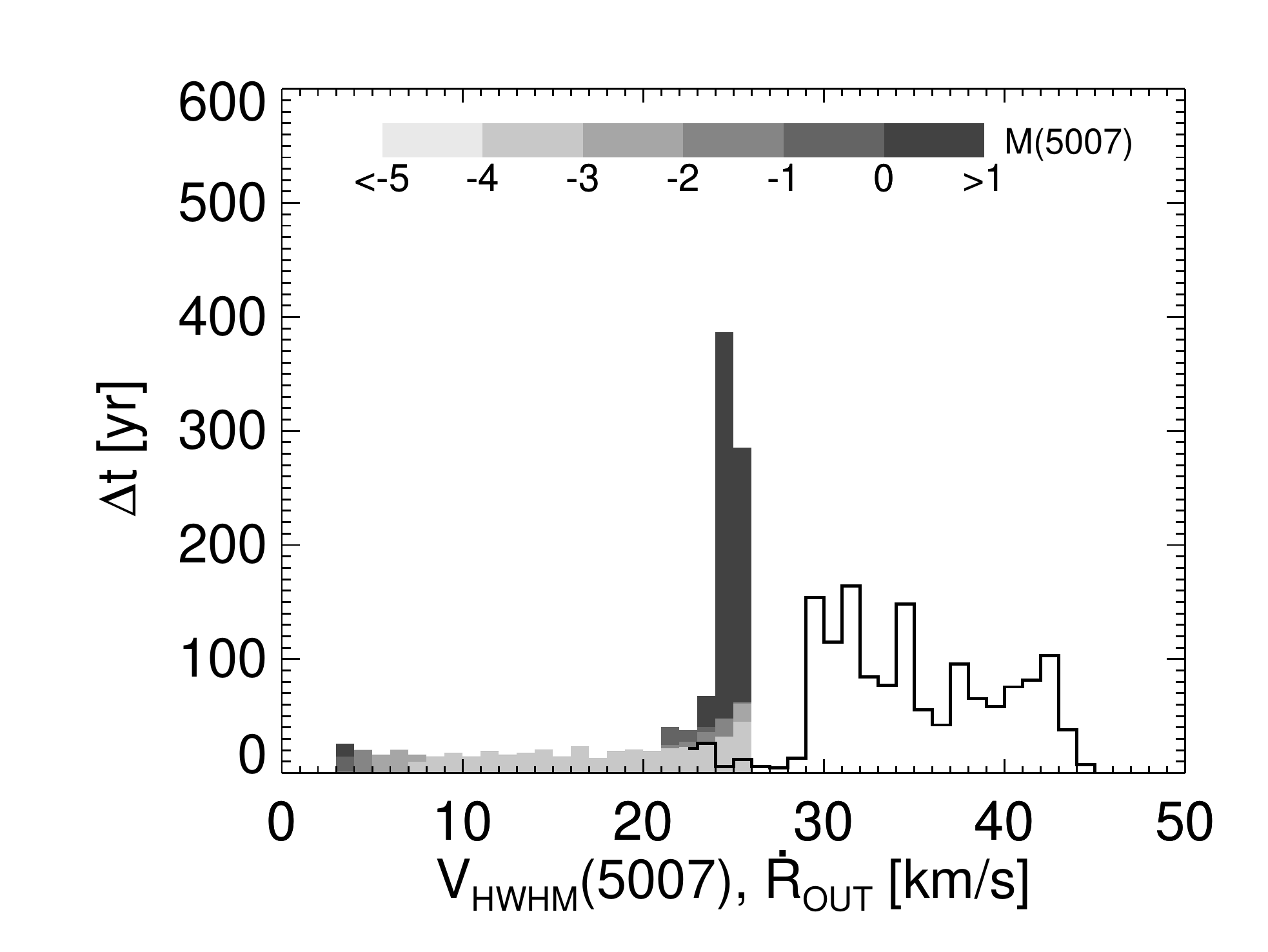}
\vskip-2mm
\caption{The same as in Fig.\,\ref{histo.0625} but for the three 0.696 \Msun\
         sequences with $Z=Z_{\rm GD}$ (\emph{left}), $Z=Z_{\rm GD}/3$ (
	 \emph{middle}), and $Z=Z_{\rm GD}/10$ (\emph{right}) whose brightness
	 evolution is also seen in Fig.~\ref{vel.richer}.
	 The maximum post-AGB age considered is 1200 years.
	 The $\Delta t$ for the faintest magnitudes are lower limits only because
	 of this time limit.
        }
\label{histo.0696}
\end{figure*}

  We start with showing the evolution of our
  models in the $V_{\rm HWHM}$--$M(5007)$ plane in Fig.~\ref{vel.richer} which
  is adapted from Fig.~4 in \citet{richer.06} and contains also the same data
  from different galaxies and/or parts of them.  
\changedIII{All the relevant data are now also available from \citet{richeretal.10b}.} 
  The theoretical tracks 
  of our two most massive central stars (0.696 and 0.625 \Msun) surrounded with PNe models 
  with different metallicities (${Z=Z_{\rm GD},\ Z_{\rm GD}/3,\ Z_{\rm GD}/10}$)
  suggest that such a diagram is not well suited to study and compare
  expansion properties of bright PNe in different environments since
  all tracks cover or embrace the observational data 
  with no obvious trend with metal content.
  Moreover, at their maximum $M(5007)$ brightness, all models, independently
  of their metallicity (cf. Fig. \ref{0.625.vel}), provide practically the same 
  halfwidth velocity of ${V_{\rm HWHM}\simeq 18}$ \kms, 
  in astonishingly good agreement with the observations!

  The theoretically expected distribution of the HWHM velocities depends on the 
  evolutionary speed through the  $V_{\rm HWHM}$--$M(5007)$ plane
  and the central-star mass distribution.
  In order to illustrate the former better, we show in
  Figs.~\ref{histo.0625} and \ref{histo.0696} histograms which give the visibility time
  $\Delta t$ spent by a particular model within a velocity bin of
  $\Delta V_{\rm HWHM} =1$ \kms. Thus, $\Delta t$ is a proxy for the expected
  number of PNe per velocity interval of 1 \kms.
  These histograms combine the information contained in Figs.~\ref{0.625.vel}
  and \ref{vel.richer}.

  In particular, Fig.~\ref{histo.0625} displays the results for the 0.625 \Msun\
  sequences.  The total simulation time shown in these figures is much longer
  than seen in Fig.~\ref{vel.richer}, and the tracks extend thus to fainter magnitudes.
  The $V_{\rm HWHM}$ bins are additionally broken down into magnitude intervals to 
  account for the fact that the velocity is not a monotonic function of $M(5007)$
  (Fig.~\ref{vel.richer}). For instance, the low-velocity stage is passed rather
  quickly, which may explain the paucity of low-velocity objects seen in 
  Fig. \ref{vel.richer}. 

  We show, for comparison, also the
  $\dot{R}_{\rm out}$ distributions.  They appear bimodal where higher
  velocities belong to the high-luminosity phase as the central star crosses
  the HR diagram, and  lower velocities to the recombination stage as the stellar
  luminosity fades.

   The histograms seen in Fig.\,\ref{histo.0696} belong to three simulations
   with the 0.696 \Msun\ central-star model.  Although the time scales are shorter,
   the general behaviour is the same as already discussed above.
   The histograms shown in Figs.~\ref{histo.0625} and \ref{histo.0696} demonstrate
   clearly that in a magnitude limited sample in which only the brightest PNe
   are contained all $V_{\rm HWHM}$ are quite low and cluster around 20~\kms.
   Very low \changed{halfwidth} velocities (${<\!10}$ \kms) occur only for faint models 
   at the beginning of their brightness evolution (cf. Fig.~\ref{0.625.vel}).  
   Considering only models
   within the brightest bin of each sequence, the means of their halfwidth velocities
   are between 18 and 20 \kms, i.e. virtually \emph{independent} of their metal content.

   From the discussion of the Figs.\,\ref{vel.richer}--\ref{histo.0696} 
   we see that our models are in excellent agreement with 
   measurements of bright PNe in distant stellar systems.
   The $V_{\rm HWHM}$ distribution of bright Galactic bulge PNe peaks between
   15 and 20 \kms\ \citep[][Fig.~6 therein]{gesicki.00}, and the 11 bright
   Virgo cluster PNe with spectroscopically resolved \oiii\ 5007 \AA\ lines
   have $V_{\rm HWHM}$ between 12 and 22 \kms, with a mean
   value of 16.5~\kms\ \citep{arnaboldi.08}.  The bright PNe of the Local Group of
   galaxies plotted in Fig.~4 of \citet{richer.06} cluster in the
   brightness range of $M(5007)\simeq -4.2\ldots -3.5$, and their eye-estimated
   mean $V_{\rm HWHM}$ is ${\simeq\!18}$~\kms.  
\changedIII{\citet{richeretal.10b} argued again that the mean 5007 \AA\ line width is not
            dependent upon maximum \oiii\ luminosity, metallicity or age
            of the progenitor population.} 
   All these findings are well explained by our hydrodynamical models, especially 
   by the 0.625 \Msun\ sequences (Fig.~\ref{vel.richer}).

\begin{figure}[t]
\includegraphics[width=\linewidth]{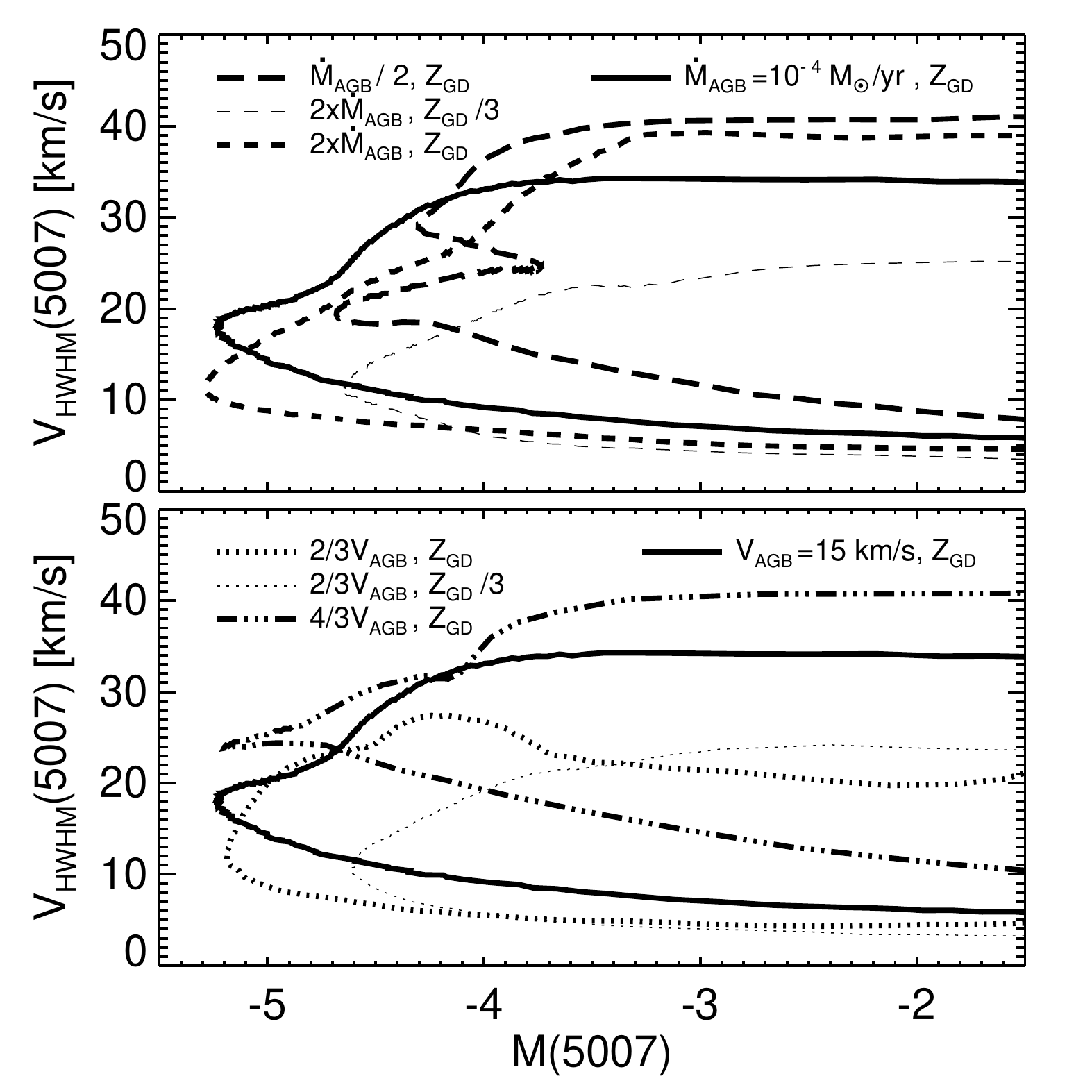}
\caption{$V_{\rm HWHM}(5007)$ vs. $M(5007)$ for 0.696 \Msun\ sequences (${\alpha=2}$) with 
         the AGB mass-loss rates (\emph{top}) and AGB wind velocities (\emph{bottom}) varied,
         as indicated in the legends.  In the top panel ${V_{\rm agb}= 15}$ \kms\ holds
         in all cases, whereas we have always ${\dot{M}_{\rm agb}=1\times10^{-4}}$ \Mdot\
         in the bottom panel.   Our standard 0.696 \Msun\ sequence
         with ${\dot{M}_{\rm agb}=1\times10^{-4}}$ \Mdot and
         ${V_{\rm agb}= 15}$ \kms\ serves in both panels as a reference (solid lines).  
         Evolution proceeds always from low to high velocities.
        }
\label{loss.init}
\end{figure}

  All sequences discussed so far have the same initial model, viz. they start with an 
  assumed constant AGB mass-loss rate of ${1\times10^{-4}}$ \Mdot\ and a constant outflow
  velocity of 15 \kms.  Here we briefly investigate how the above findings depend on initial
  mass-loss rate and wind velocity.  For this purpose we computed additional sequences for
  the 0.696 \Msun\ post-AGB stellar model (${\alpha=2}$): The initial mass-loss rate of 
  $1\times10^{-4}$ \Mdot\
  was doubled/halved and the initial wind velocity of 15 \kms\ lowered/increased by 5 \kms. 
  The results are shown in Fig. \ref{loss.init}.

  The interpretation of the influence of the AGB wind velocity is simple:
  the initial velocity enters as a corresponding up and downshift of
  $V_{\rm HWHM}(5007)$, rather independently of the metal content 
  (Fig. \ref{loss.init}, bottom, sequences $2/3V_{\rm agb}$ with $Z_{\rm GD}$ and
  $Z_{\rm GD}/3$, and $4/3V_{\rm agb},\,Z_{\rm GD}$).
  AGB mass-loss rate differences have a more subtle impact.
  The increased $\dot{M}_{\rm agb}$ leads to a corresponding lower value of 
  $V_{\rm HWHM}(5007)$ at the luminosity maximum, as expected. A lower 
  $\dot{M}_{\rm agb}$, however, does not change the expansion much at
  maximum brightness (Fig. \ref{loss.init}, top). 
  The reason is that, because of the less dense AGB wind, 
  the nebular shell becomes, in this particular case, optically thin well \emph{before} 
  its maximum possible \oiii\ line luminosity is reached.
  The \oiii\ brightness becomes temporarily lower until the shell recombines and
  becomes again optically thick, and hence brighter.  
\changed{Concerning the expansion behaviour at maximum brightness,} we see also that the 
  metal content is virtually of no influence (Fig. \ref{loss.init}, top, sequences 
  $2\dot{M}_{\rm agb},\,Z_{\rm GD}$ and $2\dot{M}_{\rm agb},\,Z_{\rm GD}/3$).

  For completeness we add the discussion of two sequences with initial
  envelopes generated by radiation-hydrodynamics simulation along the upper AGB
  (sequences Nos. 6 and 6a of \citetalias{schoenetal.07}).  
  Both sequences have
  normal metallicities, i.e. $Z=Z_{\rm GD}$, and their $V_{\rm HWHM}$ evolution with
  $M(5007)$ \changed{has already been displayed in Fig. \ref{vel.richer}.  
  The corresponding histograms are seen in Fig.~\ref{histo.hydro}. }

\begin{figure}[t]
\centering
\includegraphics[width=0.66\linewidth]{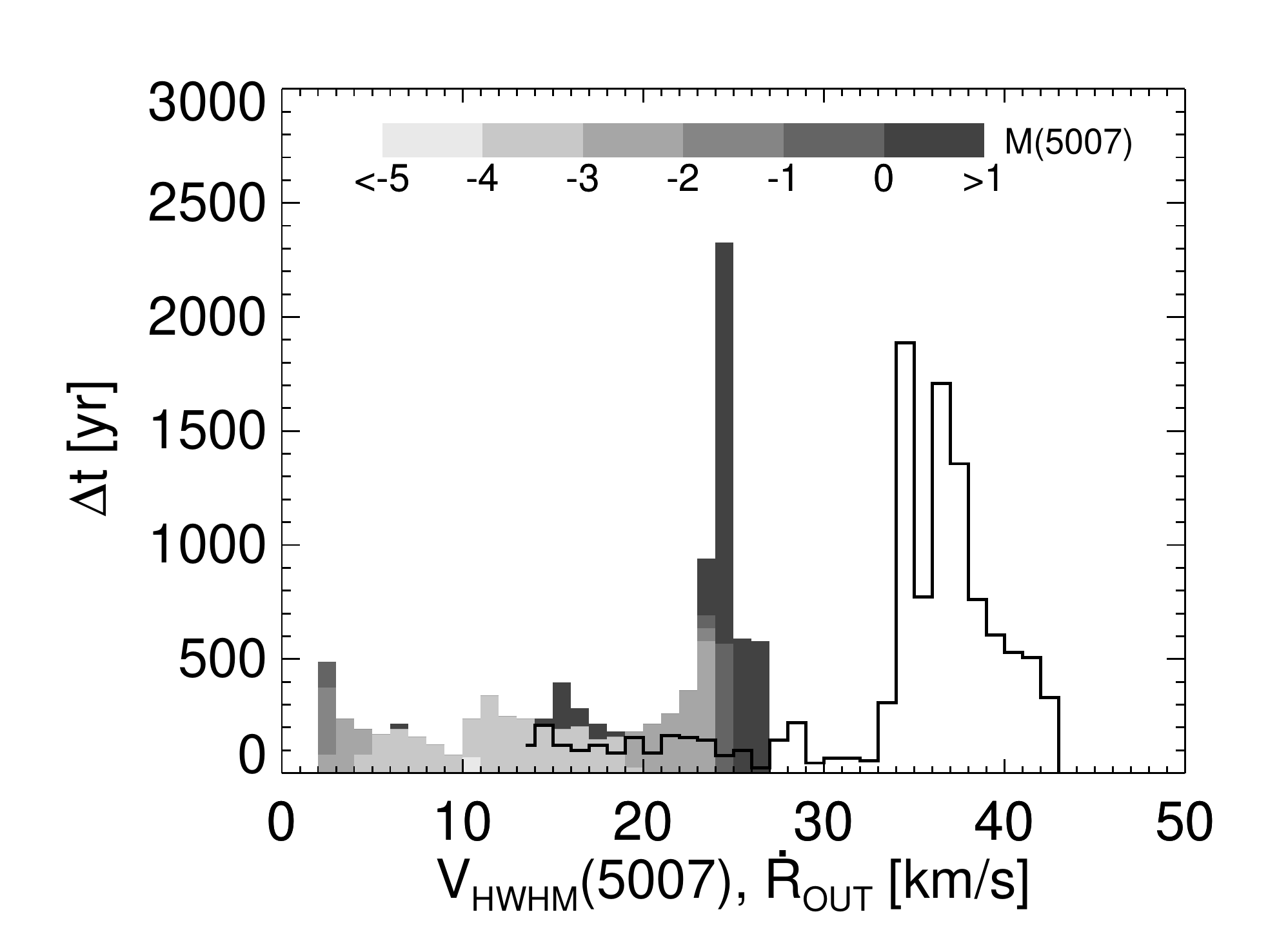}
\includegraphics[width=0.66\linewidth]{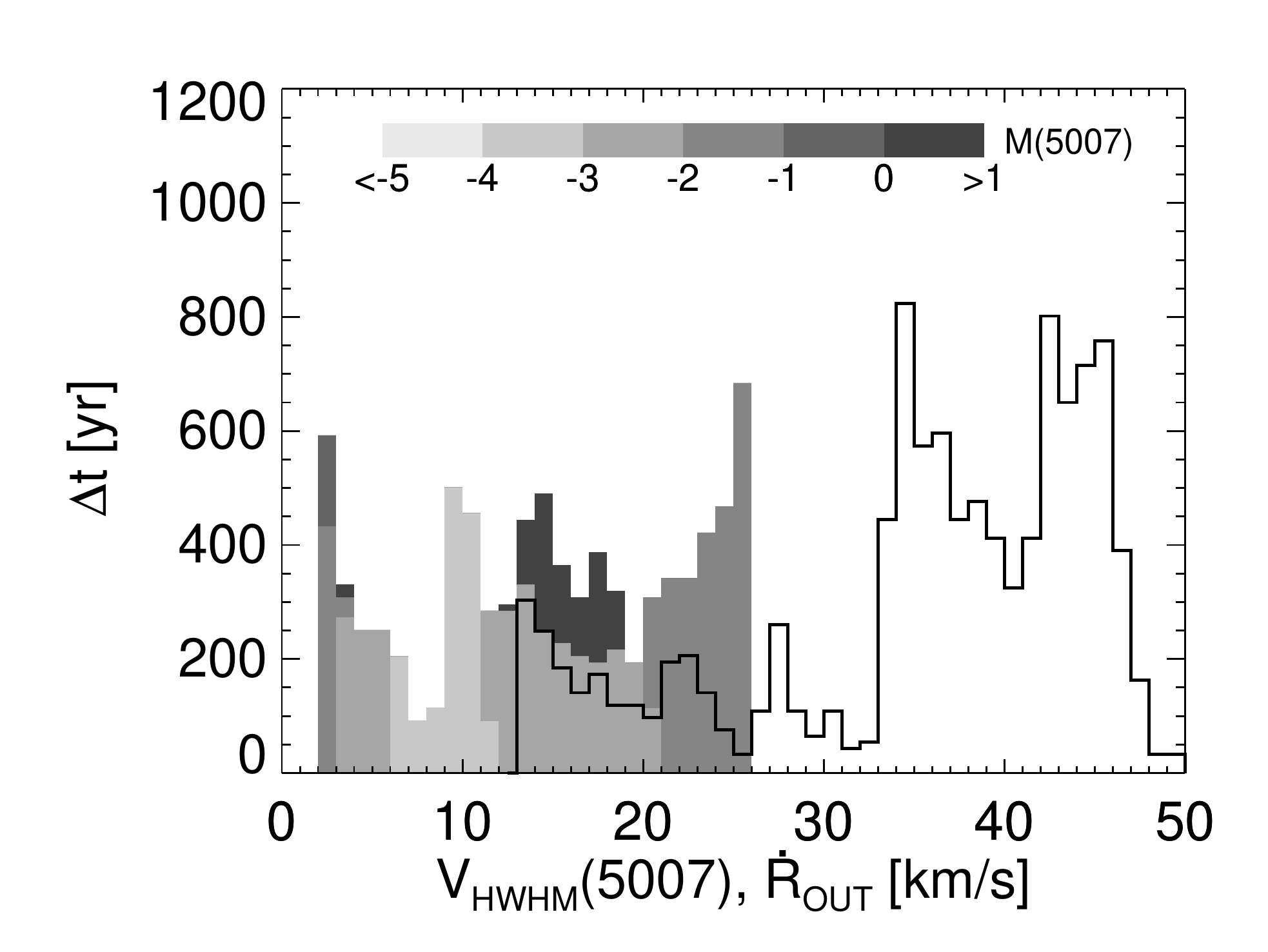}
\caption{Same as in Fig.~\ref{histo.0625} but for the sequences with
         0.605 \Msun\ (\emph{top}) and  0.595 \Msun\ (\emph{bottom}) already presented 
         in Fig. \ref{vel.richer}. 
	 The post-AGB age ranges considered are 12\,000 years for both sequences.
        }
\label{histo.hydro}
\end{figure}

\changed{The behaviour of these sequences is somewhat different:  
         At ${V_{\rm HWHM}(5007)\simeq 10\ldots 11}$ \kms\ the models become optically thin 
         and reach then also their maximum \oiii\ brightness.} 
   The HWHM velocity further increases steadily
   during the optically thin part of evolution until recombination begins.
   The total range of $V_{\rm HWHM}(5007)$ within the brightest 3 magnitudes
   is for both sequences ${\simeq\! 4}$ to ${\simeq\! 25}$ \kms.
  On average, the real
  expansion, as determined by $\dot{R}_{\rm out}$, is again considerably higher,
  viz. nearly a factor of two (Fig.~\ref{histo.hydro}).

  Since the \oiii\ brightness depends on the nebular metal (oxygen) content, it might be more
  useful to plot  $V_{\rm HWHM}(5007)$ against the H$\beta$\ brightness in order to
  get rid of metallicity effects in both the observation and theory.  We expect
  that in a  $V_{\rm HWHM}$--$M({\rm H\beta})$ plane the tracks of models which differ 
  only in their metal content will be very similar, at least during the early evolution.
  Figure \ref{vel.hbeta} confirms our conjecture: there is a rather tight correlation
  between  $M({\rm H\beta})$ and $V_{\rm HWHM}$, and at maximum H$\beta$\ brightness
  (${M({\rm H\beta})\simeq -2}$) the HWHM velocities are ${\simeq\!16\pm2}$ \kms\ for 
  5007 \AA\ and ${\simeq\!20\pm2}$ for H$\beta$.
  Larger differences between the tracks occur during later phases when recombination 
  determines the ionisation structure of our nebular models for ages above 3500 years
  (cf. Fig. \ref{0.625.vel}).

  The moderate variation of $V_{\rm HWHM}({\rm H}\beta)$ is easily explained: The hydrogen 
  line emerges from the entire ionised region, and its line width is controlled by the 
  fast moving gas behind the outer shock (cf. Fig. \ref{0.625}, top panels) together with 
  the (comparatively) high thermal broadening ($\approx\!10$ \kms).  In the beginning,
  thermal broadening dominates, but later expansion broadening becomes more important.
  Thermal broadening of oxygen lines is four times less, and the velocity
  evolution of, e.g., \oiii\ 5007 \AA\ can reasonably be traced by the line width.
  Very early, the O$^{2+}$ zone is restricted to the innermost nebular regions only, close 
  to the contact discontinuity where the gas is nearly stalling.  Hence, $V_{\rm HWHM}(5007)$
  is very low, much lower than the AGB wind velocity: only 4--5 \kms\ as compared to
  ${V_{\rm agb}= 15}$ \kms.  Later, $V_{\rm HWHM}(5007)$ steadily increases mainly because
  the O$^{2+}$ zone extends until it embraces the whole nebula and approaches finally 
  $V_{\rm HWHM}({\rm H}\beta)$.

\begin{figure}[t]
\vskip -4mm
\includegraphics[width=\linewidth]
                       {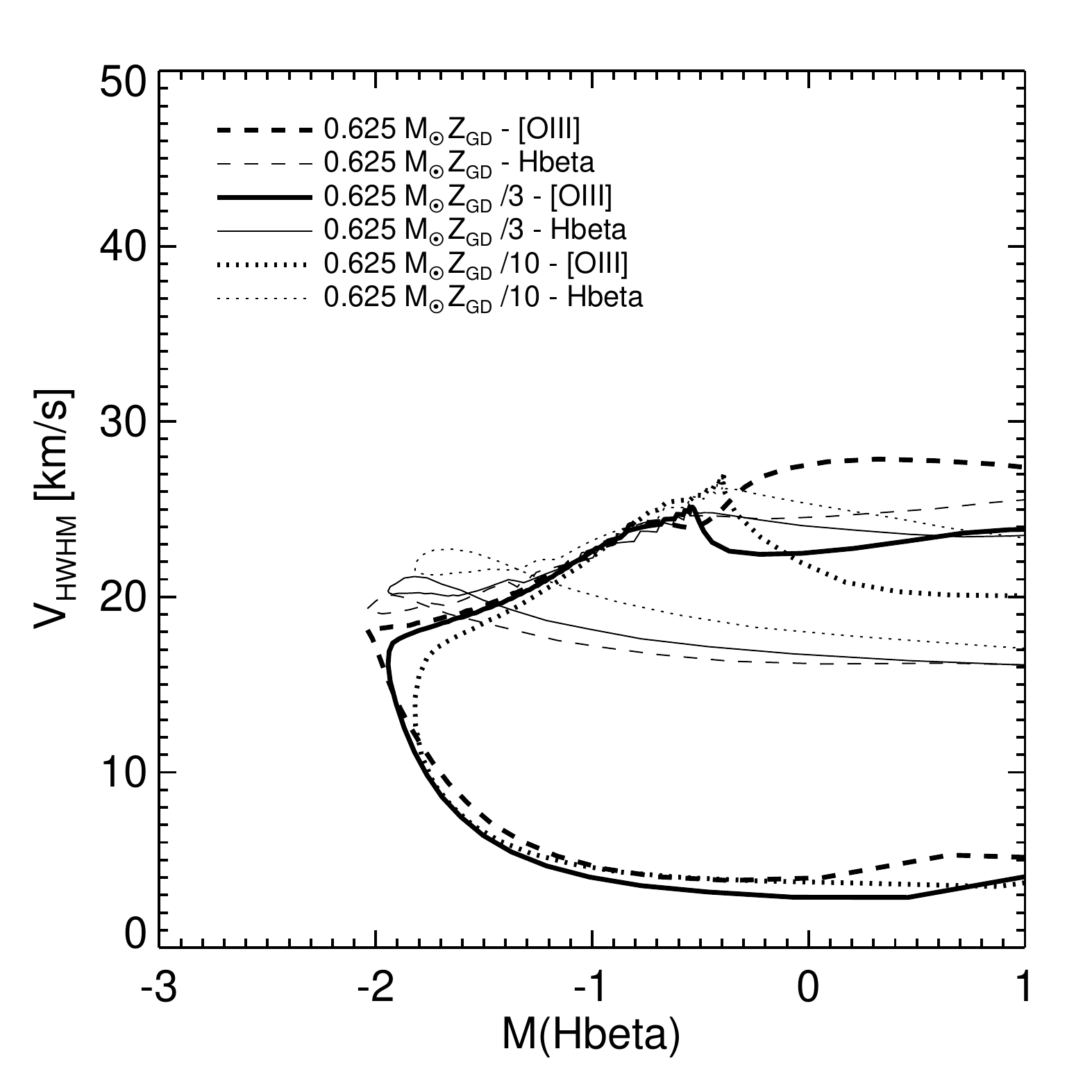}
\vskip-2mm
\caption{Run of $V_{\rm HWHM}(5007)$ and $V_{\rm HWHM}({\rm H}\beta)$  vs. $M({\rm H\beta})$ 
        for the different 0.625 \Msun\ sequences.  Evolution proceeds in all cases from faint 
        magnitudes and low velocities towards faint magnitudes and higher velocities.
        }
\label{vel.hbeta}
\end{figure}

   Our model simulations suggest that large differences between velocities measured  
   from the widths of hydrogen lines and lines of heavier ions are not only the result 
   of different thermal broadening. Rather, these differences also point to the existence of 
   a substantial positive velocity gradient within all PNe.  Such velocity 
   gradients, as they are predicted by hydrodynamical simulations, have indeed been 
   found for a few galactic PNe by means of a careful
   analysis of high-resolution slit spectrograms \citep[cf., e.g.,][]{gesetal.96, gesetal.98}.

\begin{figure}[t]
\includegraphics[bb= 0.8cm 0.4cm 15.4cm 15.0cm, width=0.98\linewidth, clip]
                {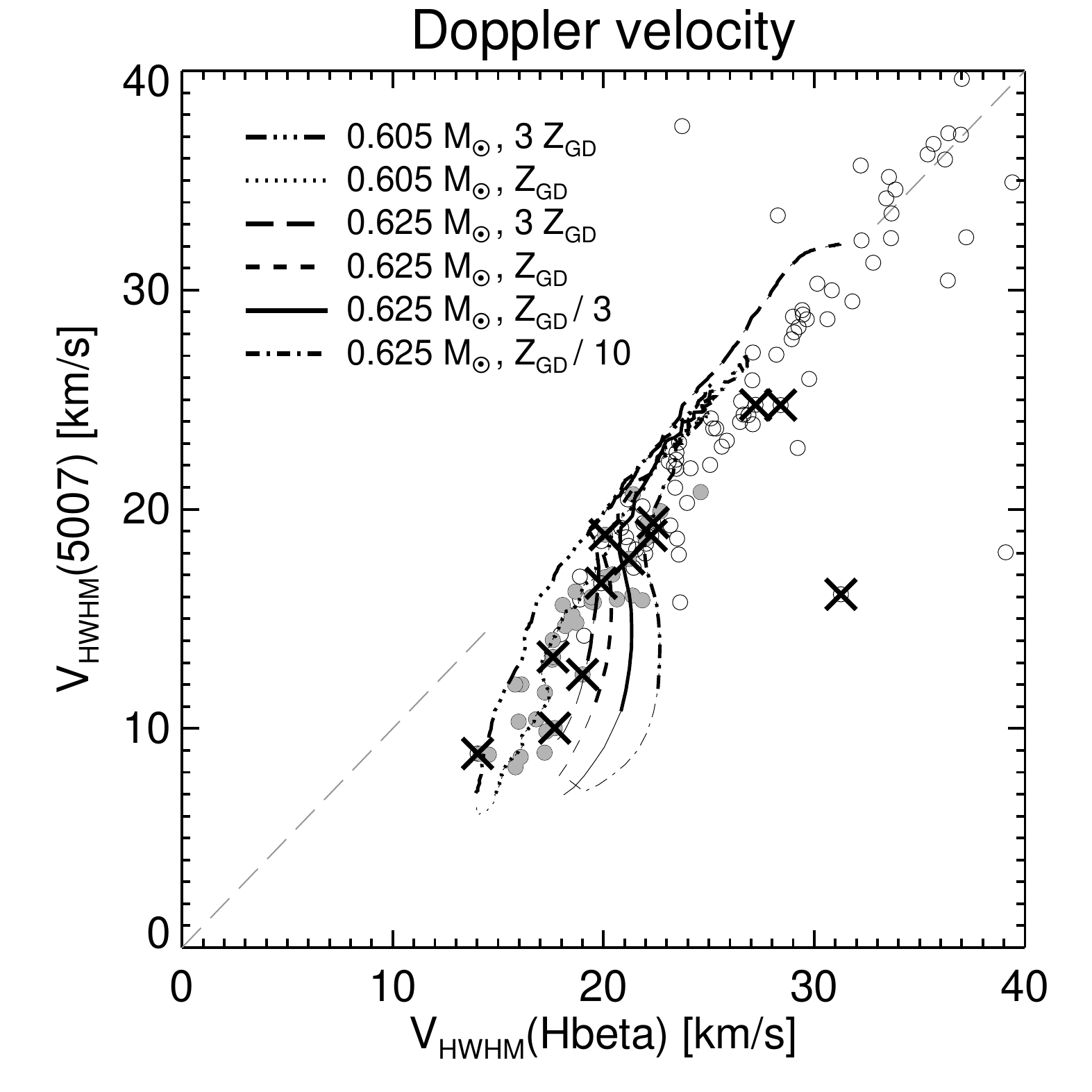}
\caption{$V_{\rm HWHM}(5007)$ vs. $V_{\rm HWHM}({\rm H}\beta)$ for four 0.625 \Msun\ and 
         two 0.605 \Msun\ sequences with different metallicities as indicated in the legend.
         Evolution proceeds always from low towards higher velocities.  Plotted are only
         the evolutionary phases with ${\teff \le T_{\rm max}}$. 
         The models are compared with recent HWHM velocity  measurements of Milky Way bulge 
         objects, either with apparently unsplit line profiles (filled symbols) or with
         split profiles (open symbols), provided to us by M. Richer (see also 
         \citealt{richeretal.08, richeretal.09, richeretal.10}), and assuming that the 
\changed{halfwidth velocities of \hb\ and \ha\ are the same. Only objects with 5007 \AA\
         line strengths greater than \hb\ are considered.}
\changed{Accordingly, thin line sections at low velocities indicate models for which 
         ${L(5007)\le L({\rm H}\beta)}$.}   The model profiles are
         broadened by a Gaussian with FWHM of 11~\kms\ in order to comply with the 
         spectral resolution of the observations. Crossed symbols mark objects which are known 
         to posses [WC]-type central stars \citep[see][]{goretal.04}.                
        }
\label{vel.o3.hbeta}
\end{figure}

  Recently, \citet{richeretal.09} presented \changed{halfwidth velocities}
  of Galactic bulge PNe based on both H$\alpha$ and \oiii\ 5007 \AA.
  These measurements are
  shown in Fig. \ref{vel.o3.hbeta}, together with selected model sequences relevant for bright
  bulge objects of the Milky Way.  One sees that the HWHM velocities predicted by our models 
  are in excellent agreement with measurements of the slowly expanding objects with unsplit 
  lines, which are obviously also the younger ones of this sample (filled symbols).  
  The agreement becomes poorer for a majority of the larger, spatially resolved PNe 
  with split lines (open symbols).   

  However, for a full appreciation of Fig. \ref{vel.o3.hbeta} we have to consider that
  all Galactic bulge objects are to some extent spatially resolved.  The typical 
  coverage of the slit is between 30\% and 50\% for the slowly expanding objects 
  (filled circles in Fig.~\ref{vel.o3.hbeta}) and usually less for the faster expanding 
  PNe with doubly peaked profiles.  Going from low to high spatial resolution increases
  the width of the line profiles of both the \oiii\  and H$\beta$ line by very similar
  amounts: In the low velocity regime we estimated from model calculations typical values 
  of 3 \kms\ for 5007 \AA\ and 4 \kms\ for H$\beta$.  For doubly peaked profiles this
  shift is larger, and this effect may explain the slight offset to the right that is seen 
  at higher velocities between the observations and our models (open symbols in 
  Fig. \ref{vel.o3.hbeta}).  

  Despite of the small inconsistency between the velocities derived from our (unresolved) 
  theoretical line profiles and those measured by \citet{richeretal.09, richeretal.10}, 
  the interpretation
  of Fig. \ref{vel.o3.hbeta} is not really affected: We see how the ionisation stratification
  develops with time (or velocity) in nebular structures with positive velocity gradients
  in the same way as appropriate hydrodynamical models predict.
\changed{Note also that the two metal-poor sequences ($Z_{\rm GD}/3,\ Z_{\rm GD}/10$)
         do not predict the correct relation between hydrogen and oxygen line widths and
         are too faint in 5007 \AA\ at low velocities.}

  Spectroscopy done by \citet{goretal.04} revealed that the sample of bulge PNe used
  by \citet{richeretal.09, richeretal.10} contains several  
  objects with WC central stars, which are marked in Fig.~\ref{vel.o3.hbeta} by crossed symbols. 
  Remarkably, their velocity behaviour appears to be the same as that of their `normal' 
  counterparts (apart from the outlier). Of course, the number of WC-type objects in 
  this Richer sample is too small in order to allow a more definitive statement.

\begin{figure}[t]
\includegraphics[bb= 0.5cm 0.4cm 15.4cm 15.2cm, width=0.99\linewidth, clip]
                {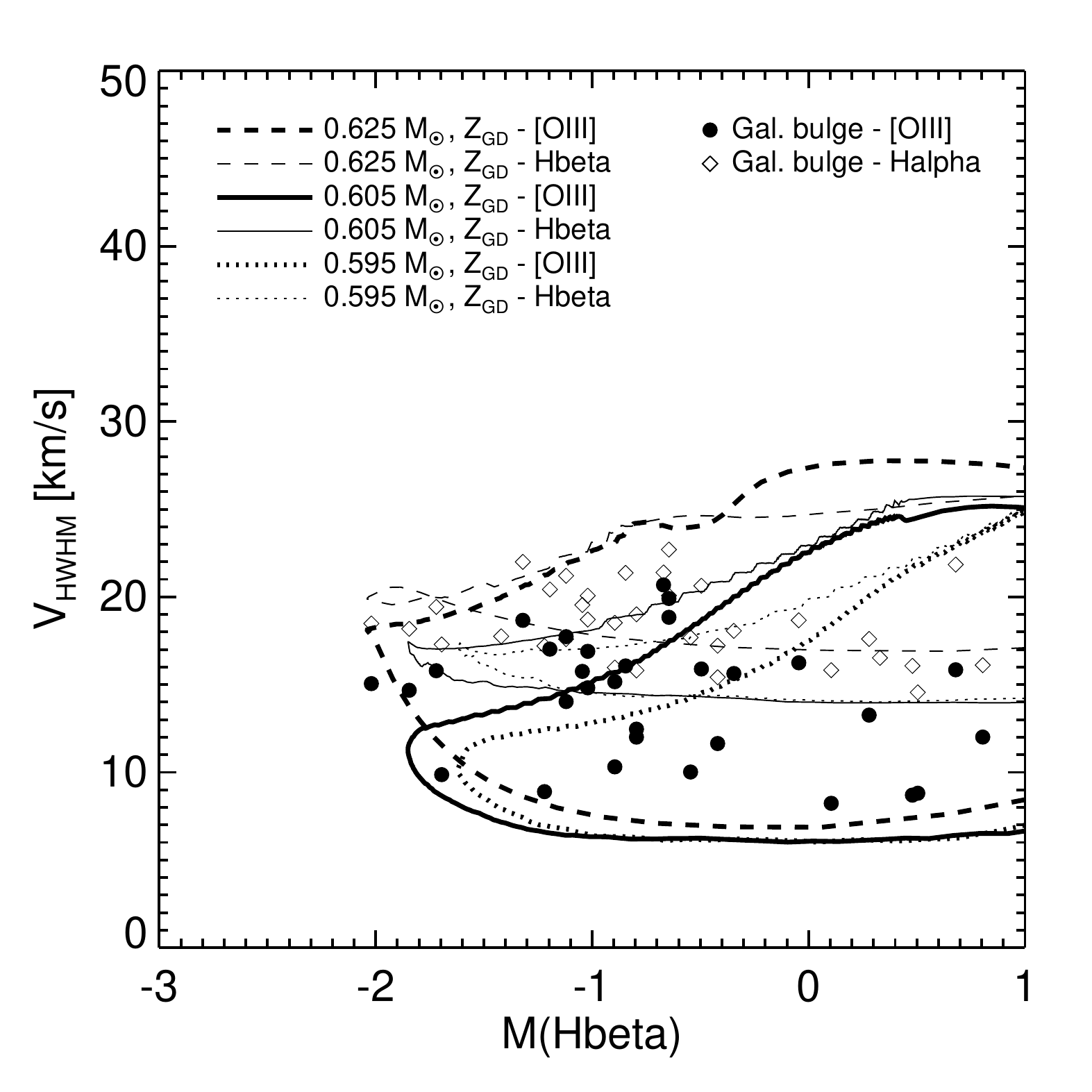}
\caption{\label{richer.bulge}
\changed{Similar as in Fig. \ref{vel.hbeta} but now for $V_{\rm HWHM}(5007)$ and 
        $V_{\rm HWHM}({\rm H}\beta)$  vs. $M({\rm H\beta})$ predicted
        from model sequences with different central stars and the same metallicity.  
        Evolution proceeds again from faint  magnitudes and low velocities towards faint
        magnitudes and higher velocities. These theoretical predictions are again
        compared with the Galactic bulge sample of \citet{richeretal.09, richeretal.10}
        already used in Fig. \ref{vel.o3.hbeta}. Plotted are only objects with unsplit
        lines, and thus the region containing higher velocities at low brightness is
        devoid of objects. Each object (except for a few cases) is plotted twice, with its
        $V_{\rm HWHM}(5007)$ (filled symbols) and 
        $V_{\rm HWHM}({\rm H}\beta)\ (= V_{\rm HWHM}({\rm H}\alpha))$ (open symbols).
        We assumed a distance of 7.5 kpc for all objects.  The theoretical profiles are
        broadened by a Gaussian with FWHM of 11 \kms\ in order to resemble the observations.} }
\end{figure}

\begin{figure*}[t]
\vskip-3mm
\sidecaption
\includegraphics[width=0.37\textwidth]{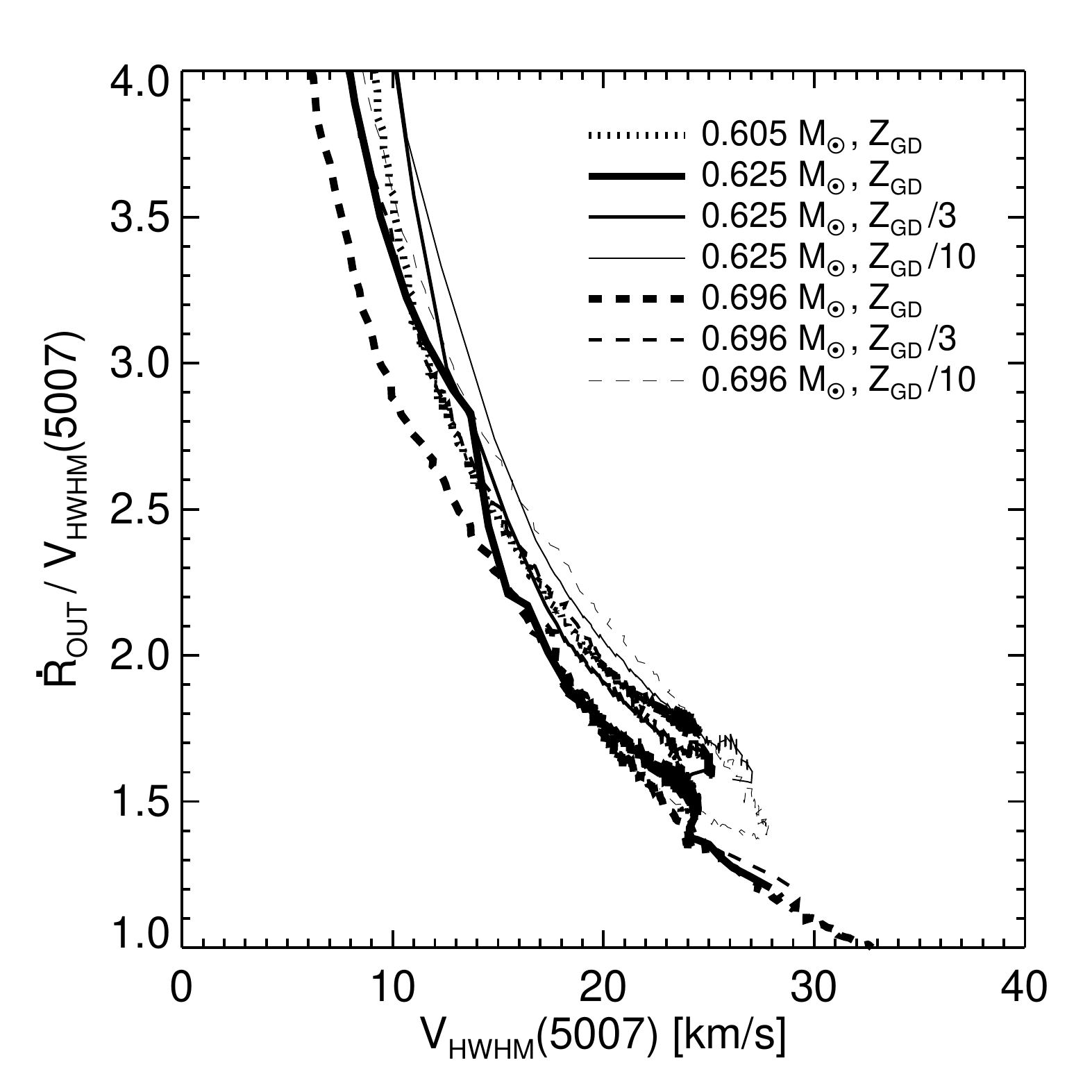}
\includegraphics[width=0.37\textwidth]{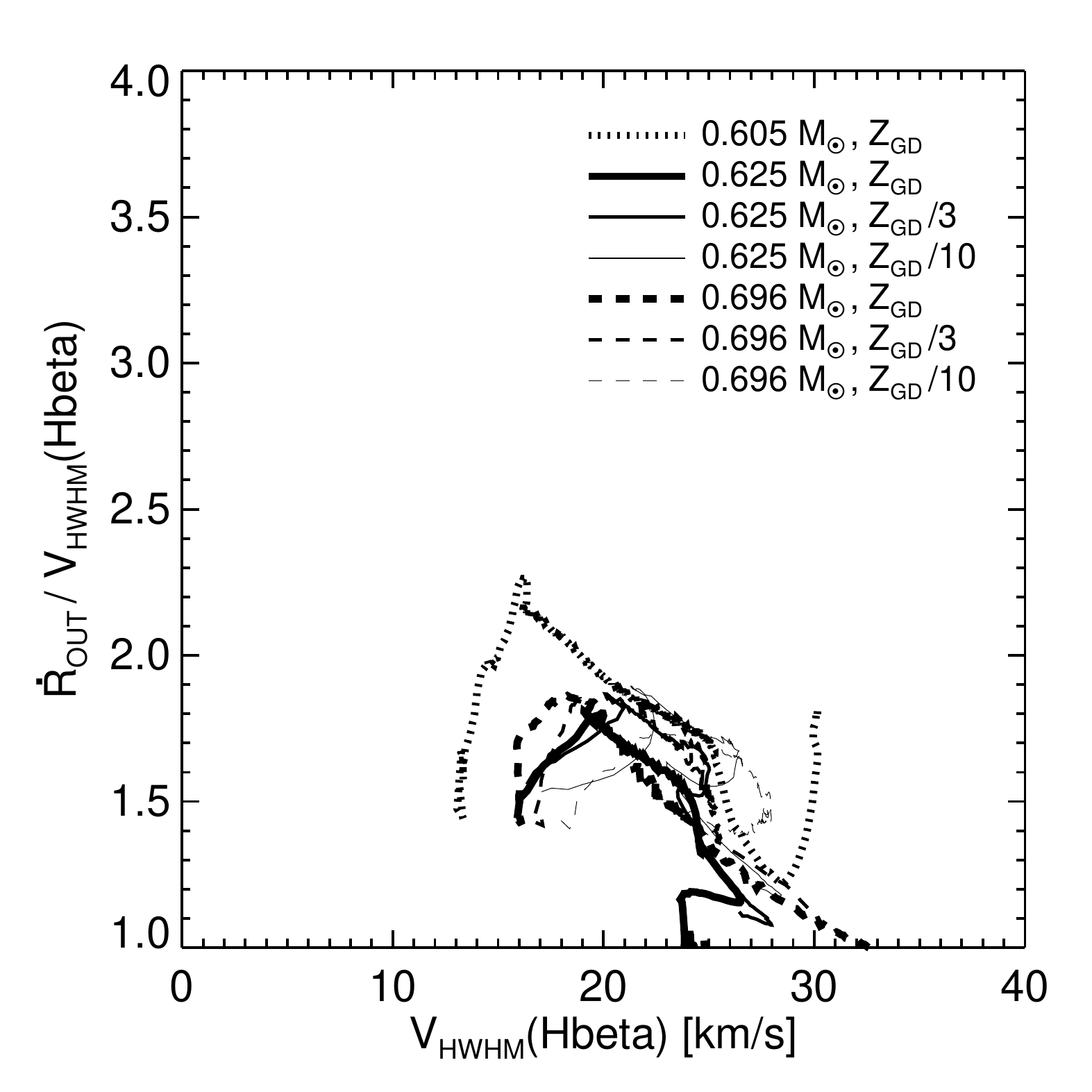}
\caption{\changed{$\dot{R}_{\rm out}/V_{\rm HWHM}$ vs. $V_{\rm HWHM}$ for \oiii\ 5007 \AA\ 
        (\emph{left}) and H$\beta$ (\emph{right}) and for the different types of hydrodynamical 
         sequences. The evolution proceeds from low to high line widths, or $V_{\rm HWHM}$.}
        }
\label{factor2}
\end{figure*}

\changed{Richer's linewidths measurements of PNe of the Galactic bulge are finally plotted
  in Fig. \ref{richer.bulge} over the \hb\ absolute brightnesses, assuming a distance 
  to the bulge of 7.5 kpc and selecting only the younger objects with unsplit lines. 
  There is again close agreement with the predictions of our 
  model sequences.  Central stars of about 0.6 \Msun\ appear to be sufficient to
  explain the observed maximum \hb\ brightness achieved by the PN population of the 
  Galactic bulge.}
  
  We can summarise the results of this section as follows:  
\begin{itemize}
\item
  Measuring the expansion of bright extragalactic PNe
  by means of the HWHM method, the only possible one for distant objects
  which cannot be spatially resolved, is not very fruitful for more detailed studies.
  Although there exists an evolution of $V_{\rm HWHM}$ with time in, e.g. \oiii\ lines, 
  the dependence on metallicity is extremely small and can virtually be neglected.   
  The time evolution of $V_{\rm HWHM}$ does by {no means} reflect the real expansion
  (cf. also Fig. \ref{0.625.vel}).
  Typical HWHM velocities at maximum nebular brightness are $\la$20 \kms, for all
  metallicities considered here.
  Lower speeds seen during the early PN evolution are 
  entirely due to ionisation stratification in conjunction with a positive radial velocity 
  gradient.  However, the \emph{real expansion speed depends on metallicity
  and can be larger by factors of two or more} (Figs.~\ref{0.625.vel}, \ref{histo.0625}, 
  and \ref{histo.0696}; see also the next section).
\item
  Based on our hydrodynamical simulations reported here we state further that the  
  very narrow range of \changed{halfwidth} velocities ($V_{\rm HWHM}$) measured for 
  the brightest PNe of galaxies with quite different properties as reported by 
  \citet[][Fig. 1 therein]{richer.07} 
  leads to only one conclusion:  \emph{Stellar evolution at the tip of the AGB occurs  
  independently of the properties of the underlying stellar population}. 
  The close agreement of observations with our model predictions
  implies that the initial condition chosen by us at the tip of the AGB for, e.g., a 
  0.625 \Msun\ stellar remnant, viz. ${\dot{M}_{\rm agb} = 10^{-4}}$ \Mdot\
  with ${V_{\rm agb} = 15}$ \kms, are very close to reality.  
\item
  The  narrow range of observed \changed{halfwidth} velocities indicates further that everywhere the 
  same main process for formation, shaping, and evolution of PNe is at work: \emph{heating 
  of sufficiently dense circumstellar wind envelopes by photoionisation, setting up an  
  expanding shock wave which defines the outer edge of the PN}.
  Wind interaction and/or magnetic fields, although important in single cases and/or 
  during certain phases, is obviously \emph{not} the main shaping agent, contrary to 
  common wisdom.
\end{itemize}

\subsection{The real expansion}
\label{real}

  Our simulations demonstrate clearly that it is impossible to derive the real expansion 
  properties of PNe from a simple measurement of the halfwidths of spatially unresolved 
  line profiles. 
\changed{Because of the ionisation and/or density structures together 
         with projection effects the contribution of the fastest moving matter  
         to (integrated) line profiles is strongly reduced.}

  However, we can use our hydrodynamic models to quantify the relation between the 
  real expansion, as defined by the propagation of the outer shock, and the
  spectroscopically measured expansion deduced from the HWHM of an emission line
  profile, which is in practice that of \oiii\ 5007 \AA, or 
  also that of H$\alpha$ or H$\beta$.
  Figure~\ref{factor2} illustrates how the ratio between shock speed and 
\changed{$V_{\rm HWHM}$ of \oiii\ 5007 \AA\ and H$\beta$ (or H$\alpha$) changes during 
         the nebular evolution, as based on our simulations.}
  We see that \emph{all} types of models behave in the same way: The ratio between both
  velocities \changed{generally} decreases with time and comes close to
  unity during the end of evolution.

  The behaviour of $\dot{R}_{\rm out}$ and $V_{\rm HWHM}$ seen in Fig.~\ref{factor2}
  can be explained as follows:.  First of all, we recall that $\dot{R}_{\rm out}$
  is determined by the density gradient of the (fully ionised) ambient medium \emph{and} 
  the sound speed
  of the nebular gas behind the shock front \citepalias[cf.][]{schoenetal.05a}. 
  $V_{\rm HWHM}$,
  however, contains all the globally averaged information of gas motions \emph{and} the
  density distribution inside the nebula measurable via the Doppler effect.  If the
  nebula has a stratified ionisation structure, $V_{\rm HWHM}$ depends also on the
  ion used.  

  For instance, during the early ionising phase of the PN evolution the 
  ratio between $\dot{R}_{\rm out}$ and $V_{\rm HWHM}$
  is quite large (greater than 3 for ${V_{\rm HWHM}(5007)\la 10}$ \kms), since O$^{2+}$ is
  restricted only to the \changed{strongly decelerated} inner regions 
  (cf. Fig. \ref{HWHM}, right panel). 
\changed{The situation is different for hydrogen lines (Fig. \ref{factor2}, right panel):
         Since hydrogen ionisation starts while the matter is still moving 
         with the AGB-wind velocity, $V_{\rm HWHM}$ starts with similar values of, 
         e.g., 10--15 \kms. The ratio of $\dot{R}_{\rm out}$ and 
         $V_{\rm HWHM}$ does not exceed a factor of about two.}

  With progressing ionisation, $V_{\rm HWHM}(5007)$ catches up later in the evolution 
  because progressively
  faster moving matter from the outer regions behind the shock contributes to the
  integrated line profile.  During the recombination phase it may happen that the
  shock is being decelerated so much that $\dot{R}_{\rm out}$ may fall even below
  $V_{\rm HWHM}$, i.e. $\dot{R}_{\rm out}$ becomes lower than the flow speed of 
  the nebular gas (cf. Fig.~\ref{factor2}, left, the 0.696 \Msun, $Z_{\rm GD}$ models
  with $V_{\rm HWHM}(5007)> 30$ \kms). 

  We see \changed{especially in the left panel of} Fig.~\ref{factor2} also  that for 
  fixed $V_{\rm HWHM}$ and for the same configurations (central star mass and initial
  envelope model) the correction for
  $\dot{R}_{\rm out}$ increases somewhat if the metallicity decreases.
  The shock propagation speed is controlled by the electron temperature and makes
  metal-poor and hotter nebula more dilute in their outer regions.  The inner
  denser regions, however, are less accelerated because of the reduced wind power.
  Altogether, the width of the integrated line profile does not change to the same
  extent (full ionisation provided, cf. Fig.~\ref{0.625.vel}).
\changed{-- Such a metallicity influence on the
         line width is not seen for hydrogen lines (Fig.~\ref{factor2}, right),
         most likely because of the generally larger thermal contribution to the
         line broadening.}

  The interesting phase from an observational point of view occurs when nebulae
  evolve through their brightest stage (in \oiii).  We have seen above that this
  happens for $V_{\rm HWHM}(5007) \simeq 15\ldots 25$ \kms, which implies correction
  factors of 1.5\ldots 2.5 to get the real expansion speeds.  A mean correction factor 
  of two appears therefore to be a reasonable choice for the brightest nebulae of a sample.
\changed{Figure \ref{factor2} (right panel) suggests that two is also a quite useful 
         mean correction factor for hydrogen lines.}

\subsection{The case of the Magellanic Cloud PNe}
\label{MC}

   The Magellanic Clouds (MC) contain a
   rather large population of planetary nebulae which can also be studied down to
   fainter magnitudes than it is possible in most other extragalactic systems.
   Thus, the MC sample of PNe is well suited for studying aspects of PN evolution in
   an extragalactic system.

   \citet{dop.85, dop.88} used in their seminal work of the Magellanic Cloud
   PNe a somewhat different definition \changed{to measure the expansion} 
   of a spherical gaseous shell:  
   ${V_{\rm exp}=0.911\,V_{\rm FWHM}}$, where $V_{\rm FWHM}$ is twice the
   HWHM velocity.  Assuming a Gaussian shape for the intrinsic profiles, 
   this is equivalent of $V_{\rm exp}$ being half of the full width at 10\% maximum 
   line intensity, $V_{\rm HW\%10M}$, or
   ${V_{\rm exp}=1.82\,V_{\rm HWHM}}$, with $V_{\rm HWHM}$ being the 
   velocity deduced from the HWHM of the line profile.  Thus, Dopita et al.'s
   \changed{``expansion''} velocities, $V_{\rm exp}$,
   are higher by nearly a factor of two as compared to the more widespread
   use of $V_{\rm HWHM}$.   The advantage of this definition is, however, that
   the \changed{such measured velocity is much closer to the maximum gas velocity
   behind the outer shock (see the discussion in Sect. \ref{exp.vel} and 
   Table \ref{tab.625}). }

\begin{figure}[t]
\includegraphics[width= \linewidth]{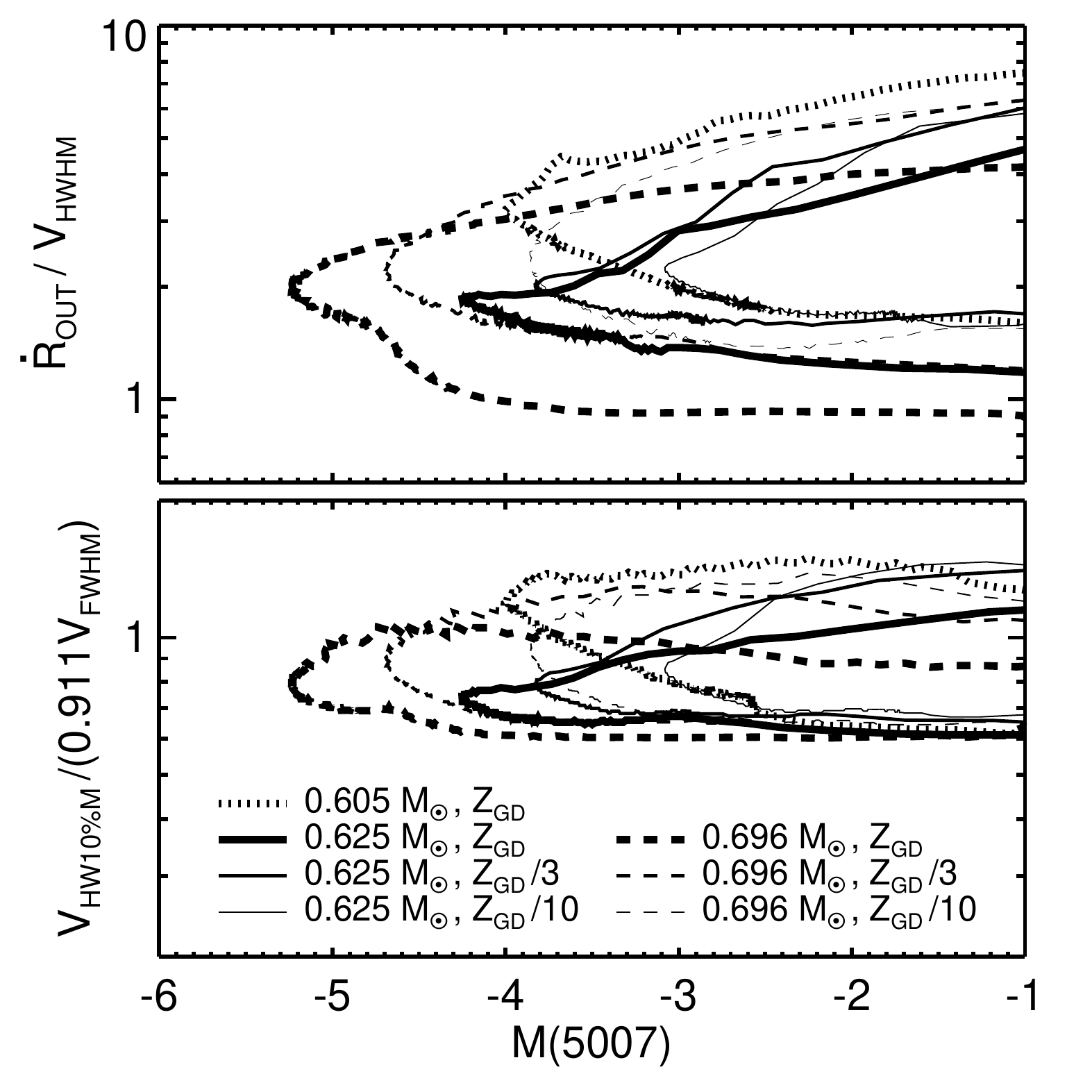}
\vskip-2mm
\caption{Evolution of $\dot{R}_{\rm out}/V_{\rm HWHM}$ (\emph{top}) and
         $V_{\rm HW10\%M}/0.911V_{\rm FWHM}$ (\emph{bottom})
         vs. absolute 5007 \AA\ brightness, $M(5007)$,
         for different nebular model sequences.
	 In both panels, evolution proceeds from faint magnitudes at the
	 upper right towards faint magnitudes at the lower right.
        }
\label{Gaussian}
\end{figure}

   \citet{dop.85, dop.88} did not measure $V_{\rm HW\%10M}$ 
   directly, but used ${V_{\rm exp}=1.82\,V_{\rm HWHM}}$ instead, thereby implying
   implicitly a Gaussian shape for the intrinsic line profile.  In reality, however,
   the spatially integrated line profiles are by no means
   Gauss functions (cf. Fig.~\ref{0.625}, bottom panels).  A moderate broadening
   by finite spectral resolution does not change this.  Additionally, the shape
   of the line profile changes with time because the density distribution
   and the velocity field develop as the model traverses the Hertzsprung-Russell
   diagram.

   The whole situation is illustrated in Fig.~\ref{Gaussian} where the relations between
   $\dot{R}_{\rm out}$ and $V_{\rm HWHM}$ (top panel) and between $V_{\rm HW10\%M}$ and
   $0.911V_{\rm FWHM}$ (bottom panel) are displayed as a function of the nebular brightness
   for the different hydrodynamical PN sequences used here.  $V_{\rm HW10\%M}$
   expresses the velocity derived from the 10\% level of the line profile.
   For a Gaussian we expect ${V_{\rm HW10\%M}/0.911V_{\rm FWHM} = 1}$ (see above).
   From Fig.~\ref{Gaussian} (bottom panel) we see that the line profiles
   deviate considerably from a Gaussian shape, mainly during the late evolution
   when the models become \changed{more extended. }

   The top panel of Fig.~\ref{Gaussian} is equivalent to Fig.~\ref{factor2} but
   shows more clearly how the correction factor, to be applied to $V_{\rm HWHM}$,
   depends on the nebular brightness.  One sees again that a factor of two is quite 
   reasonable for most models while evolving through their brightest stage.  
   One sees also more clearly that only for the metal-rich model
   sequence around 0.696~\Msun\ this factor falls below unity along the fading
   branch of evolution, as seen already above (Sect. \ref{real}).

\begin{figure*}[t]
\includegraphics*[bb= 0.5cm 9.9cm 28cm 20cm, width=\textwidth]
                 {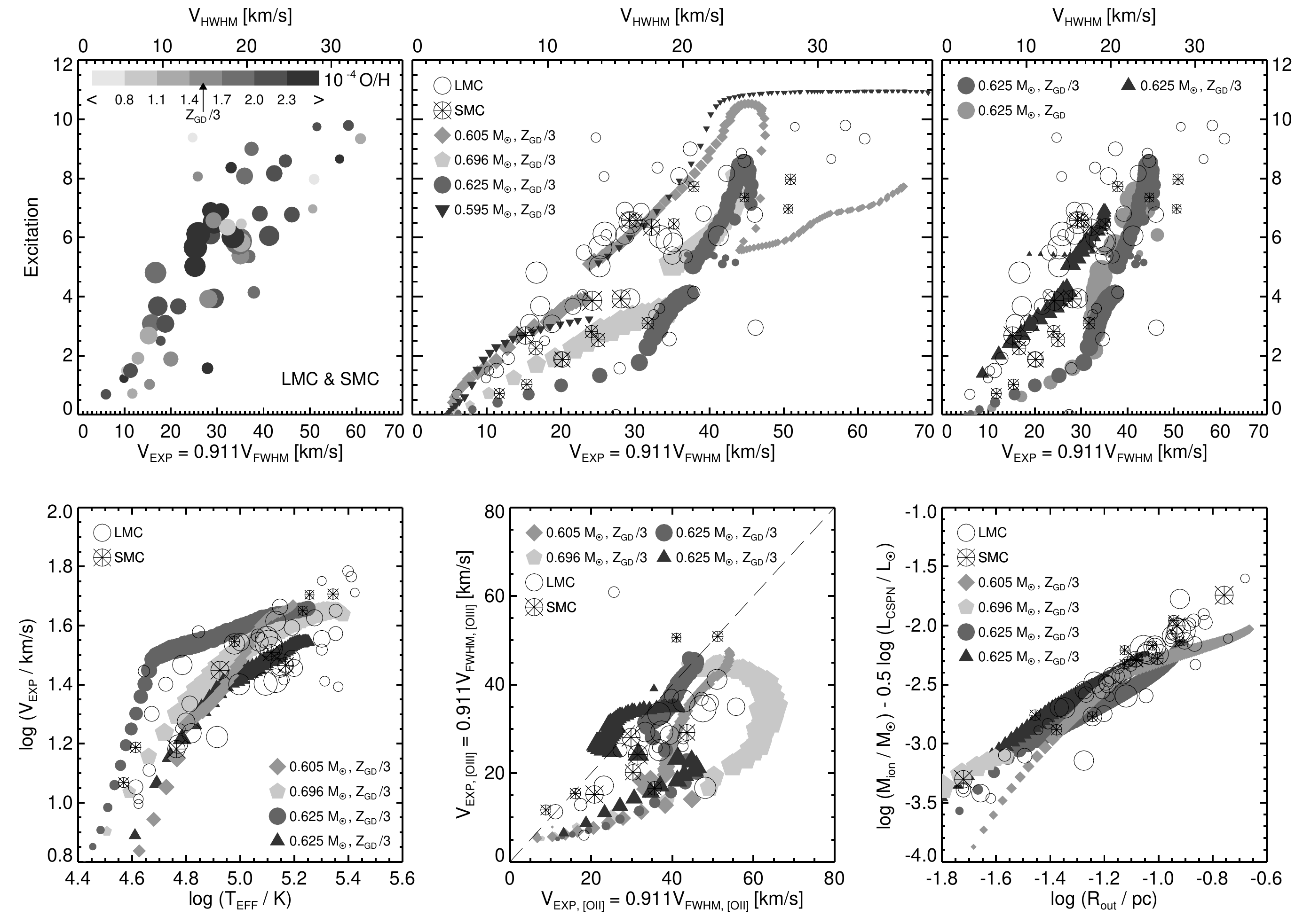}
\caption{\label{dopita.1}  
\changed{Nebular excitations, $E$, vs. 5007 \AA\ halfwidth velocities, 
         $V_{\rm exp}=0.911V_{\rm FWHM}$, for 
         Magellanic Clouds PNe \citep{dop.85, dop.88} and our hydrodynamical simulations.
         Observed objects and models are broken down into the same $M(5007)$
	 brightness classes, as indicated by the symbol sizes: the larger the symbol,
	 the brighter the object or model, respectively. The approximate brightness range 
         covers 4 magnitudes. The top abscissa is labelled in 
         terms of $V_{\rm HWHM}$ for easier comparison with previous figures.
         \emph{Left panel}: LMC and SMC PNe additionally broken down into classes of 
                            different oxygen 
                            abundances \citep[from][]{dop.91, dop.91b}, visualised 
                            by various grey shades. The darkest grey corresponds to 
                            ${\ge \!Z_{\rm GD}/2}$, the lightest grey to
                            ${\le \!Z_{\rm GD}/5}$.  The vertical arrow indicates the
                            oxygen abundance corresponding to $Z_{\rm GD}/3$.
         \emph{Middle panel}: comparison of observations with the model sequences as indicated.
	                      The gaps seen in the model sequences are artefacts due to the 
                              definition of $E$ combined with the lower oxygen abundance
                             of the Clouds (Eqs.~\ref{exc}, see text there). Note that the 
                             spacings of the models within a sequence does by no means 
                             reflect the speed of evolution.
         \emph{Right panel}: A 0.625 \Msun\ model sequence (${\alpha=2}$, $Z = Z_{\rm GD}/3$) 
                             with different initial
                             parameters, ${\dot{M}_{\rm agb}= 0.5\times 10^{-4}}$ \Mdot\ and 
                             ${V_{\rm agb}=7.5}$ \kms\ (filled triangles), compared with the
                             0.625 \Msun\ $Z_{\rm GD}/3$ sequence from the middle panel and 
                             a metal richer (${Z=Z_{\rm GD}}$) 0.625 \Msun\ sequence (see legend).}  
        }
\end{figure*}

\subsubsection{Nebular excitation and expansion}

   From the rather large samples of LMC and SMC PNe
   \citet{dop.85} and \citet{dop.88} found a positive correlation of
   $V_{\rm exp}$ with the nebular excitation parameter $E$, defined as
\begin{displaymath}
 E = 0.45\,\left[X(5007)/X({\rm H}\beta)\right], \ \ \      \hspace{1.8cm} 0.0<E<5.0\,,
\end{displaymath}
\begin{equation}   \label{exc}
 E = 5.54\,\left[X(4686)/X({\rm H}\beta) + 0.78\right], \ \hspace{1.0cm} 5.0\le E < 12.0\,,
\end{equation}
   where $X$ stands for line luminosities or fluxes, respectively \citep{dop.92}.
   This correlation must be interpreted as the evolution of the nebular expansion
   speed with time or stellar effective temperature.

\changed{Equations (\ref{exc}) constitute a purely empirical definition of the nebular
         excitation state in close resemblance to the original \citet{morgan.84}
         classification scheme \citep[cf.][]{dop.90}. Although the low excitation 
         branch, ${E<5}$, depends directly on the oxygen  abundance, 
         Eqs. (\ref{exc}) are quite useful as a proxy for the nebular evolution. }
   However, $E$ is, like the brightness, neither a unique function
   of time nor of stellar temperature 
   \citepalias[see, e.g., Fig.~15 in][]{schoenetal.07}. 

   To perform a new
   interpretation by means of our hydrodynamic sequences, we proceeded as follows:
   Firstly, we computed the excitation parameters $E$ of LMC and SMC PNe from the
   line fluxes available in the literature by means of Eq. (\ref{exc})
   and used $V_{\rm exp}$ as determined by
   \citet{dop.85} and \citet{dop.88}.  We discarded objects whose line
   profiles could only be fitted by a double Gaussian since they have most likely
   an extreme bipolar shape not suitable for comparison with spherical models.
   We did then the same with our nebular models, 
\changed{i.e we derived $E$ from the model's line strengths and determined $V_{\rm exp}$
         from the line profiles according to the \citeauthor{dop.85} prescription given 
         above and applying an appropriate broadening (Gaussian with a FWHM of 10 \kms)
         to simulate instrumental broadening.}

\changed{The result is shown in Fig.~\ref{dopita.1} which provides not only velocity
         and brightness information of \citet{dop.85, dop.88}, but also oxygen abundances
         determined by photoionisation modelling of \citet{dop.91, dop.91b}. 
         The left panel of Fig.~\ref{dopita.1} displays the Magellanic Clouds objects,
         broken down into classes of different $M(5007)$ brightnesses and oxygen abundances
         (relative to hydrogen), as indicated by the symbol sizes and their grey values.         
         This panel is a new}
         version of Fig.~4 in \citet{dop.88} with more accurate excitation values.

   There is a broad positive correlation between excitation class and expansion velocity 
   of the Magellanic Clouds objects
   in the sense that there are no low velocities for high excitation and no
   high velocities for very low excitation.  
\changed{The distribution is very narrow at low velocities (${V_{\rm exp}\la 15}$ \kms),
         but becomes rather wide at higher velocities or excitations, independently of the
         oxygen abundance.  One notices, however, that the brightest objects have 
         preferentially also the highest amount of oxygen.} 

\changed{The left panel of Fig.~\ref{dopita.1} shows again what is already evident from 
         our Fig. \ref{vel.richer} and Fig. 4 in \citet{dop.92}: 
   The most luminous objects, which accumulate around ${E\simeq6}$ (big symbols in 
   Fig.~\ref{dopita.1}), have rather modest expansion rates (in terms of $V_{\rm HWHM}$,
   top abscissa) between 15 and 20 \kms.}
   The fainter objects have, in general, either low (young objects) or high
   expansion rates (older objects).   Obviously PNe in the Magellanic Clouds 
\changed{show the same expansion behaviour as} 
   PNe in the Galactic bulge, the Local Group, and the Virgo cluster.

\changed{The middle panel of Fig.~\ref{dopita.1} compares the observations with 
         the predictions of our hydrodynamical simulations in terms of excitation, 
         brightness, and halfwidth velocities.}
   We used the slightly metal-deficient models with ${Z=Z_{\rm GD}/3}$ in order to
   take the metallicities of both Clouds roughly into account.  Note that
   the excitation $E$, if computed according to Eqs. (\ref{exc}), does not reach
   the limit $E=5$ from below because of the reduced oxygen
   content, creating thereby an artificial gap in the presentations of 
   Fig.~\ref{dopita.1}.

   For a full appreciation of this figure it is important to realise that
   the excitation $E$ does not only depend on the stellar temperature but also on
   the nebular density.  Lower densities favour higher excitations for a given
   stellar temperature.  Therefore, the models around the 0.605 \Msun\ central star,
   which are less dense than the models around the 0.696 \Msun\ star and become
   optically thin during the course of evolution, reach higher excitations than the
   0.696 \Msun\ models 
\changed{In spite of the fact that for the latter the central star achieves
         much higher temperatures (e.g. 240\,000 K instead of only 160\,000 K for the
         0.605 \Msun\ case).}  
   We showed in
   \citetalias{schoenetal.07} (Figs. 13 and 14 therein) that the highest
   degree of excitation (${E\approx 10}$) is only reached by optically very thin nebulae.
   An example is the ${\alpha=3}$, 0.595 \Msun\
   sequence shown in Fig.~\ref{dopita.1} (middle):  It does not recombine much and 
   reaches the highest expansion rates at the highest possible excitation level, as
   given by the He/H ratio which is 0.11 for our models.

\changed{Figure \ref{dopita.1} demonstrates that all objects start their expansion
         (at ${E\simeq0}$), as the models do, very slowly, with}
\changed{$V_{\rm HWHM}(5007)$ of only a few \kms.}
\changed{Since our models have AGB-wind flows with 10 or 15 \kms, these low values  
         are thus \emph{no} indication for a similarily low AGB-wind velocity.  We have 
         already explained above this phenomenon as the consequence of the ionisation 
         stratification (of oxygen) in combination with a steep velocity gradient within 
         the ionised shell during the early evolution of a PN (cf. Fig. \ref{HWHM}).}

\begin{figure*}[!th]
\includegraphics*[bb= 0.3cm 0.2cm 28cm 9.3cm, width=\textwidth]
                  {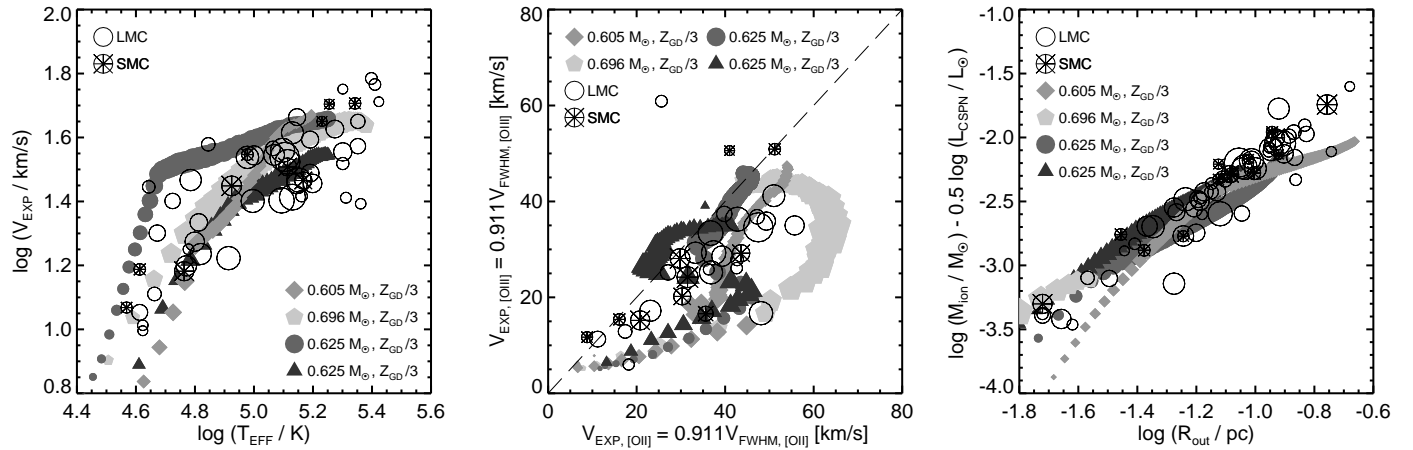}
\caption{\label{dopita.2}  
\changed{Various correlations for the same objects and model sequences as in 
         Fig. \ref{dopita.1}. The meaning of the symbols (PNe and models) is the same 
         as in Fig.~\ref{dopita.1}.  The triangles refer always to the new 0.625 \Msun\
         sequence introduced in the right panel of Fig. \ref{dopita.1}. All sequences
         are only plotted until maximum stellar temperature is reached, in order to
         avoid some confusion, and also because they become too faint. 
         \emph{Left panel}: nebular expansion from the \oiii\ 5007 \AA\ line width vs.
                            stellar temperature for Magellanic Cloud PNe compared with
                            the hydrodynamical model sequences.  The observed stellar 
                            temperatures are from the extensive photoionisation modelling 
                            by \citet{dop.91, dop.91b}. 
         \emph{Middle panel}: correlation between HWHM velocities derived from
                            [\ion{O}{ii}] 3726\,\AA\  and [\ion{O}{iii}] 5007\,\AA\ 
                             according to \citet{dop.88} and the hydrodynamical models. 
         \emph{Right panel}: \changedIII{reduced} nebular mass vs. radius relations for
                             Magellanic Clouds PNe and our hydrodynamical models. Data
                             again from \citet{dop.91, dop.91b}.} 
         }
\end{figure*}

   During the course of evolution the \changed{line widths increase},
   with some spread in terms of excitation and $V_{\rm HWHM}$ which is
   caused by the different combinations of central star and circumstellar envelope.
   Obviously, Fig. \ref{dopita.1} (middle) implies again that for all observed PNe
   the \emph{same mechanism for their formation and expansion must be at
   work, which, according to our models, is heating by ionisation and the immediate 
   creation of an expanding shell \changed{bounded by a shock front}. }

   During recombination, $E$ drops to intermediate values although the star remains
   very hot (${\teff \ga 120\,000}$ K).  The ionisation becomes stratified, and the O$^{2+}$
   is confined to the inner regions closer to the hot central star.  The outer shock
   is decelerated because of the post-shock pressure drop due to recombination
   while the inner region is still accelerated by the thermal pressure of
   the hot bubble  (shocked stellar wind gas) and somewhat later, by reionisation.
   Because of this, $V_{\rm HWHM}$ 
\changed{may decreases during recombination, but continues later to increase further}
   (cf. 0.605 \Msun\ sequence in Fig.~\ref{dopita.1}, middle panel).

\changed{From Fig. \ref{dopita.1} (middle panel) we see, however, that our model sequences 
         do not match 
         the observations perfectly, which, of course, is of no surprise since the
         models are not tuned to match objects of the Magellanic Clouds.  
         For the most relevant sequence, 
         0.625 \Msun\ and ${Z=Z_{\rm GD}/3}$, $E$ falls behind the $V_{\rm HWHM}$
         evolution, while the 
         0.605 \Msun\ sequence has the correct expansion property but remains 
         a little bit too faint at medium excitations. The 0.595 \Msun\ sequence stays
         always far too faint (cf. their small symbols) and is thus omitted for
         further considerations.
         The 0.696 \Msun\ sequence becomes somewhat too bright, although the $V_{\rm HWHM}$
         evolution is marginally right.}

\changed{We have already seen in Fig. \ref{loss.init} that a variation of the initially 
         chosen AGB-wind velocity leads to a corresponding variation of the PN expansion
         rate.  Hence, we expect that models with different AGB-wind velocities would
         be shifted horizontally in a diagram such as displayed in Fig. \ref{dopita.1}.
         To test this possibility, we used the 0.625 \Msun, ${Z=Z_{\rm GD}/3}$
         sequence with ${\dot{M}_{\rm agb}= 0.5\times 10^{-4}}$ \Mdot\ and 
         ${V_{\rm agb}=7.5}$ \kms. This combination of mass-loss rate and AGB-wind
         velocity ensures that the initial circumstellar density profile is the same as
         in the comparison sequence. However, because of the more slowly moving AGB material, 
         also $V_{\rm HWHM}$ increases now more in pace with $E$,
         although objects with the highest excitations are now not covered 
        (Fig. \ref{dopita.1}, right panel). } 

\changed{
 For comparison, we replotted here
         also the 0.625 \Msun\ sequence from the middle panel, supplemented by a corresponding
         metal richer sequence with $Z=Z_{\rm GD}$. One sees that just changing the metallicity
         will not lead necessarily to a better match with the observations.}

\changed{One concludes from this exercise that a mixture of models with somewhat lower 
         final mass-loss rates and lower AGB wind speeds and of models as they are used here
         would lead to a good match with the observed HWHM velocity-excitation 
         correlation of PNe in the MCs.}

\subsubsection{Stellar temperature and nebular expansion}

   Because of the dependence of $E$ on both stellar temperature and (mean)
   nebular density, it is not surprising that the observed sample cannot be
   described by a single evolutionary track.  A more direct comparison can be
   made if we correlate $V_{\rm exp}\ (= 0.991V_{\rm HWHM}$) of the nebula with the 
   effective temperature
   of the central star \citep[cf. Fig.~6 in][]{dop.91}.  We used the same sample
   shown in Fig.~\ref{dopita.1} and selected all those objects for which
   \citet{dop.91, dop.91b} provided stellar temperatures, based on detailed 
   photoionisation models.

  The use of the stellar temperature \changed{(if accurately known)} 
  as a proxy for the time evolution across
  the Hertzsprung-Russell diagram is useful since (i) it is a
  distance-independent quantity, and (ii) its range depends weakly on the
  stellar mass.  Furthermore, the two driving forces for the nebulae, the
  radiation field and the wind, depend directly on the stellar
  effective temperature.

\changed{Figure \ref{dopita.2} (left panel) is an adaption of the \citeauthor{dop.91}
         figure mentioned above and demonstrates again that the signatures of nebular 
         expansion, as measured by a suitable line width,} evolve with time (or stellar 
         effective temperature), and that this time evolution is correctly described 
         by our hydrodynamical model sequences.}
  Note that the theoretical sequences shown in Fig.~\ref{dopita.2} cover very
  different time spans, ranging from about 10\,000 years for 0.605 \Msun, through
  about 4000 years for 0.625 \Msun, to about 1000 years only for 0.696 \Msun.
  Yet the sequences are nearly `degenerate' if the stellar temperature is used as a
  proxy for the evolutionary progress across the Hertzsprung-Russell diagram. 

  A similar correlation \changed{between nebular expansion and stellar temperature} 
  was also found
  for Galactic PNe with double shell structures in our \citetalias{schoenetal.05a}
  where typical flow speeds of shell and rim gas were measured from spatially
  resolved high-resolution echelle spectra.

\subsubsection{Nebular ionisation stratification and expansion}  

  An indication of the ionisation structure and the velocity field of a PN can be gained 
  by measuring the HWHM velocities of different ions of the same element, viz.
  of O$^+$ and O$^{2+}$ \citep[e.g.][their Fig. 3]{{dop.88}}.  These authors found that
  velocities derived using O$^+$ are in most cases higher than those from O$^{2+}$.  
  We replotted their
  data in Fig. \ref{dopita.2} (middle), supplemented again by our dynamical models in 
  order to check whether our models are able to predict the same behaviour.

  We see from Fig.\,\ref{dopita.2} (middle) that the observed trend of the  expan\-sion 
\changed{signatures are, indeed, quite} well matched by our models, 
\changedIII{although the \oiii\ velocities of the younger models are somewhat lower 
            than observed.}
  At the beginning of the evolution all models behave similarily, with
  ${V_{\rm HWHM}([\ion{O}{ii}])> V_{\rm HWHM}([\ion{O}{iii}])}$, 
  because of the radial ionisation stratification of oxygen
  \emph{and} the positive velocity gradient. Later, if the models are
  optically thin and O$^{2+}$ becomes the main ionisation stage throughout the 
  model, both ions, O$^+$ and O$^{2+}$, trace about the same nebular regions, 
  albeit with different fractions, and hence
  ${V_{\rm HWHM}([\ion{O}{ii}])\simeq V_{\rm HWHM}([\ion{O}{iii}])}$.
  This transition occurs quite rapidly and produces the left turn of the model tracks  
  seen in Fig.\ \ref{dopita.2} (middle).  

  Note that the models around the 0.696 \Msun\
  central star remain optically thick throughout the whole evolution, 
\changed{with $V_{\rm HWHM}([\ion{O}{ii}])$ always greater than 
         $V_{\rm HWHM}([\ion{O}{iii}])$ until the star begins to fade.}
\changed{This sequence with its quite massive central star model, 0.696 \Msun, is
         obviously not an option for PNe in the Magellanic Clouds, also because, as we
         mentioned already above, the models become too bright.}

  The interpretation of Fig. \ref{dopita.2} (middle) is the following:  If 
  $V_{\rm HWHM}([\ion{O}{ii}]$ is significantly greater than $V_{\rm HWHM}([\ion{O}{iii}])$,
  the object in question is ionisation stratified and optically thick, irrespective of 
  the amount of $V_{\rm HWHM}$.  If both velocities are similar, the object is optically
  thin, which is the case for most of the MC objects plotted in 
  Fig. \ref{dopita.2} (middle).  
\changed{It appears, however, that there exists often a tendency for 
         $V_{\rm HWHM}([\ion{O}{ii}])$ of being greater than 
         $V_{\rm HWHM}([\ion{O}{iii}])$ by ${\approx\!10}$ \kms, on the average. This 
         behaviour is best matched by our 0.605 \Msun\ sequence during the optically thin
         phase. }

\subsubsection{Evolution of nebular masses}
\label{neb.mass}

\changed{As PNe expand into the AGB matter, their mass \emph{must}
         increase with time. This has been convincingly demonstrated by 
         \citet{dop.91, dop.91b} in their study of the Magellanic Cloud PNe. 
         The mass increase with time (or radius) is ruled by the speed of the (outer)
         shock and the density profile of the former AGB wind matter.
         A decrease of the ionised mass is possible during the recombination stage as
         the central star fades.} 
         
\changed{In the right panel of Fig. \ref{dopita.2} we compare the \citet{dop.91, dop.91b}
         data with our model predictions. 
         \changedIII{We used the \emph{reduced} mass as defined by \citeauthor{dop.91b}.}
         Before interpreting this figure one should say
         that nebular masses are difficult to determine, even if the distances are well known.
         Uncertainties of electron densities and radii can add up to quite large mass errors.
         Having this in mind one can only state that our simulations provide a rather 
         satisfactory agreement with the observations. }

\subsubsection{Nebular brightness and expansion}
\label{bright.expan}

\begin{figure*}[t]
\vskip-2mm
\sidecaption
\includegraphics[width=0.37\textwidth]{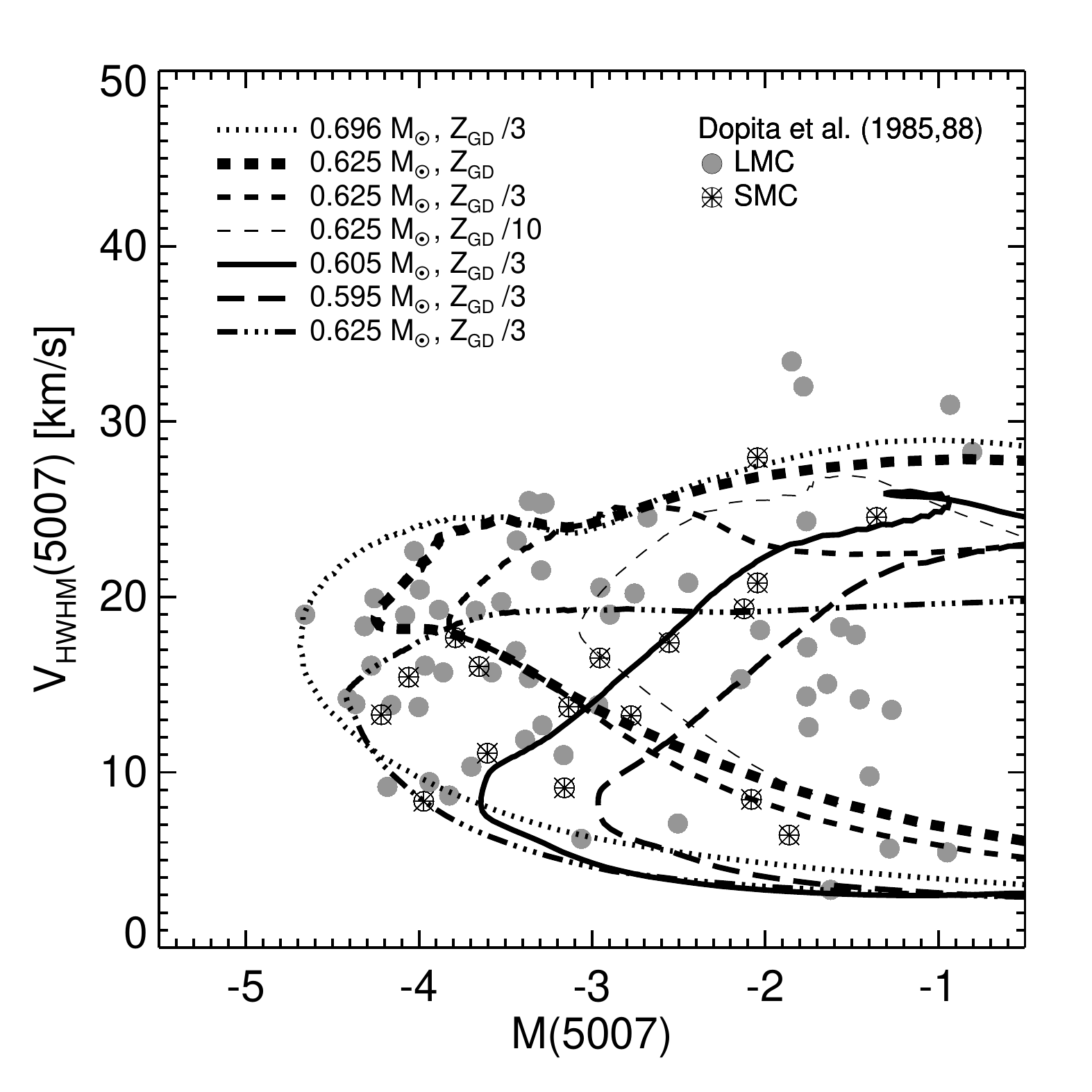}
\includegraphics[width=0.37\textwidth]{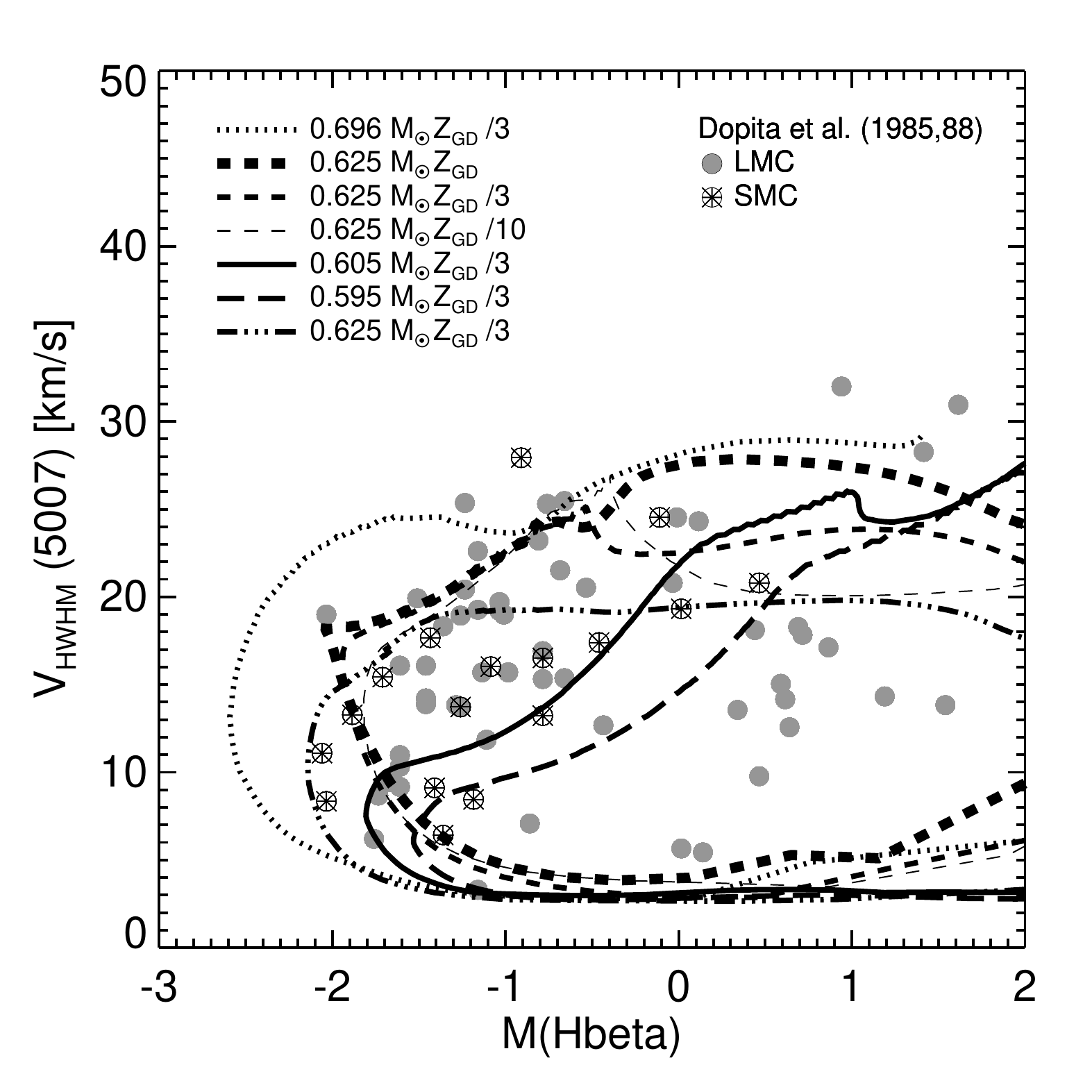}
\caption{Comparison of HWHM velocities vs. absolute magnitudes of Magellanic 
         Cloud PNe with different model predictions as indicated in the legends. 
\changed{The dashed-triple-dotted line correspond to the ${V_{\rm agb}=7.5}$ \kms\
         sequence.}
         The velocity measurements are again from \citet{dop.85, dop.88}, 
         but here $V_{\rm HWHM}$ is shown, instead of their $V_{\rm exp}$, 
         against $M(5007)$ (\emph{left}) and $M(\mbox{H}\beta)$ (\emph{right}).
\changed{The 5007 \AA\ fluxes are from \citet{jacoby.90}.}
         True distance moduli and extinction values $c$ used are 18.37 and 0.28 for the 
         LMC and 18.72 and 0.26 for the SMC \citep{dop.92}.
        }
\label{richer.MC}
\end{figure*}

  We conclude our treatment of  Magellanic Cloud PNe by applying
  $V_{\rm HWHM}$-magnitude diagrams already introduced and discussed before 
  in Sect. \ref{expansion.distant}. We thus collected the velocity information on
  MC PNe from \citet{dop.85} and \citet{dop.88}, converted their `expansion velocity`
  values, $V_{\rm exp}$, in order to be consistent with, e.g., Figs. \ref{vel.richer}
  and \ref{vel.hbeta},
  to the ones used here (${V_{\rm HWHM} = V_{\rm exp}/1.822}$), and plotted
  $V_{\rm HWHM}(5007)$ against $M(5007)$ and $M({\rm H}\beta)$, respectively 
  (Fig. \ref{richer.MC}).  
\changed{The model sequences selected are the same as used in the previous figures,
         supplemented by sequences with lower ($Z_{\rm GD}/10$) and higher
         metallicity ($Z_{\rm GD}$).}
\changed{Note that no unique relation exists between $M(5007)$ and 
         $M({\rm H}\beta)$: Each sequence has its own relation, depending on central star
         mass and corresponding nebular ionisation.}

\changed{The left panel of Fig. \ref{richer.MC} is the same as Fig. Fig. \ref{vel.richer} 
         but only for Magellanic Clouds objects, the right panel is a virtually 
         metal-independent presentation. This is seen by the close similarity between the
         ${Z=Z_{\rm GD}}$, $Z_{\rm GD}/3$, and $Z_{\rm GD}/10$ 0.625 \Msun\ sequences.}
  Again, the observed distributions of the MC objects are very well matched by our
  model sequences.  Several facts are evident from Fig. \ref{richer.MC}:
\begin{itemize}
\item  The brightest objects are obviously explained by models
       with more massive central stars (${\simeq\!0.6}$--0.7 \Msun) and optically 
       thick (or partly thick) nebulae, while objects with medium brightnesses
       need optically thin models with less massive
       central stars ($\la\!0.6$ \Msun).
\item  \changed{The $M(5007)$ peak brightness of the SMC sample appears to be fainter
                than that of the LMC by about 0.2 mag, obviously because of the lower
                LMC metallicity (more precisely, oxygen abundance).}
       This difference is not seen for $M({\rm H}\beta)$, as one would expect.  
       The bright cut-off occurs at $M({\rm H}\beta)\simeq -2$ mag for both clouds.
\item  \changed{The model sequence with ${V_{\rm agb}=7.5}$ \kms\ can explain the
        bright objects with low $V_{\rm HWHM}(5007)$,
       but fails to explain those PNe with $V_{\rm HWHM}(5007)\ga 20$ \kms.}
\item  The mean $V_{\rm HWHM}(5007)$ velocities for all objects of the \citeauthor{dop.85}
       samples are $\simeq$17 \kms\ (LMC) and $\simeq$15 \kms\ (SMC).  Considering 
       that the SMC sample is rather small and that the (formal) standard deviations are 
       about 6 \kms, we conclude that both mean values are virtually equal within
       the errors, in good agreement with our simulations.
\item  Looking only at objects within the brightest magnitude interval, those
       which are the brightest ones in H$\beta$\ \changed{have, on the average,
       a \emph{lower} halfwidth velocity} than the corresponding \oiii\ bright objects:  
       The mean $V_{\rm HWHM}(5007)$ 
       for the former is about 12 \kms\ (right panel of Fig. \ref{richer.MC}), 
       while the mean for the latter is about 18 \kms\ (left panel).
       The reason for this difference is simply the fact that (i) a PN reaches
       its maximum \hb\ brightness well before the maximum in 5007 \AA\ is
       reached \citepalias[compare Figs. 9 and 11 in][]{schoenetal.07}, at a phase 
       where the expansion is still rather slow, and that (ii) most PNe become optically
       thin and thus fainter in \hb\ shortly afterwards.  We remind the reader that the
       nebular expansion rate generally increases with time, \changed{especially during the
       early, optically thick stage. }
\end{itemize}

\section{Discussion}\label{disc}
  The work presented here is the first where modern radiation-hydrodynamics simulations
  of the evolution of planetary nebulae are systematically applied for objects in
  distant stellar populations.  To this end,
  we present a set of 1D-radiation-hydrodynamics simulations which are 
  especially tailored to study the formation and evolution of planetary nebulae 
  in galaxies and to interprete the observations.   
  The model sequences cover a wide range of metallicity in terms of abundance
  distributions scaled with respect to that which is typical for the Galactic disk.
  The different abundance patterns generally found within a galaxy are thus neglected here.
  Important metal-dependent processes (e.g. line cooling) are considered in detail, 
  but the dependence of the central-star wind on metallicity, which is of  
  relevance for the PN shaping process, could only be treated approximately. 
  Direct consequences of the stellar chemistry on details of the post-AGB evolution
  are not considered. 

  The initial conditions in terms of AGB mass-loss rate and wind speed, and the
  range of masses considered for the central stars, are chosen 
  such that (i) the emerging model structures mimic closely the observations known
  for Galactic disk objects, and that (ii) the bright end of the PN luminosity function
  can be explained by optically thick model nebulae around central-star models with
  $\simeq$0.63 \Msun.  Typical values needed for the final AGB mass-loss rates are 
  ${\dot{M}_{\rm agb}\simeq10^{-4}}$ \Mdot\ with wind velocities between about 
  10--15 \kms.   Values of about $10^{-4}$ \Mdot\ for the final AGB mass-loss rates 
  from four well-known Galactic PNe have recently been derived by \citet{sandetal.08}
  using integral field spectroscopy of the respective PN haloes.

  In order to ease the comparison of our models with distant PNe which are usually 
  (except for the Galactic bulge) not spatially resolved, observable quantities are 
  integrated over the whole model nebula.  Emphasis is given to the extraction
  of velocity information from the half widths of emission line profiles 
  ($V_{\rm HWHM}$).
  \citet{MoSt.08} showed recently by means of non-spherical model nebulae that  
  HWHM-velocities derived from the profiles of volume integrated
  emission lines are expected to be a robust measurement in 
  the sense that these velocities do not depend much on morphology and orientation
  of the model, and also not much on the details of the velocity law.
  Spherical models as they are used here appear thus fully justified for deriving
  mean properties of a larger sample of objects.

  We emphasise that all existing efforts to model 
  axisymmetric PN structures by fast collimated winds (jets) and/or magnetic fields 
  are using initial AGB wind 
  configurations that are far from reality, at least for the typical PNe 
  discussed here.  The initial AGB-wind densities employed are always 
  very low (typical values correspond to ${\dot{M}_{\rm agb} \approx 10^{-6}}$ 
  \Mdot) so that thermal pressure is dynamically unimportant \citep[see, e.g.,][]{AS.08}.
  However, as we have mentioned above, typical PN structures require final AGB-wind 
  mass-loss rates of about $10^{-4}$ \Mdot, higher by about two orders of magnitude than 
  assumed in current jet modelling.   Additionally, the inner boundary conditions
  are fixed, i.e. no allowance for an evolving star and wind is made, although
  the simulation times often become comparable to the star's transition time from the
  tip of the AGB towards the PN region ($\approx\! 10^3$ years).

  We close the discussion with the warning that our simulations are exclusively based
  on the properties of hydrogen-burning central-star models and should not be used
  to explain the evolution of PNe around hydrogen-deficient (i.e. Wolf-Rayet) central
  stars.  The samples used here for comparisons contain certainly an unknown fraction
  of such objects because it is not possible or feasible to sort them out.
  The mere fact that they do not seem to disturb the correlations found in this work
  may indicate that the fraction of PNe with Wolf-Rayet central stars 
  is rather small and/or that their evolution is not much different from that of their
  counterparts with a hydrogen-rich central star.  There exists also the possibility
  that a different evolution of Wolf-Rayet central stars is hidden because of 
  inappropriately chosen observables such as $V_{\rm HWHM}$.

\section{Conclusions}\label{concl}

  Our extensive radiation-hydrodynamics simulations presented in this work show clearly
  that, contrary to common belief, wind interaction is not the main driving force for 
  the formation and expansion of the bulk matter in planetary nebulae around
  hydrogen-burning central stars:  
  The metal-poorest models with the weakest stellar wind expand fastest.
  Instead it is the pressure gradient built up by ionisation and heating of the
  circumstellar gas that drives a D-type shock through the ambient medium and creates
  thereby an expanding circumstellar structure consisting of different shells,
  called a PN.  So to speak, it is the
  radiation field of the star, not its wind, that is the main agent for formation
  and evolution of a PN.  The overall shape of a PN is then controlled by the matter
  distribution set up by the wind \emph{before} ionisation starts.
  Possible processes have been proposed by \citet{speck.06}.   The central-star wind
  is only necessary to prevent the circumstellar matter from
  falling back to the stellar surface and may compress the inner nebula to a
  denser shell, called the rim, when the wind power becomes strong enough at higher
  metallicities. 

  Although these wind compressed rims are often the most
  conspicuous parts of a PN, they contain only a very small fraction of the total
  nebular mass.  The outer boundary of the rim, i.e. the rim's shock, separates then
  the bulk of nebular matter which is thermally expanding.
  Using a stellar wind under conditions typical for Galactic disk objects, the rim gains
  no more than about 20\% of the total ionised nebular matter at the end of the
  horizontal evolution.  Only once recombination in the shell has started, the 
  (ionised) mass contribution of the rim becomes much higher and may even become
  dominant.  At low metallicities, 
  the rim becomes unimportant at all and loses its typical signature. 

  Based on our extensive modelling and existing observations 
  the basic properties of PNe in different environments can be summarised as follows: 
\begin{itemize}
  \item  The metal content has a profound influence on the expansion properties and on
         the development of the overall morphology.  For low metallicities, the expansion
         is fast due to the high electron temperatures while the wind interaction 
         generated from the central star is weak, leading to an only relatively low
         bubble pressure; hence low-metallicity PNe are expected to be rather extended. 
         On the other hand, at high metallicities, expansion
         is slow and wind pressure strong; hence we see a much (geometrically) thinner 
         structure with a prominent rim which is clearly separated from the outer
         shell matter.  According to our models, the transition where the stellar wind
         becomes \changed{less important} for shaping the inner parts of a PN occurs 
 \changed{at about or slightly below $Z_{\rm GD}/10$, depending on the initial parameters of
          the models.}

  \item  \changed{We showed that low-metallicity PNe are prone to deviations from thermal 
                  equilibrium because of their reduced line-cooling efficiency and fast
                  expansion.  This may already occur at about 1/10 of the solar
                  (or Galactic disk) abundance distribution.
                 Studies employing classical photoionisation codes may become problematic
                 in these cases.}

  \item  In contrast to thermal equilibrium, the ionisation/recom\-bi\-nation equilibrium  
         is \changed{always} rather well established, except possibly for a very brief 
         phase during the rapid stellar fading after the `horizontal' evolution through 
         the HR diagram. 
  \changed{Independent of metallicity, the halo of a PN is generally far from thermal
         \emph{and} ionisation equilibrium.}

 \item   Although the influence of metallicity on expansion property and density distribution
         is considerable, the half widths of (volume) integrated emission line profiles are 
         rather insensitive to the metal content.  The increase of the HWHM velocity, 
         $V_{\rm HWHM}$ (in, e.g., the 5007 \AA\ \oiii\ line) is initially due to
         the growing O$^{2+}$ zone encompassing faster moving gas from the outer 
         nebular regions, but later due to the generally increasing expansion speed 
         as a consequence of sound speed behaviour
         and upstream radial density gradient. The evolution with time, or brightness, is such
         that all optically thick (or nearly optically thick) models used in this study 
         reach, irrespective of their 
         metallicity, a HWHM velocity of about 18 \kms\ at their maximum \oiii\ brightness.
         Up and down shifts of this value are possible by changing the original
         AGB wind speed accordingly.
        
         Our model predictions are in remarkable agreement with recent observations
         presented by \changedIII{\citet[][]{richeretal.10b}} which show 
         a very limited range of 
         the mean HWHM velocity, $\simeq$15--20 \kms, for a sample of the brightest
         PNe drawn from galaxies of the Local Group with quite different ages and 
         metallicities.  According to our modelling, these brightest, optically thick 
         PNe should have typical central-star masses of ${\simeq\!0.65}$ \Msun, with 
         final AGB mass-loss rates of ${\simeq\!10^{-4}}$ \Mdot.

         The few investigated bright PNe in the Virgo cluster show the same behaviour
         as those in the Local Group: their mean HWHM expansion 
         (in 5007 \AA) is $\simeq$16.5 \kms, implying that also the same parameters for
         the central stars and the final mass-loss rates hold for the bright PNe
         in the Virgo cluster.

 \item   It could be shown by our simulations that a more sensitive method to determine
         the PN kinematics must rely on measuring the expansion from the outermost wings
         of the line profile, as already proposed by \citet{dop.85, dop.88}.  Only in
         this case it is possible to get information about the fastest moving parts of
         a PN, usually those immediately behind the shock whose speed is also much more 
         sensitive to the metal content than the HWHM velocity.  This method requires, of
         course, high-quality line profiles which are still difficult to get for more distant
         stellar systems. 

 \item   The run of \changed{nebular expansion} with evolution (i.e. with time) 
         cannot be determined by correlating a \changed{velocity measurement} 
         just with nebular (absolute) magnitude.  One needs 
         instead a parameter which depends directly on the evolutionary status of the 
         central star, e.g. $\teff$.  If $\teff$ is not available, or too difficult to
         determine, the nebular excitation, $E$, can be used as an distant independent 
         proxy of the stellar effective temperature.

         The $V_{\rm HWHM}$ evolution seen for the PNe in the Magellanic Clouds, 
         if plotted over $E$ or $\teff$, are fully explained by our model sequences with the
         appropriate metallicity and with central star masses between 
         $\approx$0.6 \ldots 0.7 \Msun: $V_{\rm HWHM}(5007)$ ranges in both clouds from 
         about 5 to about 30 \kms, but the brightest PNe (in 5007 \AA) have 
         HWHM velocities between 15 and 20 \kms\ only.\footnote{The Magellanic Clouds 
         are included in the \changedIII{\citet{richeretal.10b}} sample of galaxies.}
         Close to maximum \hb\ the range is wider, $\simeq$5 \ldots $\simeq$20 \kms, 
         with a mean value of about 12 \kms, again in full agreement with our model
         predictions. 

         A similar velocity evolution with stellar effective temperature (i.e. with time) 
         was also found for Galactic disk objects; see  
\changed{\citetalias[][Fig. 12 therein]{schoenetal.05a}, and also \citet{SS.06}.} 
         A direct comparison, however, is
         impossible because the velocities measured are based on
         spatially resolved line profiles and refer to the outer parts of the shell.
         They are usually greater than the HWHM velocities of volume-integrated
         line profiles.

         Also \citet{richeretal.08} found recently the signature of a modest velocity 
         evolution by studying the Milky Way bulge PNe: the median line width in \oiii\
         5007 \AA\ is 17.2 \kms\ for the cooler central stars, but 23.7 \kms\ for the
         hotter, more evolved ones.  The standard deviations of both distributions are 
         5.1 and 7.6 \kms, respectively.  Note however that most of the bulge PNe are
         to some extent spatially resolved.

 \item   Another important result is that the HWHM velocities always \emph{underestimate}
         the true expansion speed of a planetary nebula.  The amount depends on the
         evolutionary stage, but for objects near their maximum \oiii\ brightness,
         the real speed is about a factor of two higher (cf. Fig. \ref{factor2}).

 \item   The nearly identical expansion behaviour, 
\changed{(in terms of the line width $V_{\rm HWHM}$)} 
         observed for the bright PNe in stellar 
         populations which differ in age and metallicity allows the conclusion
         that \changedIII{most likely}
         only one basic (and rather simple) physical mechanism is responsible 
         for driving the formation and expansion!   According to our models this 
         is always photoionisation by the stellar radiation field. 
 \changedIII{The stellar wind, as it is modelled here, does not play a dominant r\^ole.
         Also, it does not appear very likely that}         
         more `exotic' scenarios like disk accretion and
         jets, or binary and common envelope evolution which are heavily debated,
 \changedIII{are} of major importance, at least for the  bright PNe with more massive
         central stars and envelopes.        
\end{itemize}

   It has already be mentioned in the discussion above that our models
   of planetary nebulae are entirely based on radiation-hydrodynamics simulations of 
   wind envelopes around hydrogen-burning central-star models.  Some of these conclusions 
   may not apply for planetary nebulae around hydrogen-poor, i.e. Wolf-Rayet, central stars. 
\changedIII{For instance, the winds of Wolf-Rayet central stars are much more powerful 
            which makes them very important for the formation and evolution of PNe.}}
   However, since the formation and evolution of hydrogen-poor central stars is still 
   unclear, any detailed  simulations of the sort as reported here are premature.

\begin{acknowledgements}
  C. S. acknowledges support by DFG grant SCHO 394/26. We thank Dr. A. Monreal-Ibero
  for providing us with the observational data for NGC 4361.  We are especially grateful to 
  Dr. M. Richer who provided us with data necessary to compare our models with the
  observations. Dr. Richer acted also as referee, and we are especially grateful for
  his constructive criticism which led to a significant improvement of our presentation.
\end{acknowledgements}

{}

\end{document}